%% file: thesis.tex
\documentclass[12pt,a4paper]{report}
\usepackage{amsmath,amssymb}
\usepackage{fixmath}
\usepackage{graphicx}
\usepackage{epsfig}
\usepackage[font=footnotesize,labelfont={sf,bf},textfont=sf]{caption}
\usepackage[usenames]{color}
\usepackage[perpage,symbol*]{footmisc}

\DefineFNsymbols*{alg}{\dagger\ddagger\S\P\|{**}{\dagger\dagger}{\ddagger\ddagger}{\S\S}{\P\P}}
\setfnsymbol{alg}
\usepackage{amsfonts}
\def\slash#1{\setbox0=\hbox{$#1$}#1\hskip-\wd0\hbox to\wd0{\hss\sl/\/\hss}}
\usepackage{epsf, cite}
\usepackage[all]{xy}
\usepackage{latexsym}


\let\non\nonumber

\def\lbldef#1#2{\expandafter\gdef\csname #1\endcsname {#2}}

\def\href#1#2{#2}

\newcommand{\beq}{\begin{equation}}
\newcommand{\eeq}{\end{equation}}
\def\wv{worldvolume }

\def\bea{\begin{eqnarray}}
\def\eea{\end{eqnarray}}
\newcommand{\pd}[2]{\frac{\partial{#1}}{\partial{#2}}}

\def\ct{\tilde{C}}

\def\gs{\,\raise.15ex\hbox{/}\mkern-11.5mu G} 
\def\cs{\,\raise.15ex\hbox{/}\mkern-11.5mu C} 

\def\lt{\lfloor}
\def\rt{\rfloor}
\newcommand{\tb}[3]{[X^{#1},X^{#2},X^{#3}]}
\newcommand{\gtb}[3]{[H^*,X^{#1},X^{#2},X^{#3}]}
\newcommand{\fbr}[5]{\lt X^{#1},X^{#2},\tb{#3}{#4}{#5}\rt}

\def\onalign#1{\leavevmode\vtop{\baselineskip=0pt\lineskip=-1.2ex
\ialign{##\crcr#1\crcr}}}
\def\ellbar{\onalign{\hidewidth$ \ell$\hidewidth\cr\cr$\mathchar'26$}}

\def\p{\partial}

\def\cPr{ { {\cal{P}}_{\cal{R}} }  }
\def\Pr{ { {\cal{P}}_{\cal{R}} }  }
\def\Prp{ { {\cal{P}}_{\cal{R^+}} }  }
\def\Prm{  {\cal{P}}_{ {\cal{R^-}} }  }
\def\cPpm{  {\cal{P}}_{ {\cal{R^\pm}} }  }
\def\cRp{ { { \cal{R}}^+}  }
\def\cRm{ { {\cal{R}}^-}  }
\def\cRpm{ { {\cal{R}}^\pm}  }
\def\cRmp{ { {\cal{R}}^\mp}  }
\def\cR{ {\cal{R}}  }

\def\Pb{\bar{P}}
\def\Gb{\bar{\Gamma}}
\def\cPb{  {\bar{\cal{P}}}  }
\def\cRb{  {\bar{\cal{R}}}  }
\def\eb{\bar{e}}
\def\gb{\bar{G}}
\def\gh{\hat{G}}
\def\fs4{ fuzzy $S^{4}$}
\def\fs3{ fuzzy $S^{3}$}
\def\fs2{ fuzzy $S^{2}$}
\def\mnc{$Mat_{N}(\mathbb{C})$} 
\def\mncs{$Mat_{N}(\mathbb{C})$ }
\newcommand{\mc}[1]{$Mat_{#1}(\mathbb{C})$ }
\newcommand{\mcp}[1]{$Mat_{#1}(\mathbb{C})$}

\def\anst{${\cal A}_n(S^3)$ }
\def\anstp{${\cal A}_n(S^3)$}
\def\ansf{${\cal A}_n(S^4)$ }
\def\ansfp{${\cal A}_n(S^4)$}

\providecommand{\openone}{\leavevmode\hbox{\small1\kern-3.8pt\normalsize1}}

\def\m{\begin{matrix}}
\def\em{\end{matrix}}

\def\rep{representation }
\def\reps{representations }
\def\irrep{irreducible representation }
\def\irreps{irreducible representations }

\def\irrepsp{irreducible representations}

\def\lpb{\ellbar_{pl}}

\def\gne{G^{(11)}_N}

\def\xt{\tilde{X}}
\def\mt{\tilde{m}}
\def\wt{\tilde{w}}
\def\ut{\tilde{u}}
\def\vt{\tilde{v}}
\def\Lt{\tilde{L}}
\def\nt{\tilde{n}}
\def\ti{\tilde{I}}
\def\tj{\tilde{J}}
\def\ct{\tilde{c}}

\def\zb{\bar{z}}
\def\twb{\bar{\tau}}
\def\Z{{\mathbb Z}}
\def\mod{\ \mbox{mod}\ }
\def\Pit{\tilde{\Pi}}
\def\P{\mathcal P}
\def\Ph{\hat{\mathcal P}}
\def\ph{\hat{P}}
\def\Qh{\hat{Q}}
\def\Phih{\hat{\Phi}}
\def\P{\mathcal P}
\def\D{\mathcal D}
\def\H{\mathcal H}
\def\V{\mathcal V}
\def\X{\mathbb X}
\def\A{\mathcal A}
\def\L{\mathcal L}

\def\bx{\square}

\begin{document}

\title{\LARGE{{\bf Aspects of M-Theory Brane Interactions and String Theory Symmetries}}}

\author{
\Large{Neil Barclay Copland}\\[20pt]\Large{Trinity College, Cambridge}
\\[35pt] \includegraphics[width=1.8cm]{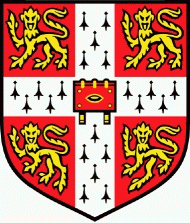}
\\[25pt] \Large{A Dissertation Submitted for the Degree of}\\
\Large{Doctor of Philosophy}
\\[20pt]
\Large{Department of Applied Maths and Theoretical Physics}\\
\Large{University of Cambridge}}

\date{\Large{\today}}

\maketitle

\pagenumbering{roman}

\section*{Summary}

\input{summary}


\newpage
\section*{Declaration}
This dissertation is my own work and contains nothing which is the outcome of work done in collaboration with others, except as specified in the text. It is based on the research presented in the papers \cite{bermancopland,BC2,BC3} written with David Berman. A plan of the structure of the thesis, including details of major references for each chapter is included at the end of Chapter 1. This dissertation has not been submitted in whole or in part for a degree at this or any other institution. 
\\
\\
\\
\\
Neil Copland

\cleardoublepage
\newpage
\section*{Acknowledgements}
\input{ack}
\newpage
\
\vskip 8cm 
\begin{center}
Not only is the universe stranger than we imagine, \\
it is stranger than we can imagine.\\
\it{
\qquad\qquad\qquad\qquad- Arthur Stanley Eddington}
\end{center}
\addtolength{\parskip}{-0.2cm}
\tableofcontents
\newpage
\addtolength{\parskip}{0.2cm}

\newpage
\pagenumbering{arabic}

\input{ch1}

\input{ch2}
\input{ch3}
\input{ch4}
\input{ch5}
\input{ch6}
\input{ch7}
\input{ch8}

\input{ch9}

\input{conc}

\renewcommand\bibname{References}

%


\addcontentsline{toc}{chapter}{Appendix}\appendix
\setcounter{section}{0}
\input{app1}
\input{app2}
\addcontentsline{toc}{chapter}{References}
\addtolength{\baselineskip}{-0.13cm}
\input{bib}
\addtolength{\baselineskip}{0.13cm}
\end{document}

%% file: summary.tex
The dissertation consists of two parts. The first presents an account
of the effective worldvolume description of $N$ coincident M2-branes
ending on an M5-brane in M-theory. It begins by reviewing M-theory,
the various viewpoints of coincident D-strings ending on a D3-brane in
type-IIB string theory and the M5 picture of the M2-M5 brane system
with which we are concerned. It then turns to Basu and Harvey's recent
description of the worldvolume theory of the M2-branes in terms of a
Bogomol'nyi equation, and its solution via a fuzzy
(three-) funnel. Tests of the consistency of this picture are then performed and many of the issues with it are addressed. First of all, the
picture is successfully generalised to describe M2-branes ending on
calibrated intersections of M5-branes. This is followed by a discussion
of how a refinement of the fuzzy three-sphere algebra used can lead to
the correct $N^{3/2}$ scaling of degrees of freedom for this system. A
reduction of this Basu-Harvey picture to the D1-string picture of the
D1-D3 intersection is then performed via constructing a reduction of
the fuzzy-three sphere to the fuzzy two-sphere. Along the way, a number of observations are made concerning the multiple M2-brane action and supersymmetry. 

The second part of the dissertation begins with a review of the doubled formalism of Hull, in which T-duality for torus fibrations has a clear geometric interpretation after the fibres are doubled. A constraint is needed to preserve the correct number of degrees of freedom and  this constraint is interpreted as a chirality constraint on the doubled co-ordinates. A holomorphic factorisation argument is then used to demonstrate quantum equivalence with the standard formalism by deriving the partition function, including instanton and oscillator sectors.

%% file: ack.tex
I am very grateful to my supervisor David Berman for always having time for me, for collaboration on the work contained in the thesis, for helping me through the vicissitudes of a PhD and for sharing his insight into the subtle mysteries of the five-brane.  For discussions related to work in this thesis I would like to thank Anirban Basu, M\aa ns Henningson, Neil Lambert, Costis Papageorgakis, Sanjaye Ramgoolam and David Tong. I would like to thank those other students in the department who have shared some part of the journey, and with whom I have shared discussions mathematical or otherwise; Andy, David, Hari, James, Jimmy, Jo\~{a}o, Joe, Julian, Kerim, Nagi, Paul and especially Bella for enthusiastic proof-reading. I would also like to thank Andrew Goodwin for many coffee sessions, semi-colons and \TeX\ tips. 

I am grateful to PPARC for funding my studentship and to Trinity College and DAMTP for providing travel expenses. 

Great thanks also go to my family, who have been nothing but supportive despite still not knowing what on earth it is I do. I am sure they would have supported me whatever path I chose. Greatest thanks of all to Sarah who has been there with love and support throughout, and who will never let me forget that there is more to life than squiggles.

%% file: ch1.tex
\chapter{Introduction}

String theory is seen as the leading contender for the quantum theory of gravity that fully describes our world. The fundamental building blocks are tiny strings. The particles of gravity and the standard model may be found as different modes of open and closed strings. The theory also contains D-branes, extended objects upon which the open strings can end. D-branes of the same kind can be stacked on top of each other, and these coincident stacks have their own effective theories on their worldvolumes.

The success of string theory is that it is a well-defined anomaly-free quantum theory which contains General Relativity as well as higher-order corrections. Another attraction of string theory is its uniqueness. After generalising away from the notion of point particles to extended strings almost all aspects of the theory are fixed by requiring consistency. However, for supersymmetric strings this includes fixing the dimension of space-time as ten. This obviously contradicts current observations, but the appearance of a four-dimensional theory can be achieved by compactifying on special manifolds. Unfortunately current methods of doing this lose much of the uniqueness and elegance of the theory. There are alternate ways of getting a four-dimensional theory, such as the `braneworld' scenario where our universe resides on the four-dimensional worldvolume of a three-brane.

Although these phenomenological approaches are important, we will be concerned with the complementary approach of addressing the deeper understanding of the fundamental theory. This can often utilise and provoke discoveries in complex and diverse areas of mathematics. 

String theory is defined as a perturbative expansion, and although such perturbative theories can yield extremely accurate results when the expansion parameter is small, without evaluating the whole infinite expansion we cannot know the whole picture. Also, though we have touched upon the uniqueness of string theory, there are actually five different consistent string theories. These issues were overcome by the discovery of M-theory, a non-perturbative 11-dimensional theory which contains the five string theories as limits. The fundamental objects of M-theory are membranes (M2-branes) which have 3-dimensional worldvolumes. There are also six-dimensional five-branes (M5-branes) which are dual to the membranes. All string theory strings and D-branes can be obtained from these M-branes (or from the geometry of compactification), making the M-branes the most fundamental objects that we have, however, there is still much to be learned about them.

In string theory it is understood that open strings stretching between D-branes become massless when the D-branes coincide. For $N$ coincident D-branes this leads to a non-Abelian gauge theory with $N^2$ degrees of freedom on the worldvolume of the branes. However, for $N$ coincident membranes the worldvolume theory is unknown, and there is no picture of what the degrees of freedom are, though scattering calculations have indicated the number of them should scale like $N^{3/2}$. Similarly little is known about the theory on coincident five-branes.

M2-branes can end on M5-branes, just as fundamental strings can end on a D-brane. These interactions can be considered from various viewpoints, including the effective theories on the different brane worldvolumes. As these theories are effective they have restricted ranges of validity, but if we take a large number of coincident membranes the regimes in which the theory on each brane is valid should overlap, allowing comparison of the single five-brane theory with the multiple membrane theory. The five-brane worldvolume picture contains a string-like soliton, which is identified with the boundary of the membrane on the five-brane. Recently Basu and Harvey put forward a proposal for the membrane worldvolume theory, giving a Bogomol'nyi equation, along with its solution in terms of a `fuzzy funnel' describing the M2-branes' worldvolume opening up onto an M5-brane\cite{BasuH}. This generalised the better understood D1-D3 intersection of IIB string theory and passed a number of consistency checks. 

However, many of these tests were satisfied by construction, and the picture had $N^2$ rather than the expected $N^{3/2}$ degrees of freedom. In the first part of this dissertation we perform a further test of the validity of the Basu-Harvey picture by extending it to calibrated intersections of five-branes, and reducing the solution to the D1-D3 case. We also argue that by using a more appropriate algebra for the fuzzy funnel we can reproduce the required scaling of degrees of freedom. Along the way we will make observations concerning possible multiple membrane actions and supersymmetry, as well as providing a reduction of the fuzzy three-sphere to the fuzzy two-sphere.

The development of M-Theory was partly led by the discovery of dualities which related the five consistent versions of string theory to each other. S-duality can relate one of the theories at strong coupling to itself at weak coupling, allowing access to regimes where perturbation theory is not normally valid. T-duality, on the other hand, can relate theories compactified on different manifolds to each other. A simple example is that type IIA string theory compactified on a circle of radius $R$ describes the same physics as type IIB string theory on a circle of radius $1/R$, after the exchange of momentum modes (which are quantised in a periodic direction) with winding modes (where the string wraps around the periodic dimension). These winding modes do not exist in ordinary field theory, so T-duality is an important extra feature of string theory. The connections provided by these dualities were suggestive of an underlying theory, subsuming the five consistent versions of string theory.

It is T-duality that we will be concerned with in the second part of the dissertation. More specifically, the doubled formalism introduced by Hull\cite{Hull:2004in}. In this formalism the number of fibres of a torus bundle are doubled in a particular way. For the above example of a periodic direction of radius $R$, a dual co-ordinate of radius $1/R$ is introduced. T-duality would then interchange these two co-ordinates. However, this involves a doubling of the degrees of freedom and a constraint must be introduced to preserve the correct counting. In a certain basis the constraint is a chirality constraint on the co-ordinates, and further, on a Euclideanised worldsheet it can be interpreted as a holomorphicity constraint. The formalism gives a more geometric picture of how T-duality works, and is useful for describing T-folds, which are generalisations of manifolds where the transition functions are allowed to include T-duality transformations. These T-folds are an interesting new source of string compactifications.

Hull showed classical equivalence of the formalism to the standard formulation\cite{Hull:2004in} and then quantum equivalence by gauging a symmetry associated to a conserved current\cite{Hull:2006va} (a current whose vanishing implied the constraint). Here we demonstrate quantum equivalence by calculating the doubled partition function and using the interpretation of the constraint as a holomorphicity one to apply it using holomorphic factorisation techniques. We find that we are required to include a topological term and to use an unconventional normalisation, both of which were also needed by Hull to show quantum equivalence.

If string theory does describe our world, M-theory branes would be the most fundamental building blocks we know of, and knowledge of how they interact would be vitally important to our understanding of the world at the most basic level. Similarly, symmetries have been an immensely important cornerstone of modern physics, and are extremely important to string and M-theory. A deeper understanding of T-duality symmetry would give much insight into stringy effects not known in field theory, as well as new 4-dimensional theories through compactifications. Full understanding of these M-brane worldvolume actions and T-duality symmetries are important steps to understanding the true nature of string and M-theory.

The structure of this dissertation is as follows. In Chapter \ref{Ch:Mth} we introduce M-theory, including membranes, five-branes and its relation to string theory and dualities. This is followed by Chapter \ref{Ch:insect} in which we examine branes ending on other branes from various perspectives, in particular the fuzzy funnel description of D1-branes ending on a D3-brane\cite{CMT} and the extension to calibrated intersections of D3-branes\cite{CL}. Chapter \ref{BasuHarvey} is an exposition of Basu and Harvey's proposed equation for membranes ending on a five-brane and the properties of its solution\cite{BasuH}. In Chapter \ref{Ch:calib} we generalise their work to calibrated intersections of five-branes, as first reported in \cite{bermancopland}, and Chapter \ref{Ch:FSH} details how a projection to fuzzy spherical harmonics on the fuzzy funnel can give the desired scaling of degrees of freedom for coincident membranes\cite{BC2}. The first part concludes with details of the reduction to the D1-D3 system in Chapter \ref{Ch:fuzred} (also based on \cite{BC2}). The second part of the dissertation begins with a description of Hull's doubled formalism\cite{Hull:2004in,Hull:2006va} in Chapter \ref{Ch:DF}. Then in Chapter \ref{Ch:Holo} we show quantum equivalence of the doubled formalism to the standard formulation for a periodic Boson of constant radius $R$ using holomorphic factorisation techniques, as performed in \cite{BC3}. The brief final chapter contains conclusions and suggestions for future work. Also included is a large appendix detailing the construction and various aspects of fuzzy spheres.

%% file: ch2.tex
\chapter{M-theory}\label{Ch:Mth}

After making the conceptual leaps to allow more than four space-time dimensions and more than one worldvolume dimension (that is, stepping up from a 1-dimensional particle worldline to a 2-dimensional string worldsheet) it is natural to ask if there are any higher dimensional objects in different dimensional spacetimes that we can describe.

A powerful constraint is supersymmetry. Demanding worldvolume supersymmetry on these higher dimensional ``branes" constrains which branes are allowed in which dimension, and more importantly if we desire no massless particles of spin greater than 2 then spacetime supersymmetry fixes the maximal dimension of spacetime as 11.

Supergravity can be formulated in 11 dimensions, and it has $N=1$ supersymmetry. From it all lower dimensional supergravities can be obtained by dimensional reduction. This 11-dimensional supergravity has membranes as its fundamental extended objects. These membranes are dual to solitonic five-branes. The theory also contains pp-wave and Kaluza-Klein monopole solutions. 

Membrane and five-brane worldvolume actions can be formulated, though writing a covariant action for the five-brane proved difficult due to the two-form field with anti-self-dual field strength which is part of the tensor multiplet propagating on its worldvolume.

Importantly, the double dimensional reduction of the 11-dimensional supermembrane is the fundamental string in type IIA string theory, as can be seen by reducing the worldvolume action. In fact all string theory D-branes can be found by dimensionally reducing one of the 11-dimensional branes or the Kaluza-Klein monopole.
Many connections and dualities were found between the five consistent string theories and 11-dimensional supergravity,  leading to the conclusion than these were all aspects of the same theory, which was non-perturbative and 11-dimensional. The low energy effective action of this theory is that of 11-dimensional supergravity, and perturbatively expanding around certain points in its moduli space leads to the 10-dimensional string theories. This non-perturbative theory is the mysterious, and mysteriously named, M-theory. It is conjectured to be the well defined quantum theory,  which includes gravity, that describes our world. Although some details are known, there is very little knowledge of what the microscopic theory should be. It is our aim to shed some more light on this theory.

\section{11-Dimensional Supergravity}

The field content of 11-dimensional supergravity\cite{CJS} consists of the metric $g_{\mu\nu}$, a rank 3 anti-symmetric tensor field $C_{\mu\nu\rho}$ and a 32 component Majorana gravitino $\Psi^\alpha_\mu$. These have 44, 84 and 128 physical degrees of freedom respectively. The Lagrangian is given by
\bea\label{11dsugraL}
I_{11}&=&\frac{1}{2\kappa_{11}^2}\int d^{11}x\sqrt{-g^{(11)}}\left[R-\frac{1}{2.4!}G^2-\frac{1}{2}\bar{\Psi}_\mu\Gamma^{\mu\nu\rho}D_\nu(\Omega)\Psi_\rho\right.\nonumber\\ 
& &\left.-\frac{1}{192}\left(\bar{\Psi}_\mu\Gamma^{\mu\nu\rho\lambda\sigma\tau}\Psi_\tau+12\bar{\Psi}^\nu\Gamma^{\rho\lambda}\Psi^\sigma\right)G_{\nu\rho\lambda\sigma}\right]\nonumber\\ 
& & -\frac{1}{12\kappa_{11}^2}\int C\wedge G\wedge G +\mbox{terms quartic in} \ \Psi 
\eea
where $G=dC$ is the field strength of $C$ and $\Omega_\mu^{ab}$ is the spin connection, which appears in the covariant derivative $D_\nu(\Omega)\Psi_\rho=\left(\partial_\nu-\frac{1}{4}\Omega_\nu^{ab}\Gamma_{ab}\right)\Psi_\rho$. 

The equation of motion for the 3-form potential $C$ can be re-written in the form
\beq
d(*G+\frac{1}{2}C\wedge G)=0 \qquad 
\eeq
where $*G$ is the Hodge dual of $G$. This has the form of a Bianchi identity and we can identify $*G+C\wedge G/2$ with $dC^{(6)}$ where $C^{(6)}$ is a 6-form potential and the dual of $C$. The field strength of $C^{(6)}$ is $G^{(7)}=*G=dC^{(6)}-C\wedge G/2$. The appearance of $C$ in this field strength makes  reformulation of the action in terms of only the dual field strength difficult. The existence of 3- and 6-form potentials was suggestive of extended objects with 3 and 6 space-time dimensional worldvolumes, even before the ``D-brane revolution"\cite{Polchinski:1995mt} in string theory.

\section{Reduction to the Type IIA String Theory Effective Action}\label{sec:Mred}

We can compactify 11-dimensional supergravity on a circle of fixed radius in the $x^{10}=z$ direction\cite{HuqN, GianiP}. From the 11-dimensional metric we obtain the 10-dimensional metric, a vector field and a scalar (the dilaton). The 3-form potential leads to both a 3-form and a 2-form in 10 dimensions. Using the Sherk-Schwarz reduction procedure the ansatz is
\bea \label{dimred}
g^{11}_{ab}&=&e^{-2\phi/3}g_{ab}+e^{4\phi/3}C^{(1)}_aC^{(1)}_b\qquad\qquad\, C_{abc}=C^{(3)}_{abc}\nonumber\\
g^{11}_{az}&=&e^{4\phi/3}C^{(1)}_a \qquad\qquad\qquad \qquad\qquad\quad C_{abz}=B_{ab}\nonumber\\
g^{11}_{zz}&=&e^{4\phi/3}.
\eea
$g_{ab}$ is the 10 dimensional metric with $a,b,c,\ldots$ representing 10-dimensional indices. $C^{(1)}$, $B$ and $C^{(3)}$ are one, two and three forms respectively. $\phi$ is the dilaton and we have performed a Weyl rescaling so that the resulting action is in the string frame. The resulting bosonic action is 
\bea
I_{10}&=&\frac{2\pi \lpb}{2\kappa_{11}^2}\int d ^{10}x\sqrt{-g}\left\{e^{-2\phi}\left[R(g)-4(\partial \phi)^2+\frac{1}{2.3!}H_{abc}^2\right]\right. \nonumber\\&&\left.-\left[\frac{1}{4}\left(G^{(2)}\right)^2+\frac{1}{2.4!}\left(G^{(4)}\right)^2\right]\right\}-\frac{2\pi\lpb}{2\kappa_{11}^2}\int B\wedge G^{(4)}\wedge G^{(4)}.
\eea
We have compactified on a circle of radius $\lpb=\ell_{pl}/{2\pi}$ and $G^{(2)}$, $H$ and $G^{(4)}$ are the field strengths of $C^{(1)}$, $B$ and $C^{(3)}$ respectively.

However, since the 11-dimensional metric is asymptotically flat we would also like the 10-dimensional metric to have this property. As things stand we have $g_{ab}\rightarrow e^{2\phi_o/3}\eta_{ab}$ as we go towards spatial infinity, where $\phi_0$ is the asymptotic value of the dilaton. We rescale the metric to an asymptotically flat form, and rescale other fields to remove extra factors of $e^{\phi_o}$. This requires
\bea
g_{ab}\rightarrow e^{2\phi_0/3}g_{ab}\ \qquad\qquad&& C^{(1)}_a\rightarrow e^{\phi_0/3}C^{(1)}_a\nonumber\\
B_{ab}\rightarrow e^{2\phi_0/3}B_{ab}\qquad\qquad&& C^{(3)}_{abc}\rightarrow e^{\phi_0}C^{(3)}_{abc}.
\eea
Since $g_s$, the IIA string coupling which counts loops in string amplitudes, is given by $g_s=e^{\phi_0}$, this leaves the action in the form
\bea
I_{10}&=&\frac{g_s^2}{16\pi G^{(10)}_N}\int d ^{10}x\sqrt{-g}\left\{e^{-2\phi}\left[R(g)-4(\partial \phi)^2+\frac{1}{2.3!}H_{abc}^2\right]\right.\nonumber\\&& \left.-\left[\frac{1}{4}\left(G^{(2)}\right)^2+\frac{1}{2.4!}\left(G^{(4)}\right)^2\right]\right\}\non\\ &&-\frac{g_s^2}{2.16\pi G^{(10)}_2}\int B\wedge G^{(4)}\wedge G^{(4)},
\eea
where we have also used the relation
\beq\label{consts}
G^{(10)}_N=\frac{G^{(11)}_N}{2\pi\lpb g_s^{2/3}}.
\eeq
We have fixed $z$ on a circle of radius $\lpb$, but the radius of the eleventh dimension measured at infinity is naturally measured in the 11-dimensional metric:
\beq\label{r11}
R_{11}=\frac{1}{2\pi}\lim_{r\rightarrow\infty}\int\sqrt{|g_{zz}|}dz=\lpb e^{2\phi_o/3}=\lpb g_s^{2/3}.
\eeq
This relation is extremely important in M-theory and it reduces (\ref{consts}) to the standard Kaluza-Klein form
\beq
G^{(10)}_N=\frac{G^{(11)}_N}{V_{11}},
\eeq
where $V_{11}=2\pi R_{11}$ is the volume of the internal space.

Standard formulae in 10 and 11 dimensions give us that $G_N^{(10)}=8\pi^6g_s^2(\alpha')^4$ and $\gne=\frac{(\lpb)^9}{32\pi^2}$, which leads to the relation $\lpb= \ell_s g_s^{1/3}$. Thus we can write the following relations between the constants in 11-dimensions and those of IIA string theory:
\bea
\ell_{pl}=2\pi \ell_s g_s^{1/3},\\
R_{11}=\ell_s g_s.
\eea
We can see from this second relation that as we go to strong coupling we are going to the decompactification limit; i.e. towards the 11-dimensional theory. It is also useful to express $R_{11}$ in units of the 11-dimensional Planck length (divided by $2\pi$), allowing (\ref{r11}) to be rewritten as
\beq\label{R11}
R_{11}=g_s^{2/3}.
\eeq

\section{The M2-brane}

Objects with three-dimensional worldvolumes were investigated as long ago as 1962 by Dirac\cite{Dirac}. The 3-form potential of 11-dimensional supergravity is suggestive of coupling to such a membrane and indeed membrane solutions of 11-dimensional supergravity were found\cite{Duff:1990xz}. These solutions break many of the symmetries of the vacuum, including supersymmetries, and these broken symmetries give rise to Goldstone modes living on the extended object. The broken supersymmetries give rise to Goldstone Fermions, and in fact it is requiring worldvolume supersymmetry that allows us to determine which branes are allowed in which dimension. Thus if we look for an object in 11-dimensional supergravity that breaks half the supersymmetry we have 8 Fermionic degrees of freedom on the worldvolume. These can be matched with 8 Bosonic degrees of freedom coming from Goldstone scalars resulting from broken translation invariance in the directions transverse to the extended object; this implies a 3-dimensional worldvolume. A supersymmetric worldvolume action for the membrane has been found. As with the supergravity action we can dimensionally reduce on a circle to ten dimensions. This time there are two possibilities: allow the membrane to wrap around the circle which leads to the effective action for the IIA string (double dimensional reduction, so called because the worldvolume and spacetime dimension are both decreased), or take the circle to be in one of the transverse directions, which leads to a D2-brane in IIA string theory (direct dimensional reduction).

\subsection{The Membrane Solution of 11-Dimensional Supergravity}

There is an extremal membrane solution of 11-dimensional supergravity\cite{Duff:1990xz} whose form is given by
\bea
ds^2&=&H^{-2/3}\eta_{\mu\nu}db^\mu  dx^\nu +H^{1/3}\delta_{pq}dy^p dy^q, \nonumber\\
C&=&\pm \frac{1}{3!}H^{-1}\epsilon_{\mu\nu\rho}dx^\mu dx^\nu dx^\rho, \qquad \mbox{where}\ H=1+\left(\frac{R}{\rho}\right)^6.
\eea
The indices are split into $\mu,\nu,\ldots=0,1,2$ and $p,q,\ldots=3,4,\ldots,10$ and $\rho=\sqrt{\delta_{pq}y^py^q}$ is the transverse radius. $H$ has the harmonic property $\delta^{pq}\partial_p\partial_qH=0$.

\subsection{The Supermembrane Action}

The supermembrane action in 11-dimensions was constructed in \cite{Bergshoeff:1987cm}. It is given by
\beq
S=\int d^3\xi \left(\frac{1}{2}\sqrt{-g}g^{ij}E_i^{\;A}E_j^{\;B}\eta_{AB}+\epsilon^{ijk}E_i^{\;A}E_j^{\;B}E_k^{\;C}B_{CBA}-\frac{1}{2}\sqrt{-g}\right).
\eeq
$i$ labels worldvolume co-ordinates $0,1,2$ with metric $g_{ij}$ of signature $(-,+,+)$. $B$ is the super 3-form \cite{Brink:1980a,Cremmer:1980ru}, its part with 3 bosonic indices is just $C$ which appears in the Lagrangian (\ref{11dsugraL}). $E_i^{\;A}=(\partial_i Z^M)E_M^{\;A}$ where $Z^M$ are superspace co-ordinates and $E_M^{\;A}(Z)$ is the supervielbein. 

This action is invariant under a Fermionic kappa symmetry that means half of the Fermionic degrees of freedom are redundant and can be gauge fixed. The bosonic part of the action appears below. Though the single membrane action is known, that of multiple membranes is unknown. The action for $N$ coincident D-branes is the Born-Infeld action where the fields become $N\times N$ matrix valued. The $N^2$ large-$N$ scaling of the degrees of freedom is explained in terms of massless fundamental strings stretching between the branes. An $N^{3/2}$ scaling is expected for coincident membranes, but there is no picture of what these degrees of freedom are.

\subsection{Reduction to the Superstring in 10 Dimensions}

By doubly dimensionally reducing the membrane action one can obtain the superstring effective action in 10 dimensions\cite{Duff:1987bx}. To avoid an unenlightening proliferation of superspace indices we shall start from the bosonic sector of the supermembrane action which is given by

\beq
S=\int d^3\xi \left(\frac{1}{2}\sqrt{-
\hat{\gamma}}\hat{\gamma}^{ij}\partial_i x^{\hat{m}}\partial_jx^{\hat{n}}\hat{g}_{\hat{m}\hat{n}}-\frac{1}{6}\epsilon^{ijk}\partial_ix^{\hat{m}}\partial_jx^{\hat{n}}\partial_kx^{\hat{p}}C_{\hat{m}\hat{n}\hat{p}}-\frac{1}{2}\sqrt{-\hat{\gamma}}\right),
\eeq
where $\hat{\gamma}^{ij}$ is the worldvolume metric and $g_{\hat{m}\hat{n}}$ is the background metric.

Splitting the co-ordinates as $\xi^i=(\sigma^a,\rho)$ for $a=1,2$ and $x^{\hat{m}}=(x^m,z)$ for $m=0,1,\dots,9$ we make the gauge choice $z=\rho$ and demand $\partial_\rho x^m=0$, $\partial_z \hat{g}^{\hat{m}\hat{n}}=0$ and $\partial_z C_{\hat{m}\hat{n}\hat{p}}=0$. We can then make a reduction ansatz equivalent to (\ref{dimred})
\bea
\hat{g}_{mn}&=&e^{-2\phi/3}g_{mn}+e^{4\phi/3}C^{(1)}_mC^{(1)}_n,\qquad\qquad C_{mnp}=C^{(3)}_{mnp}\, ,\nonumber\\
\hat{g}_{mz}&=&e^{4\phi/3}C^{(1)}_a, \qquad\qquad\qquad\qquad\qquad\quad\, C_{mnz}=B_{mn}\, ,\nonumber\\
\hat{g}_{zz}&=&e^{4\phi/3},
\eea
which implies that $\sqrt{-\hat{g}}=\sqrt{-g}$. It can be shown that substitution into the field equations leads to the string equation of motion one would expect from 
\beq
S=\int d^2\sigma \left(\frac{1}{2}\sqrt{-\gamma}\gamma^{ab}\partial_a x^{m}\partial_bx^{n}g_{mn}-\frac{1}{2}\epsilon^{ij}\partial_a x^{m}\partial_b x^{n}B_{mn}\right).
\eeq
($C^{(3)}, C^{(1)}$ and $\phi$ have decoupled here but persist in the Fermionic sector.) The $x^z$ component of the equations of motion yields an identity which confirms consistency. In fact substituting into the action directly yields a 2-dimensional action equivalent to that of the string. This can be extended to the full supersymmetric case which yields the superspace action of the type IIA superstring coupled to IIA supergravity. Since the IIA superstring is known to be a consistent quantum theory this gives hope that there should be a theory of membranes in 11-dimensions which is also consistent. Notice that the membrane is not conformally invariant but leads to the conformally invariant superstring. 

\subsection{Reduction to a D2-brane}\label{dual}

The membrane in 11 dimensions can be reduced to the D2-brane in 10 dimensions by compactifying on a circle in one of the transverse directions. In fact this was how the Fermionic part of the D2-brane action was obtained\cite{Townsend:1995af}. The relationship follows quickly upon dualising one of the scalars on the M2-worldvolume. This yields a vector field in 3 dimensions (which has one physical degree of freedom), and reduces the number of transverse scalars to seven. This maintains the balance of Fermionic and bosonic degrees of freedom on the worldvolume. Starting from the D2-brane and reversing the process illustrates the hidden 11-dimensional Lorentz invariance of string theory.

\section{The M5-brane}

Matching degrees of freedom between transverse scalars and Fermions would lead to the conclusion that the membrane was the only extended object in 11-dimensions, but there are other worldvolume fields that should be considered. Broken gauge symmetries can lead to tensor fields propagating on worldvolumes, and indeed the broken gauge symmetry of the 3-form field of 11-dimensional supergravity leads to a 2-form with anti-self-dual field strength on a five-brane worldvolume. This gives 3 physical degrees of freedom, which, along with the 5 transverse scalars gives 8 Bosonic degrees of freedom to match the Fermionic degrees of freedom coming from breaking half the supersymmetry. This is a $(0,2)$ tensor multiplet on the worldvolume, giving a superconformal theory in 6-dimensions.

The five-brane solution was first found in supergravity by Gueven\cite{Gueven} and after some difficulty its worldvolume action and field equations were later found (the difficulties stem from the problem of writing an action involving the self-dual field strength). The five-brane can be reduced to the D4-brane of IIA string theory by double dimensional reduction, and it can also be reduced to the NS5-brane by direct dimensional reduction. The 6-dimensional worldvolume of the five-brane allows chiral Fermions, which leads to an anomaly that must be cancelled.

\subsection{The M5-brane Solution to Supergravity}

The extreme M5-brane solution takes a similar form to that of the membrane
\bea
ds^2&=&H^{-1/3}\eta_{\mu\nu}dx^\mu  dx^\nu +H^{2/3}\delta_{mn}dy^m dy^n,\nonumber\\
G&=&*_y dH,\qquad\mbox{where}\ H=1+\left(\frac{R}{\rho}\right)^3.
\eea
In defining $G$ we have used $*_y$, the hodge star in the transverse directions. Again the indices are split, into $\mu,\nu,\ldots=0,1,\ldots, 5$ and $m,n,\dots=6,7,\ldots,10$ and $\rho=\sqrt{\delta_{mn}y^my^n}$ is the transverse radius.

The membrane is an ``electric" singular solution to the supergravity equations coupled to a membrane source. It has a Noether electric charge given by 
\beq
Q=\frac{1}{\sqrt{2}}\int_{S^7}(*G+\frac{1}{2}C\wedge G)=\sqrt{2}\kappa_{11}T_3.
\eeq
The five-brane, however, is a solitonic solution with  topological magnetic charge given by 
\beq
P=\frac{1}{\sqrt{2}\kappa_{11}}\int_{S^4}G=\sqrt{2}\kappa_{11}T_6.
\eeq
These charges obey a higher dimensional analogue of Dirac quantisation given by $QP=2\pi n$ for integer $n$, or equivalently $2\kappa_{11}^2 T_3T_6=2\pi n$. Along with the relation $T_6=\frac{1}{2\pi}T_3^2$, which can be deduced from the quantisation of the periods of $C$, this implies we have only one independent dimensionful parameter in 11 dimensions.

\subsection{The M5-brane Worldvolume Action and Reduction}

There are difficulties formulating the worldvolume action for a five-brane as it must contain the 2-form tensor field with anti-self-dual field strength which is part of the tensor multiplet of the effective theory. Two approaches can be taken: one is to introduce an auxiliary field to ensure that the generalised self-duality condition appears as an equation of motion\cite{PST,BLNPST}, and the other is to formulate the action in such a way that 6-dimensional general covariance is not manifest\cite{PerryS, APPS1}. Alternatively one can work without an action and use the equations of motion obtained via the superembedding formalism\cite{d11p5,cov5eom}.

A starting point for deriving the action with non-manifest covariance was ensuring the correct dimensional reduction to a four-brane. This made the covariance in five of the dimensions obvious, but to prove it in the fifth spatial direction required more work. We single out the $x^5$ direction as different and write the indices $\hat{\mu}=(\mu,5)$. The anti-self-dual field is represented by $B_{\mu\nu}$ which is a 5d anti-symmetric tensor with 5d curl $H_{\mu\nu\rho}=3\partial_{[\mu}B_{\nu\rho]}$ and dual $\bar{H}^{\mu\nu}=\frac{1}{6}\epsilon^{\mu\nu\rho\lambda\sigma}H_{\rho\lambda\sigma}$. The metric also splits into $G_{\mu\nu}$, $G_{\mu5}$ and $G_{55}$, with $G_5$ being the 5-dimensional determinant. The Bosonic Lagrangian can then be written as
\beq
L=-\sqrt{-\mbox{det}(G_{\hat{\mu}\hat{\nu}}+iG_{\hat{\mu}\rho}G_{\hat{\nu}\lambda}\bar{H}^{\rho\lambda}/\sqrt{-G_5})}-\frac{1}{4}\bar{H}^{\mu\nu}\partial_5B_{\mu\nu}+\frac{1}{8}\epsilon_{\mu\nu\rho\lambda\sigma}\frac{G^{5\rho}}{G^{55}}\bar{H}^{\mu\nu}\bar{H}^{\lambda\rho},
\eeq
note the Born-Infeld and Wess-Zumino like terms. 

In the PST approach\cite{PST,BLNPST} $B$ has additional $B_{\mu 5}$ components and there is an auxiliary field, $a$. However there are also extra gauge freedoms and one can set $B_{\mu5}=0$ and make a simple choice for $a$ so that the action becomes equivalent to the above. Both versions of the action can be supersymmetrised into a kappa symmetric form. 

Similarly to the membrane case, double dimensional reduction on a circle gives a IIA string theory object, here the four-brane. At first a four-brane with an anti-symmetric tensor field is found, but analogously to the membrane-D2 reduction there is a worldvolume duality transformation that yields the standard D4 action with a worldvolume vector field\cite{APPS2}. The five-brane can also be compactified on a torus and identified with the D3-brane in IIB string theory after dualising\cite{Berman53}.

\subsection{Anomalies on the Five-brane}

The presence of chiral Fermions and the anti-self-dual field on the five-brane worldvolume leads to potential anomalies\cite{WAG}. These can be calculated in the standard way, ultimately from index theorems. The contributions from these two sources can be expressed in terms of a closed 8-form, $I_8$, which gives the anomaly 6-form via the standard descent formalism ($I_8=dI_7$, $\delta I_7=dI_6$):
\beq
I_8=\frac{1}{(2\pi)^4}\left[-\frac{1}{768}(trR^2)^2+\frac{1}{192}trR^4\right].
\eeq
This anomaly can be cancelled by adding an additional term to the 11-dimensional supergravity action
\beq
I_{11}'=T_3\int C\wedge I_8.
\eeq
This term cannot be checked with the microscopic theory in 11 dimensions due to the lack of a quantised  membrane theory, but dimensionally reducing to IIA this term is equivalent to 
\beq
I_{10}'=T_2\int B\wedge I_8.
\eeq
In fact this term is already known from a 1-loop string calculation\cite{VafaW,Duff:1995wd}. It is intriguing that this 1-loop term is related to the 11-dimensional term which is M-theoretic in origin (this correction goes like a fractional power of Newton's constant in 11-dimensions and could not be generated in perturbative 1-loop supergravity). The full anomaly story is actually more complex and there are additional normal bundle anomalies which can be cancelled by adding gravitational corrections to the Chern-Simons term\cite{Witten5,FHMM}.

\subsection{M-brane Intersections and Open Membranes}\label{sec:open}

The membranes described previously do not have to be closed, they can have a boundary\cite{Strominger:1995ac}. The membrane couples to the 3-form $C$ whose field strength $G=dC$ is invariant under $C\rightarrow C+d\Lambda$ for some 2-form $\Lambda$. However, in the presence of a boundary the minimal coupling of $C$ to the membrane leads to a term $\int_{\partial M}\Lambda$. This would break gauge invariance, but if we couple the boundary (which will be a string) to a 2-form field which varies under gauge transformations as $b\rightarrow b-\Lambda$ we can preserve the gauge invariance. Of course the five-brane worldvolume contains exactly such a 2-form and we deduce that membranes can end on five-branes, making five-branes act much like the D-branes of M-theory. The five-brane worldvolume does contain such a string soliton\cite{PerryS,HLW} and this a configuration that will be covered in depth in later chapters. The five-brane also contains a 3-brane soliton\cite{3brane}, and indeed two five-branes can intersect along three common spatial directions and preserve 1/4 supersymmetry. Examining the supersymmetry algebra of the five-brane one sees that the only central charges it could contain are a 1-form or a 3-form, consistent with there being no other (supersymmetric) intersections.

\section{Connection to other String Theories and Dualities}

We have learned how the F1-string and D2-, D4- and NS5-branes of type IIA string theory can be obtained from the 11-dimensional M2- and M5-branes. However, from a fundamental theory which unifies the five consistent string theories we should expect to find the complete complement of IIA branes, and the connection to the other four string theories should be clear. To complete the IIA picture the D0 and D6 branes are easily found from compactification. The D0-particle corresponds to one unit of the quantised momentum in the periodic 11th dimension with higher momentum states corresponding to coincident D0-particles. The D6-brane corresponds to the 11-dimensional Kaluza-Klein monopole\cite{revisited}.

\subsection{Type IIB String Theory}\label{Sec:IIAB}

There is a well-known duality between type IIA and type IIB string theories called T-duality. This duality relates IIA string theory compactified on a circle of radius R with IIB on a circle of radius $1/R$ under exchange of winding and momentum modes (for a more general review of T-duality see \cite{Giveon:1994fu}, we will also discuss it in more detail in Chapters \ref{Ch:DF} and \ref{Ch:Holo}). It then follows that IIB on a circle is equivalent to M-theory on $T^2$ under such an exchange. Letting $R_{11}$ and $R_{10}$ go to zero with a fixed ratio leads to uncompactified type IIB string theory with IIB string coupling $g_S^{(B)}=R_{11}/R_{10}$. The $SL(2,\mathbb{Z})$ symmetry of type IIB - which includes the S-duality that relates weak to strong coupling ($g_S^{(B)}\leftrightarrow 1/g_S^{(B)}$) - is just the $SL(2,\mathbb{Z})$ of reparameterisations of the torus\cite{T4lecs}. Note that the chiral type-IIB theory comes from the non-chiral 11-dimensional theory, something that had previously been forbidden by `no-go' theorems. The chirality is introduced by massive spin-2 multiplets coming from the membrane ``wrapping" modes on $T^2$\cite{BHO}.

\subsection{Heterotic and Type I Strings}

It is a more difficult proposition to obtain the heterotic string, given its different numbers of left-movers and right-movers on the worldsheet. However, by compactifying a five-brane on the 2-complex-dimensional surface $K3$ (which has topology such that it admits 19 self-dual and 3 anti-self-dual 2-forms) one gets $(19,3)$ scalars from the 2-form, $(0,8)$ Fermions and $(5,5)$ other scalars, exactly what one would expect on the heterotic string worldsheet\cite{K3}. One can also get the $E_8\times E_8$ heterotic string in 10 dimensions by compactifying M-theory on $\mathbb{R}^{10}\times S^1/\mathbb{Z}_2$\cite{HW}. Here again we obtain a chiral theory from a non-chiral one. This time previous `no-go' theorems are circumvented by compactifying on an orbifold rather that a manifold. Since $E_8\times E_8$ heterotic and $SO(32)$ heterotic are T-dual to one another, once we have the connection to one we can quickly find connections to the other.

Type I string theory comes from orbifolding type IIB and through similar arguments to those above it can be deduced that type I string theory (or rather its T-dual, type IA) is the  $R\rightarrow 0$ limit of M-theory on a cylinder of radius $R$\cite{HW}. Again this illustrates that the moduli space of vacua is in general 11-dimensional, with 10-dimensional perturbative string expansions only in certain 10-dimensional limits.

\subsection{Dualities from  M-theory}

Dualities in M-theory lead to dualities in various  lower dimensions. Membrane-five-brane duality in 11 dimensions leads to string-string duality in 6 dimensions, between fundamental F-strings and solitonic D-strings, both in the heterotic theory\cite{Duff96}. However, as in the previous section, compactifying different string theories on different manifolds can lead to a duality between the heterotic string and the IIA string\cite{HullT}. Further compactification of each theory on $T^2$ leads to a more surprising duality of dualities: the solitonic string has a non-perturbative S-duality which is a perturbative T-duality in the dual fundamental string picture\cite{Duffsw}.

\subsection{Deducing 11 Dimensions from 10}

Now that we have introduced dualities we will mention some of the results that suggested string theory was an aspect of an 11-dimensional theory. One of the papers which ignited interest in M-theory was \cite{Witten:1995ex} in which Witten studied the strong coupling behaviour of type II superstrings. For the type IIB theory the $SL(2,\mathbb{Z})$ S-duality relates behaviour at strong coupling to that at weak coupling. However, the strong coupling behaviour of the type IIA string is more complex. Type IIA string theory has a 1-form Ramond-Ramond gauge field $A$, and also a central charge $W$ in its supersymmetry algebra. Considering the type IIA effective action as compactified 11-dimensional supergravity, $A$ is the $G_{mz}$ components of the metric and $W$ is the eleventh component of the momentum. The existence of such a charge leads to a BPS inequality of the form 
\beq\label{BPSW}
M\geq\frac{c_1}{g_s}|W|
\eeq
and supermultiplets saturating this bound will have a reduced number of states.

$W$ is zero in the elementary string vacuum, but there exist classical black hole solutions carrying $W$ charge. These black holes obey an equivalent inequality to (\ref{BPSW}). Assuming that there exist BPS particles with non-zero $W$, these particles should possess a discrete spectrum of values of $W$, with quantum independent of the string coupling $g_s=e^\phi$. This implies $M=c|n|/g_s$ and the masses diverge as $g_s\rightarrow 0$, so the particles are not seen as elementary string states. However moving to strong coupling at low energy the BPS states are protected from corrections and these particles should still exist with mass approaching zero. These multiplets contain spin 2 particles and there are infinitely many multiplets at strong coupling. This strong coupling theory still has IIA supersymmetry in 10 dimensions and charged string states coupled to $A$. The only possibility is that these states are the Kaluza-Klein tower from compactification on a circle; the charge corresponds to the rotation of the additional $S^1$. As we have seen, 11-dimensional supergravity has states in correspondence with those of IIA. Comparison leads to the important relation $R_{11} =g_s^{2/3}$, obtained earlier in (\ref{R11}). Again we see type IIA strings have a hidden 11-dimensional Lorentz invariance which adds weight to the claim that the five perturbative string theories are expansions about certain 10-dimensional points in the moduli space of a non-perturbative theory which is  11-dimensional in general.

Many other reasons why an 11-dimensional fundamental theory should be expected are catalogued in \cite{DuffMth}.

\subsection{Other Interesting Aspects of M-theory}

One popular candidate for a microscopic description of M-theory is Matrix Theory\cite{BFSS}. This posits that the membrane in the infinite-momentum frame is related to the $N\rightarrow \infty$ limit of the supersymmetric quantum mechanics of $N$ $D0$-branes, itself the reduction to 0 dimensions of $d=10$ super-Yang-Mills. In Matrix Theory M-branes can be thought of as composites of D0-particles with non-commutative geometry playing an important role.

As well as the more familiar relation of string theory on $AdS_5\times S^5$ to $\mathcal{N}=4$ super-Yang-Mills, the AdS/CFT correspondence also relates M-theory on $AdS_4\times S^7$ and $AdS_7\times S^4$ to 3- and 6- dimensional conformal field theories\cite{Maldacena}. These $AdS$ spaces are realised as the near-horizon geometries of M2-branes and M5-branes respectively.  Another area of progress is a derivation of the entropy of black holes in terms of M-theory microstates\cite{blackholes}.

%% file: ch3.tex
\chapter{D1-D3 and M2-M5 Intersections}\label{Ch:insect}

In Section \ref{sec:open} we  described how an M2-brane, or a stack of coincident M2-branes, can end on a five-brane and the first part of this dissertation will be mainly concerned with this system. However, as much more is known about D-branes, in particular the non-Abelian worldvolume action of coincident D-branes, and since they are related by dimensional reduction, D-brane intersections in string theory are a useful arena for inferring properties of the M-theory intersection. In particular we focus on the D1-D3 system in IIB string theory which is related to the M2-M5 system by dimensional reduction and T-duality.

As described previously, the Goldstone modes of broken symmetries in the bulk  propagate on brane worldvolumes, leading to worldvolume actions for the effective theory on the brane. Thus we have three complementary pictures of our D1-D3 or M2-M5 intersection. Taking the D1-D3 system as an example, firstly, there should exist a supergravity solution in the 10-dimensional bulk describing the intersection (as well as, of course, a full string theory one!). There is also the worldvolume description of the D3-brane; here the endpoint of the D-strings carries magnetic charge and appears as a soliton. The geometry of the brane is distorted and the D3-brane is stretched in a spike along the D1 worldvolume direction. Finally, we can look at the D1-string worldvolume action and search for a solution corresponding to the worldvolume expanding as the string opens up onto the D3-brane. The scalars in the directions transverse to the worldvolume, yet tangent to the D3-brane, are non-zero and non-commuting. They form a so-called ``fuzzy sphere'', whose radius diverges as the D3-brane is approached. For the D1-brane the three non-zero transverse scalars, $X^i$, obey the Nahm equation
\beq\label{Nahmeq}
\frac{dX^{i}}{d\sigma} = \frac{i}{2}\epsilon_{ijk}[X^{j},X^{k}] \, ,
\eeq
where $\sigma$ is the spatial worldvolume direction of the D1-string. This is not surprising as the Nahm equation first arose as a way of classifying monopole solutions\cite{Nahm}, and in the ``dual'' D3-brane worldvolume picture the end of the D1-string is a monopole. The M2-M5 intersection has an M5-brane worldvolume interpretation as a self-dual string; however, until recently a membrane worldvolume solution was lacking. This is because the worldvolume action for coincident membranes is unknown. (For there to be an overlap in the range of validity of the different worldvolume pictures, $N$, the number of membranes, must be large.)

Once we have the D1-D3 solution, a large number of consistency checks and extensions can be performed. One important extension that will concern us is the extension to calibrated intersections of D3-branes. Here more than three scalars are active and a generalised form of the Nahm equation, (\ref{Nahmeq}), is required.

\section{D3-brane Worldvolume Picture: The BIon Spike}\label{sec:BIon}

Considering the limit of the D3-brane Born-Infeld action where gravitational effects are ignored, \cite{Callan, Gibbons} were able to find static finite energy solutions to the Born-Infeld electrodynamics corresponding to both fundamental and D-string states ending on the brane. By $SL(2,\mathbb{Z})$ invariance the self-dual D3-brane admits both.

These solutions can be easily seen by looking at the supersymmetry variation in the linear theory. This is just given by dimensionally-reduced super Yang-Mills. The gaugino supersymmetry variation is given by
\beq
\delta\chi=\Gamma^{\mu\nu}F_{\mu\nu}\epsilon\, ,
\eeq
where $\mu, \nu$ are 10-dimensional indices which split into worldvolume and transverse indices such that $F_{ab}$ is the field strength on the brane, $F_{ai}=\partial_aX^i$ and $F_{ij}=[X^i,X^j]$, where $X^i$ are the transverse co-ordinates. If we have a point-like Coulomb field
\beq
A_o=\frac{c_p}{r}\, ,
\eeq
where $c_p$ is a constant and $r$ is the radius in the D3-brane co-ordinates, it would break supersymmetry. However, if we introduce an excited transverse scalar
\beq
X^9=\frac{c_p}{r}\, ,
\eeq
then supersymmetries satisfying $(\Gamma^0+\Gamma^9)\epsilon=0$ will be preserved so that we have a half-BPS configuration. This configuration has a spike in the $X^9$ direction, supported by the electric charge. The energy of the spike is proportional to its length, so an infinite spike has infinite length, explaining why the Coulomb energy diverges. The energy  can be checked against that of a fundamental string and indeed it matches. Similarly, there is a magnetic monopole solution with 
\beq\label{BIon}
F_{\theta\phi}=-\frac{Nc_m}{r^2}\, ,\qquad X^9=\frac{Nc_m}{r}\,.
\eeq
This solution to the BPS equation is the standard BPS monopole solution with the transverse direction $X^9$ playing the role of the Higgs field. Note that these solutions can be superposed to give multiple spike solutions that do not attract or repel one an other, as the electromagnetic force balances the deformation of the brane. One can ask whether these solutions are solutions of the full theory, and indeed they are, as might be expected for BPS solutions. As well as being found in \cite{Callan, Gibbons} these were also found in \cite{HLW} by dimensional reduction from the self-dual string.

\section{D1-brane Worldvolume Picture: The Fuzzy-Funnel}

We now turn to the worldvolume picture of the coincident D1-strings. We begin by looking for static solutions to the equations of motion arising from the linearised action.
 
\subsection{Linearised Solution}\label{sec:linsol}

\begin{figure}
\begin{center}
\includegraphics[width=14.5cm]{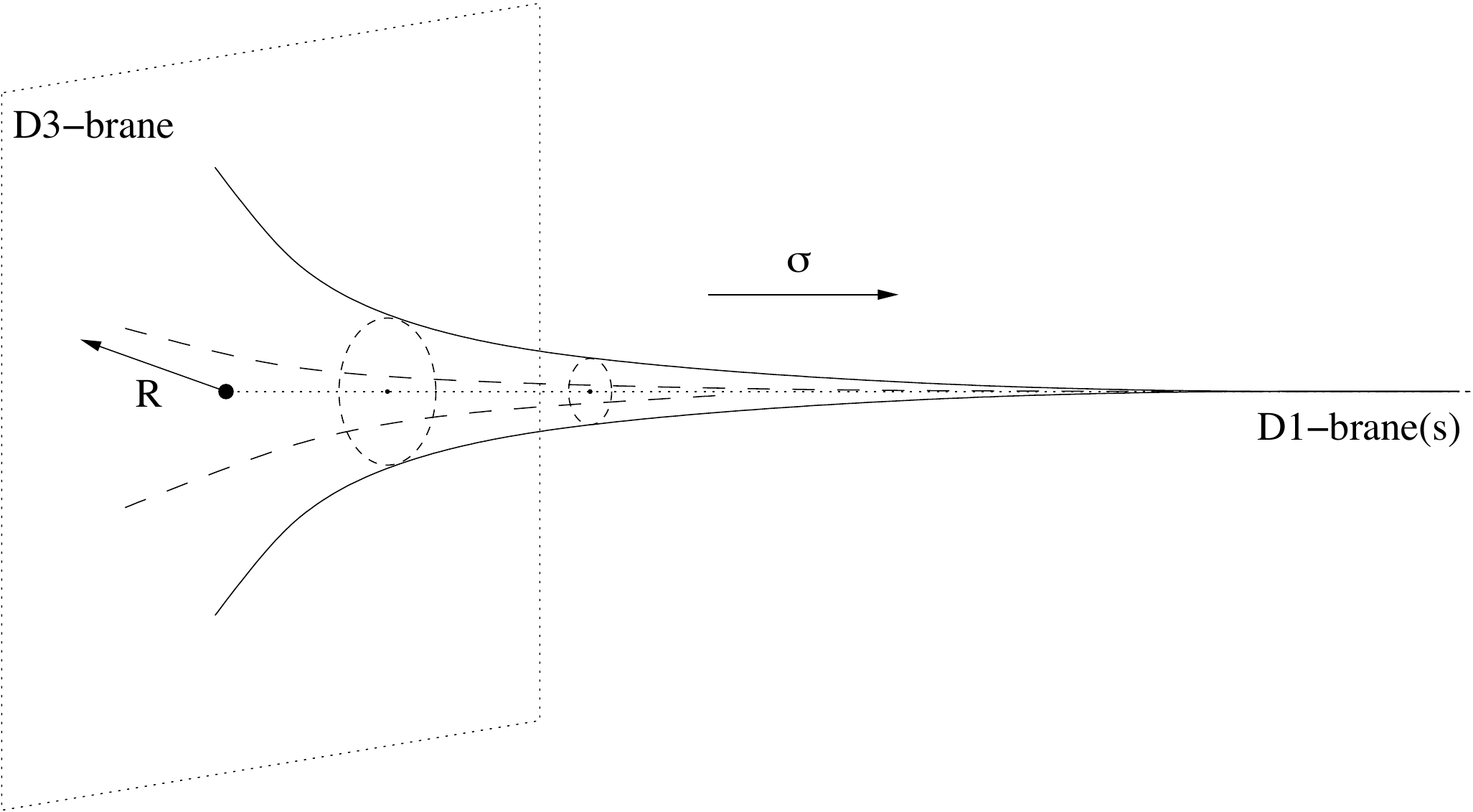}
\caption{D1-branes ending on a D3-brane. The cross-section is a sphere whose radius blows up as the D1-strings open out onto the D3-brane. The D3-brane is pulled out in a spike, the endpoint of the strings appears like a monopole in the D3-brane theory, with the transverse co-ordinate in the D1-string direction playing the role of the Higgs field.}
\label{pic:D1-D3}
\end{center}
\end{figure}

Multiple coincident D-branes are described by the Born-Infeld action. Here we consider a flat background with no gauge fields and look for a static solution. In this case we can expand the Born-Infeld action to leading order in $\lambda=2\pi\ell_s^2$ as
\beq\label{D1action}
S\simeq-T_1\int d^2\sigma\left(N+\frac{\lambda^2}{2}\left(\partial^a\Phi^i\partial_a\Phi^i+\frac{1}{2}[\Phi^i,\Phi^j][\Phi^j,\Phi^i]+\ldots\right)\right).
\eeq
As we look for static solutions the only non-zero derivatives are with respect to $\sigma$, the worldvolume spatial co-ordinate, and $\sigma=\Phi^9$ by the choice of static gauge. The equation of motion for the transverse scalars $\Phi^i$ is given by
\beq\label{D1eom}
\partial^a\partial_a\Phi^i=[\Phi^j,[\Phi^j,\Phi^i]].
\eeq
In \cite{CMT} an ansatz was made based on preserving spherical symmetry and a representation of the $SU(2)$($\cong SO(3)$) matrix algebra. This ansatz was to take three scalars to be non-zero,
\beq\label{fuzzansatz}
\Phi^i=\hat{R}(\sigma)\alpha^i\, ,\quad i=1,2,3\, ,
\eeq
where the $\alpha^i$ are $N\times N$ matrices obeying the $SU(2)$ algebra $[\alpha^i,\alpha^j]=2i\epsilon_{ijk}\alpha^k$ ($N$ is the number of coincident D-strings). These matrices are the co-ordinates of the fuzzy sphere (see Appendix \ref{app:FS2}) and the radius squared, $\Phi^i\Phi^i$, is proportional to the identity matrix times $\hat{R}(\sigma)^2$. The equation of motion is reduced to the ordinary differential equation 
\beq\label{D1soln}
\hat{R}''(\sigma)=8\hat{R}(\sigma)^3
\eeq
which has a solution
\beq\label{D1solnsoln}
\hat{R}(\sigma)=\pm\frac{1}{2(\sigma-\sigma_\infty)},
\eeq
where $\sigma_\infty$ is a constant indicating where the radius diverges (this is not the most general solution). We see that we do indeed have a ``fuzzy-funnel'' solution. The cross section is given by a fuzzy-sphere (see Appendix \ref{app:FS2}), with the radius diverging as $\sigma\rightarrow\sigma_\infty$, which we identify with the location of the D3-brane, and tapering off as $\sigma\rightarrow\infty$ (see Figure \ref{pic:D1-D3}).

\subsection{Supersymmetry}\label{sec:d1susy}

As for the D3-worldvolume picture, we can consider the equations arising from the preservation of some supersymmetry. The linearised supersymmetry variation is identical to that used in Section \ref{sec:BIon} except dimensionally reduced to two instead of four dimensions. When the gauge field is zero, $F_{ai}=\partial_aX^i$ and $F_{ij}=[X^i,X^j]$ lead to
\beq\label{susyvar}
(2\Gamma^{\sigma i}\partial_\sigma\Phi^i+i\Gamma^{ij}[\Phi^i,\Phi^j])\epsilon=0.
\eeq
If the active scalars satisfy the Nahm equation 
\beq\label{Nahmeqphi}
\frac{d\Phi^{i}}{d\sigma} = \pm\frac{i}{2}\epsilon_{ijk}[\Phi^{j},\Phi^{k}] \, 
\eeq
then the supersymmetry condition will be satisfied for spinor parameters obeying $\Gamma^{\sigma123}\epsilon=\pm\epsilon$. Indeed the solution given by (\ref{fuzzansatz}) and (\ref{D1soln}) does satisfy the Nahm equation, and the conclusion is that we have a BPS solution preserving half the supersymmetry of the D1-brane worldvolume.  As is so often the case, a BPS solution to the linear theory is a solution to the full Born-Infeld theory - as confirmed in \cite{CMT}. Note that in keeping with this, we could have obtained (\ref{Nahmeqphi}) as a Bogomol'nyi equation for the expression for the energy following from the action (\ref{D1action}) (we will do this in a more general case in equation (\ref{D1bog})).

\section{Consistency Checks of the Solution}

After checking that the fuzzy funnel solution solved the equations of motion following from the full non-linear action, \cite{CMT} performed several checks to see if this solution could be identified with the BIon spike. A physical radius can be defined via
\beq
R(\sigma)^2=\frac{\lambda^2}{N}\sum_{i=1}^3\mbox{Tr}[\Phi^i(\sigma)^2].
\eeq
We will from now on work with an irreducible $N\times N$ representation of $SU(2)$ where the quadratic Casimir is given by
\beq
\sum_{i=1}^3(\alpha^i)^2=(N^2-1)\openone_{N\times N}.
\eeq
This, combined with the solution (\ref{fuzzansatz}) and (\ref{D1solnsoln}), leads to a radial profile given by
\beq\label{D1profile}
R=\frac{N\pi\ell_s^2}{\sigma-\sigma_\infty}\sqrt{1-\frac{1}{N^2}}.
\eeq
For large $N$ this is precisely the relationship obtained for the BIon spike \cite{Callan} (c.f. (\ref{BIon})).

A further check is the energy of the configuration. The expression for the energy (which follows quickly from the action for static solutions) can be linearised using the conditions required for supersymmetry (actually the Nahm equation and the first order equation $\hat{R}'=\mp2\hat{R}^2$, which (\ref{D1solnsoln}) obeys). Using the full details of the solution this linearised energy can be rewritten in the form
\beq
E=T_3\frac{N}{\sqrt{N^2-1}}\int4\pi R^2dR\left[\left(\frac{\partial \sigma}{\partial R}\right)^2 +1\right].
\eeq
In the large-$N$ limit the $N$ dependence disappears and this is just the energy of any spherically symmetric BPS solution of the D3 worldvolume theory, the BIon spike being such a solution.

We can also look at the Chern-Simons term, which should be included as well as the Born-Infeld term in the D-brane action. Substituting the ansatz into this Chern-Simons coupling yields
\beq
\mp i\mu_3\frac{N}{\sqrt{N^2-1}}\int dt4\pi R^2dRC^{(4)}_{t123}(t,R).
\eeq
Up to the $N$ dependent factor which goes to 1 in the large-$N$ limit, this is exactly the standard coupling to the RR 4-form $C^{(4)}$ expected from our interpretation that the D1-string opens up onto a D3-brane.

These checks support the claim that the fuzzy funnel is the D1-brane worldvolume picture of the BIon spike. The three non-zero transverse scalars do not commute and form a fuzzy sphere which has a radius which diverges as we approach a 3-hypersurface which can be identified with the D3-brane. Note that this configuration with non-commutative scalars is distinct from the Myers effect\cite{Myers}, in which there is a background field.

\subsection{Further Extensions}\label{furext}

There are various further extensions performed in \cite{CMT} that serve to show the consistency of the fuzzy funnel solution. Solutions were found corresponding to double funnels suspended between two separated D3-branes. Solutions corresponding to $(p,q)$-strings (bound states of D1-strings and fundamental strings) were also obtained. It was then confirmed that the solution was consistent in the background of a stack of D3 branes. For a stack of (anti) D3-branes the solution corresponding to a D-string ending on an (anti) D3-brane was picked out, but the end point did not have to be coincident with the stack of (anti) D3-branes.

An analysis of the fluctuations of the funnel was performed, and again the agreement with similar analyses of the D3-picture of the BIon spike\cite{Callan,Kastor,Lee} was striking. The simplest fluctuations are of the form 
\beq
\delta\Phi^m(\sigma,t)=f^m(\sigma,t)\openone_N
\eeq
in directions transverse to both branes. These obey a linear equation of motion
\beq
\left(\left(1+\lambda^2\frac{(N^2-1)}{4\sigma^2}\right)\partial_t^2+\partial_\sigma^2\right)f^m=0\, ,
\eeq
which agrees with the D3-brane picture up to the familiar $\frac{N^2-1}{N^2}$ factor. This mode is the $l=0$ angular momentum mode, and higher $l$ modes can also be analysed. They will have the form
\beq
\delta\Phi^m(\sigma,t)=\sum_{l=0}^{N-1}\psi_{i_1i_2\ldots i_l}(\sigma, t)\alpha^{i_1}\alpha^{i_2}\ldots\alpha^{i_l}\, ,
\eeq
where the expansion has to stop at $N-1$ as that is the maximal number of linearly independent products of $\alpha^i$. This is an expansion in spherical harmonics. Again the fluctuations agree with those of the BIon spike, though the spike did not have the truncation at $l=N-1$. However, as we will shortly discuss, at large $l$ the D3-brane picture breaks down. The lack of truncation was an issue with the D3-brane analysis, where modes of arbitrarily high $l$ could propagate out along the spike. The analysis can also be done for the modes transverse to the string yet tangent to the D3-brane, these too agree with the dual D3-picture up to $1/N$ corrections.

\subsection{Range of Validity}\label{validity}

The two descriptions of the D1-D3 system we have considered agree remarkably well and we can check whether this should be expected given the approximations that are being made. The Born-Infeld action will receive higher derivative corrections from the $\alpha'$ expansion in string theory. In the D3-brane picture, to be able to ignore these requires $R\gg\ell_s$, which is equivalent to $\sigma\ll N\ell_s$. Similarly, for the D1-string picture we require $\sigma\gg\ell_s$, which is equivalent to $R\ll N\ell_s$. Thus there is a large range of overlap for large $N$.

Taking into account possible higher commutator corrections may reduce this range of validity to $R\ll\sqrt{N}\ell_s$, but it is suspected such corrections may vanish for BPS states such as the solutions considered so far. Gravitational effects have also been neglected, this can be justified for very weak string coupling,

\section{Extension to Calibrated Intersections and a Generalised Nahm Equation}\label{calib}

In Section \ref{sec:linsol} we searched for solutions to the Yang-Mills approximation to the Born-Infeld action by specifically looking for solutions with only three transverse scalars active. It is possible to find more complex solutions corresponding to more complicated configurations, with the D-strings ending not on a single D3-brane, but on calibrated intersection of D3-branes.

\subsection{A Generalised Nahm Equation}\label{sec:GNE}

The linearised action minus the constant piece is 
\beq\label{D1action2}
S=-T_1\lambda^2\int d^2\sigma \mbox{Tr}\left(\frac{1}{2}\partial^a\Phi^i\partial_a\Phi^i+\frac{1}{4}[\Phi^i,\Phi^j][\Phi^j,\Phi^i]\right)
\eeq
and the corresponding expression for the energy of a static configuration is
\beq\label{D1energy}
E=T_1\lambda^2\int d^2\sigma \mbox{Tr} \left(\frac{1}{2}\partial\Phi^i \partial\Phi^i +\frac{1}{4}[\Phi^i,\Phi^j][\Phi^j,\Phi^i]\right),
\eeq
were $\partial$ represents a $\sigma$ derivative. We now proceed to perform the Bogomol'nyi trick, writing the energy as a squared term plus a topological piece:
\beq\label{D1bog}
S=-T_1\lambda^2\int d^2\sigma \frac{1}{2}\mbox{Tr} \left(\left(\partial\Phi^i-\frac{1}{2}c_{ijk}[\Phi^j,\Phi^k]\right)^2 +\frac{1}{3}c_{ijk}\partial\left(\Phi^i\Phi^j\Phi^k\right)\right).
\eeq
In the case considered earlier with three active scalars this is an identity with $c_{ijk}=\epsilon_{ijk}$. The first term gives the Nahm equation (\ref{Nahmeqphi}) as the Bogomol'nyi equation. However, in the more general case where $c_{ijk}$ is a general anti-symmetric 3-tensor, we can only rewrite the action in the form (\ref{D1bog}) if we impose
\beq\label{constraint}
\frac{1}{2}c_{ijk}c_{ilm}\mbox{Tr}\left([\Phi^j,\Phi^k][\Phi^l,\Phi^m]\right)=\mbox{Tr}\left([\Phi^i,\Phi^j][\Phi^i,\Phi^j]\right).
\eeq
The Bogomol'nyi equation is then the generalised Nahm equation
\beq\label{newnahm}
\partial\Phi^i=\frac{1}{2}c_{ijk}[\Phi^j,\Phi^k]\, ,
\eeq
and when this is satisfied the energy is given by the second term in (\ref{D1bog}), a topological term which only depends on the boundary conditions for the $\Phi^i$. It should be noted that this modified Nahm equation for specific $c$ has appeared before in the pre-D-brane literature, for example \cite{Fairlieetal}.

The equation of motion is still given by (\ref{D1eom}) and upon substituting the generalised Nahm (\ref{newnahm}) twice it is equivalent to the algebraic equation
\beq\label{algeom}
\frac{1}{2}c_{ijk}c_{jlm}[[\Phi^l,\Phi^m],\Phi^k]=-[[\Phi^i,\Phi^j],\Phi^j]\, .
\eeq
Multiplying this by $\Phi^i$ and taking the trace we see that it implies the constraint (\ref{constraint}). Thus a solution to the generalised Nahm equation (\ref{newnahm}) and the algebraic equation of motion (\ref{algeom}) solves the equation of motion and (\ref{constraint}).

\subsection{Supersymmetry}
The generalised Nahm equation can be used in the linearised supersymmetry variation (\ref{susyvar}) to obtain
\beq
0=\sum_{i<j}[\Phi^i,\Phi^j]\Gamma^{ij}(1+c_{ijk}\Gamma^{ijk9})\epsilon.
\eeq
To solve this we define projectors
\beq\label{dproj}
P_{ij}=\frac{1}{2}\left(1+c_{ijk}\Gamma^{ijk9}\right)
\eeq
with no sum on $i,j$, and in all cases of relevance there will be at most one value of $k$ for which $c_{ijk}\neq 0$ for each $i,j$ pair. If we make sure that $c$ is normalised such that $c_{ijk}=\pm1$, then $P_{ij}^2=P_{ij}$. We can impose $P_{ij}\epsilon=0$ for each $i,j$ for which $c_{ijl}\neq 0$. We look for a solution that preserves some fraction of supersymmetry by imposing projectors that commute with each other. For two projectors to commute requires the two associated sets of indices, namely $i,j,k$ and $i',j',k'$, to have exactly one index in common. This will preserve $16\times2^{-m}$ supersymmetries, where $m$ is the number of {\it independent} projectors (certain sets of projectors leave other projectors automatically satisfied), provided we also satisfy 
\beq\label{brack}
\sum_{c_{ijk}=0}[\Phi^i,\Phi^j]\Gamma^{ij}\epsilon=0,
\eeq
where the sum is now over pairs $i,j$ such that $c_{ijk}=0$ for all $k$. Use of the projectors reduces this further to a set of equations relating commutators (see for example Equation (\ref{dc2})).

\subsection{Calibrations}\label{subcalib}

In the absence of fluxes, branes act to minimise their worldvolume. Requiring supersymmetry means that their worldvolumes are a special set of sub-manifolds of the target space called calibrated manifolds\cite{Harvey:1982xk}. Intersecting D-branes preserving some fraction of supersymmetry can also be analysed using calibrations\cite{GP,Jerome}. The combined worldvolume of all the intersecting branes is a single calibrated manifold. To define this manifold requires a closed form $\omega$ in the bulk (a $p$-form for intersecting $p$-branes) called the calibration form. A calibration form satisfies $P[\omega](\xi)\leq dvol(\xi)$ where $\xi$ is any tangent vector to the $p$-dimensional submanifold, $P$ represents the pullback to the worldvolume and $dvol$ is the induced volume-form on the submanifold. For a calibrated submanifold the inequality is saturated. Such a submanifold is the minimal volume element of its homology class.

The observation of \cite{CL} was that since the D1-brane breaks half the supersymmetry of the background via $\Gamma^{09}\epsilon_L=\epsilon_R$, then imposing the projectors (\ref{dproj}) is equivalent to requiring $\Gamma^{0ijk}\epsilon_L=\epsilon_R$. This is what one would expect for a D3-brane in the $i,j,k$ spatial directions. In fact given a set of projectors dictated by the non-zero components of $c_{ijk}$ then this corresponds to a calibrated intersection of D-branes given by the calibration form $c=\frac{1}{3!}c_{ijk}dx^i\wedge dx^j\wedge dx^k$. Possible calibrated intersections of D3 branes can be deduced from the M5 intersections given in \cite{GP,Jerome} and from this the form of $c$ can be deduced. Then we can look for a solution by solving the modified Nahm (\ref{newnahm}), imposing the projectors (\ref{dproj}), solving the algebraic condition on the brackets following from (\ref{brack}) and finally guaranteeing the equation of motion by solving (\ref{algeom}). In fact for all the calibrations considered, (\ref{algeom}) is automatically satisfied given the projectors and the conditions on the brackets (\ref{brack}), giving us a minimal energy solution to the equations of motion preserving half of the supersymmetry.

\subsection{The First Non-trivial Configuration and Solution}

The first non-trivial configuration, as described in \cite{CL}, corresponding to five active scalars and two intersecting D3-branes is given by
\bea\label{dc2} \begin{matrix} D3:&1&2&3&\cr D3:&1&&&4&5\cr
D1:&1&&&&&&&9\cr\end{matrix}\nonumber\\
c_{123}=c_{145}=1\quad\quad\quad\quad\qquad \nu=1/4\quad {} \eea
\bea 
{\Phi^1}' &=& [\Phi^2,\Phi^3]+[\Phi^4,\Phi^5]
\ , \nonumber\\
{\Phi^2}' &=& [\Phi^3,\Phi^1]\ ,\quad {\Phi^3}' = [\Phi^1,\Phi^2]
\ , \nonumber\\
{\Phi^4}' &=& [\Phi^5,\Phi^1]\ ,\quad {\Phi^5}' = [\Phi^1,\Phi^4]\ ,
\nonumber\\
{}[\Phi^2,\Phi^4] &=& [\Phi^3,\Phi^5]\ ,\quad 
[\Phi^2,\Phi^5]\,=\,[\Phi^4,\Phi^3]\ .
\nonumber
 \eea
The calibrated manifold here is $\mathbb{R}\times$ a complex curve and the calibration form is $dx^1$ wedged with the K\"ahler form of the complex curve.

This can be solved by a simple generalisation of the the single D3-brane case.  We take $\Phi^i=f(\sigma)A^i$, where
\bea A^1 &=& \mbox{diag}\, (\alpha^1,\alpha^1)\nonumber\\ A^2 &=& \mbox{diag}\,
(\alpha^2,0)\nonumber\\ A^3 &=& \mbox{diag}\, (\alpha^3,0)\nonumber\\ A^4 &=&
\mbox{diag}\, (0,\alpha^2)\nonumber\\ A^5 &=& \mbox{diag}\,
(0,\alpha^3)\,, 
\eea
and where the $\alpha^i$ satisfy $[\alpha^i,\alpha^j]=2i\epsilon_{ijk}\alpha^k$ as before. The same solution for $f(\sigma)$ as in the single D3-brane case suffices. The $A^i$ are $2N\times 2N$ matrices. This and solutions for the other more complex intersections of D3-branes have the correct energy and couplings for the corresponding D3-brane intersection. Moduli for this solution were also discussed in \cite{CL}, indicating possible deformations to less trivial solutions for intersecting D3-branes than those above.

\section{The Nahm Equation and Monopoles}\label{sec:Nahm}

It has already been mentioned how the Nahm equation arises as a way of constructing magnetic monopole solutions. Before \cite{CMT} the Nahm equation was discussed in the context  of D1-D3 solutions by \cite{Diaconescu}, with the issue of boundary conditions clarified by \cite{Tsimpis}. We give a very brief description of the transform from Nahm data to monopole solutions (see for example \cite{Manton}). The Nahm data consists of fields $X(\sigma)$ obeying the Nahm equation (\ref{Nahmeq}), with $X^i(\sigma)^\dagger=-X^i(\sigma)$, $X_i(-\sigma)=X^t_i(\sigma)$ and $X_i(\sigma)$ having a simple pole at the origin where the matrix residues should form an irreducible representation of $SU(2)$ (normally for the Nahm equation one considers a finite interval $[-1,1]$ with poles on each boundary and we will consider that in this section; this would be a string suspended between two D-branes and give a finite mass monopole, as opposed to a single D-brane with a semi-infinite string attached which we consider elsewhere). Given these data we can find a monopole solution by first finding solutions $v_a(\sigma)$ of the 1-dimensional Dirac equation
\beq
\left(\openone\otimes\frac{d}{d\sigma}+\left(iX^i(\sigma)-x^i\right)\tau_i\right)v_a=0.
\eeq
The $\tau^i$ are Pauli matrices and the $x^i$ will be the co-ordinates of the space containing the monopole. One must normalise so that $\int_{-1}^1d\sigma v_a^\dagger(\sigma)v_b(\sigma)=\delta_{ab}$.  Then the Higgs and gauge fields of the monopole are given by
\bea
\Phi(x)_{ab}&=&i\int_{-1}^1d\sigma \sigma v_a^\dagger(\sigma)v_b(\sigma)\nonumber\\
A_i(x)_{ab}&=&\int_{-1}^1d\sigma  v_a^\dagger(\sigma)\frac{\partial}{\partial x^i}v_b(\sigma).
\eea
The inverse transformation is via a 3-dimensional Dirac equation. It has been demonstrated how the Nahm transformations can be derived from tachyon condensation  in the D1-D3 picture\cite{Hashimoto}.

\section{Validity of Non-linear Theory for D1-D3}\label{sec:nonlin}

The full non-Abelian 
Born-Infeld action for coincident D1-branes is given by
\beq
S=-T_1\int d^2\sigma
\mbox{STr}\sqrt{-\mbox{det}(\eta_{ab}+\lambda^2\partial_a\Phi^i Q_{ij}^{-1}\partial_b\Phi^j)\mbox{det}(Q^{ij})},
\eeq
where $Q^{ij}=\delta^{ij}+i\lambda[\Phi^i,\Phi^j]$. STr denotes
the symmetrised trace prescription \cite{Tseytlin,BrecherP}. In 2 dimensions
the gauge field carries no propagating degrees of freedom and may be
completely gauged away, which is why only partial derivatives appear
in the above action. (It is known that
there is some possible ambiguity in the non-Abelian Born-Infeld theory
since the derivative approximation is not valid in a non-Abelian
theory, yet this action has been shown to possess many of the right
properties, see for example \cite{Myers,Brecher}).

For three non-zero
scalars, that depend only on $\Phi^9$, we can expand the determinant
to give an expression for the energy \beq E=T_1\int d\sigma \mbox{STr}\sqrt{I+\lambda^2\partial\Phi^i
\partial\Phi^i-\frac{1}{2}\lambda^2[\Phi^i,\Phi^j]^2-\left(\frac{1}{2}\lambda^2\epsilon_{ijk}\partial\Phi^i[\Phi^j,\Phi^k]\right)^2} \, \, .
\eeq The terms under the square root can then be rewritten using the
Nahm equation as a perfect square, so that\beq E=T_1\int d\sigma \mbox{STr}\left(I+\frac{1}{2}\lambda^2\left(\partial\Phi^i
\partial\Phi^i-\frac{1}{2}[\Phi^i,\Phi^j]^2\right)\right) \, \, ,
\eeq and we can clearly see that the energy reduces to the linear form and a solution to the Nahm equation solves the full non-linear theory.

Expanding the energy for five non-zero scalars we have 
\beq 
 E=T_1\int d^2\sigma \mbox{STr}
\sqrt{\left(I+\frac{1}{2}\lambda^2(\partial\Phi^i
\partial\Phi^i-\frac{1}{2}[\Phi^i,\Phi^j]^2)\right)^2+\lambda^6\left(\epsilon_{ijklm}\partial\Phi^i[\Phi^j,\Phi^k][\Phi^l,\Phi^m]\right)^2}\, ,
\eeq where we have used the modified Nahm equation and the associated
algebraic conditions for the configuration (\ref{dc2}) to write the first square in that form. Thus if
the epsilon term vanishes for a solution to the linear equations of
motion, it is also a solution to the full non-linear equations of
motion. One can check that for the solutions given in
\cite{CL} this is indeed the case, and thus their solutions
are again solutions of the full Born-Infeld theory. It is interesting
to note however that there do exist solutions to (\ref{newnahm})
which do not have the form described in \cite{CL}, solutions where
this second term in the energy does not vanish. These solutions
correspond to the case where the calibration is deformed away from the
flat intersection \cite{Lambert,Helling}. The non-linearity of the
brane action then plays a key role. In what follows we will restrict
ourselves to the case of flat intersecting branes so that the linear
equations will suffice.

Thus it seems that the non-linear action is such that solutions which saturate
the bound of the linear theory are also solutions of the non-linear
theory. For configurations preserving less than half of the membranes' supersymmetry the algebraic equations which are additional to the generalised Nahm type equation are actually necessary to derive a Bogomol'nyi
bound.

\section{The D1-D5 Intersection}

We have seen a solution to the D1-string worldvolume theory with five active scalars corresponding to the D-string ending on a calibrated intersection of D3-branes, but we can also look at solutions with five scalars that corresponds to the D1-string ending on a D5-brane\cite{CMT2,CMT3}. To linear order the equations of motion for the five-scalar action are the same as for three scalars. This leads to the same $R\sim\sigma^{-1}$ behaviour as in the D1-D3 case, not the $R\sim\sigma^{-1/3}$ behaviour expected for a D5-brane. To get this we need to include higher order terms. We substitute the ansatz
\beq
\Phi^i(\sigma)=\frac{R(\sigma)}{\sqrt{c}\lambda}G^i\, ,\qquad i=1,\ldots,5
\eeq
into the Born-Infeld action for five scalars which we used in Section \ref{sec:nonlin} (The $G^i$ are the fuzzy four-sphere matrices which obey $c\openone=\sum_iG^iG^i=n(n+4)$ as proven in Appendix \ref{app:FS4}, leading to equation (\ref{RS4})). This yields an action for the radial profile which gives a Bogomol'nyi bound which is saturated when 
\beq
R'(\sigma)\sim\sqrt{8R^4/(c\lambda^2)+16R^8/(c\lambda^2)^2}.
\eeq
For small $R$ the first term dominates and $R\sim\sigma^{-1}$, while for large $R$ the second term dominates and $R\sim\sigma^{-1/3}$ as required. One can check that this also has the coupling expected for $n$ D5-branes, but this configuration is not supersymmetric  and there are additional contributions to the energy. Looking at the same configuration from the D5-brane point of view, a spike solution can be found with excellent agreement in the large-$n$ limit. The D1-D7 intersection has also been discussed in \cite{Cook}.

\section{The Self-dual String}\label{sec:sds}

\begin{figure}
\begin{center}
\includegraphics[width=14.5cm]{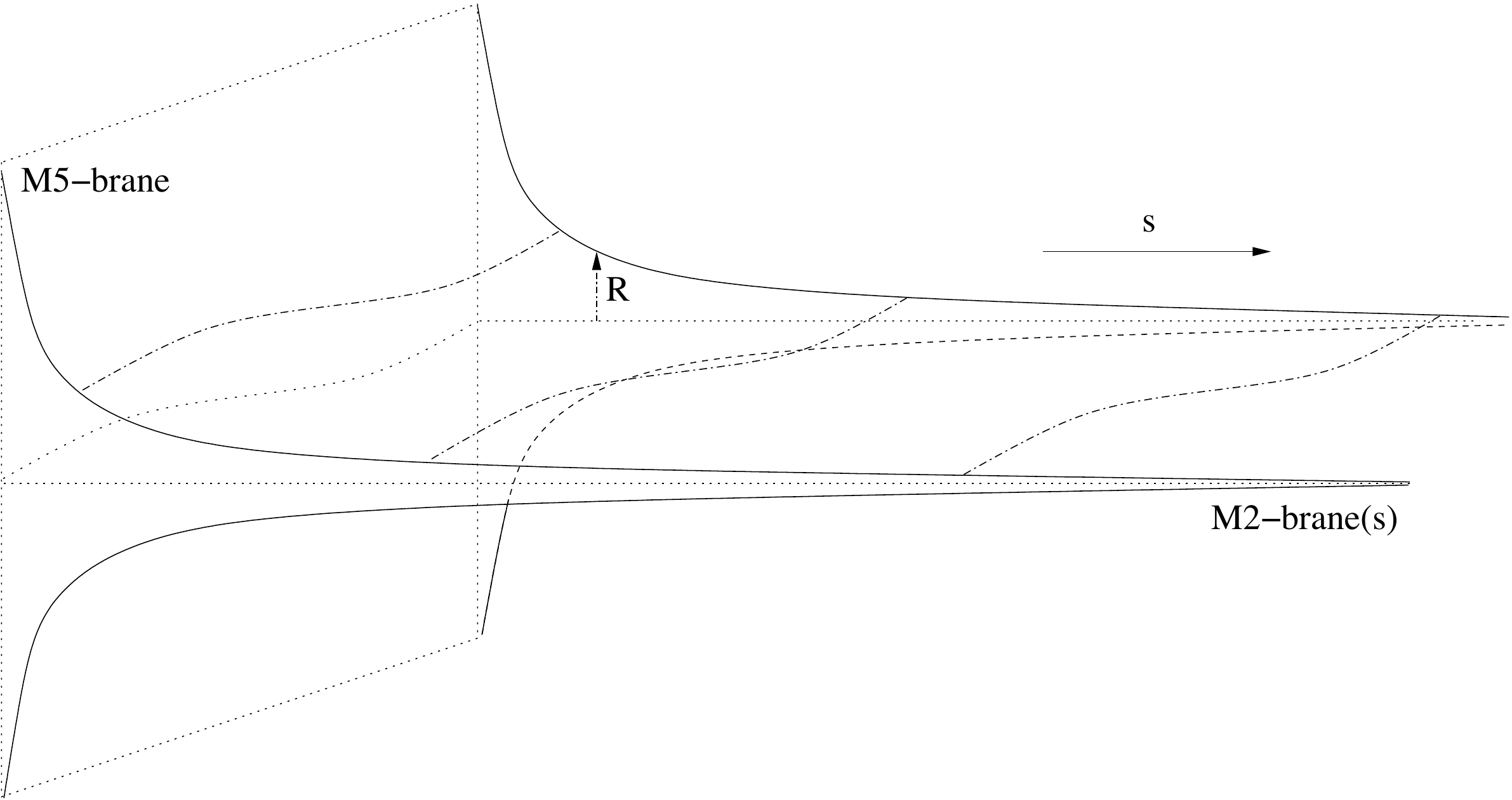}
\caption{M2-branes ending on a M5-brane, the endpoint is a string. Seven spatial dimensions are suppressed; the cross section of the `ridge' is really $S^3\times\mathbb{R}$.}
\label{pic:M2-M5}
\end{center}
\end{figure}

A membrane can end on a five-brane if the membrane boundary carries the charge of the self-dual field $B$ on the five-brane worldvolume\cite{Strominger:1995ac}. In \cite{PerryS} a solution to the field equations of $B$ was found, but none of the scalars were non-zero and it was not supersymmetric. A BPS solution was found in \cite{HLW} by considering the supersymmetry transformation of the Fermions. A representation of this configuration is given in Figure \ref{pic:M2-M5}.

The linearised supersymmetry equation is
\beq
\delta_\epsilon\Theta^j_\beta=\epsilon^{\alpha i}\left(\frac{1}{2}(\gamma^a)_{\alpha\beta}(\gamma_{\underline{b}'})_i^{\,j}\partial_aX^{\underline{b}'}-\frac{1}{6}(\gamma^{abc})_{\alpha\beta}\delta_i^{\,j}h_{abc}\right)=0,
\eeq
where $\underline{a}'$ labels transverse scalars, $a$ labels worldvolume directions, $\alpha,\beta$ are spinor indices of $spin(1,5)$ and $i,j$ are spinor indices of $USp(4)$, the spin covers of $SO(1,5)$ and $SO(5)$ respectively. 
The idea in constructing the solution is to balance the contribution of the 3-form field strength $h$ with a contribution from the scalars. The worldvolume of the string soliton can be taken to be in the $0,1$ directions with all fields independent of $x^0$ and $x^1$, and in general we can take only one of the transverse scalars $X^{5'}=\phi$ to be active. Finally the ansatz $h_{0ab}=0=h_{1ab}$ and $h_{01a}=v_a$ is made. Then we are left with the requirement
\beq
v_a=\pm\frac{1}{4}\partial_a\phi
\eeq
to satisfy supersymmetries whose parameters obey
\beq
\epsilon^{\beta j}=\pm(\gamma^{01})_{\alpha}^{\,\beta}(\gamma^5)_i^{\, j}\epsilon^{\alpha i}.
\eeq

It can be shown that this 1/4 BPS solution (1/4 with respect to 11-dimensional supersymmetry) solves the non-linear equation of motion and supersymmetry condition for the five-brane if the scalar satisfies
\beq
\delta^{mn}\partial_m\phi\partial_n\phi=0.
\eeq
Therefore the string solution is given by
\bea\label{sds}
H_{01m}&=&\pm\frac{1}{4}\partial_m\phi,\nonumber\\
H_{mnp}&=&\pm\frac{1}{4}\epsilon_{mnpq}\delta^{qr}\partial_r\phi,\nonumber\\
\phi&=&\phi_0+\frac{2Q}{|x-x_0|^2},
\eea
where we could replace $\phi$ by a more general superposition of solutions. The string has both magnetic and electric charge given by $\mp Q$. There is a conformal factor in the full equations of motion which guarantees that they are satisfied even at $x=x_0$. Hence no source is required and the solution is truly solitonic. The string soliton has 4 Bosonic and 4 Fermionic zero modes, and its own anomalies which need to be cancelled. The string soliton can be dimensionally reduced to get the BIon spike and other T-dual configurations\cite{HLW}. Properties of the self-dual string were investigated more recently in \cite{Berman}.

\newpage

%% file: ch4.tex
\chapter{The Basu-Harvey Equation and M2-M5 Intersections}\label{BasuHarvey}

This section is primarily concerned with the work of Basu and Harvey\cite{BasuH}, who put forward a proposal for the generalisation of the Nahm equation (\ref{Nahmeq}) that should arise in M2-M5 system. They also presented a solution to this equation representing coincident membranes ending on a five-brane.

As the action for multiple M2-branes is not known, it is a case of using this work to gain insight into what this action should be, rather than having an action from which to derive it. We refer to the equation they put forward as the Basu-Harvey equation, and we will see that it is deduced by requiring properties we would expect from generalising the D1-D3 case to a fuzzy three-sphere cross-section, rather than a fuzzy two-sphere. As Appendix \ref{app:FS3} describes, the fuzzy three-sphere is more complicated than the fuzzy two-sphere. In fact it is more complicated than the fuzzy four-sphere as odd spheres in general are harder to work with. 

The solution can be compared with the self-dual string of the previous section, as well as the energies expected for such a configuration and the properties of fluctuations around the solution. After these checks one can feel a bit more confident and look at some terms we might expect in multiple M2-brane action if the Basu-Harvey equations is to appear as a Bogomol'nyi equation.

The Basu-Harvey equation, and its solution in terms of the fuzzy three-sphere construction given in Appendix \ref{app:FS3}, is not without its drawbacks, not least the fact that the degrees of freedom are $N\times N$  matrices when the expectation was for $N^{3/2}$ degrees of freedom for $N$ coincident membranes.

\section{Properties Expected for the Solution}

We know that the M5 picture of the intersection is in terms of the self-dual string, described in Section \ref{sec:sds} and pictured in Figure \ref{pic:M2-M5}. Thus we expect a relationship between the radius, $R$,  in the directions transverse to the membrane yet tangent to the five-brane (i.e. $R=\sqrt{(X^1)^2+(X^2)^2+(X^3)^2+(X^4)^2}\,$), and $s$, the M2 worldvolume direction away from the five-brane, of the form 
\beq\label{ridge}
s\sim \frac{Q}{R^2},
\eeq
as implied by (\ref{sds}) (Q is the quantised charge, given by the number of membranes). As the self-dual string was a static solution with no dependence on $\sigma$, the co-ordinate along the string, the active scalars should only depend on $s$ for the membrane description as well.

In the D1-D3 case there were three active scalars transverse to the string in the directions of the D3-brane worldvolume. This meant the cross-section was a (fuzzy) two-sphere, as expected for a spike. In our M-theory system we have an extra transverse scalar and the cross-section should be a (fuzzy) three-sphere, giving the $SO(4)$ symmetry we would expect between the scalars. As well as this $SO(4)$ invariance we also demand translation invariance.

The fuzzy $S^3$ co-ordinates, $G^i$, obey the equation
\beq\label{FS3}
G^i +\frac{1}{2(n+2)}\epsilon_{ijkl}G_5G^jG^kG^l =0,
\eeq
where $G_5$ is a difference of projection operators, it obeys $\{G_5,G^i\}=0$ and $(G_5)^2=1$ (full details can be found in Appendix \ref{app:FS3}). This equation is a quantised version of a higher Poisson bracket equation obeyed by the classical 3-sphere (as $[\alpha^a,\alpha^b]=2i\epsilon_{abc}\alpha^c$ is for the two sphere; again more details are contained in Appendix \ref{app:PN}). The fuzzy three-sphere relation (\ref{FS3}) was first obtained in \cite{BasuH} and we will derive it in full in Appendix \ref{app:G4b}. It is around this relation that the Basu-Harvey equation is built.

\section{The Basu-Harvey Equation}

The Basu-Harvey equation is given by 
\beq\label{BaH}
\frac{dX^i}{d s}+\frac{M_{11}^3}{8\pi\sqrt{2N}}\frac{1}{4!}\epsilon_{ijkl}[G_5,X^j,X^k,X^l]=0.
\eeq
The anti-symmetric 4-bracket is a sum over permutations with sign, e.g.
\beq\label{N4b}
[X^1,X^2,X^3,X^4]=\sum_{\mbox{perms}\ \sigma} \mbox{sign}(\sigma)X^{\sigma(1)}X^{\sigma(2)}X^{\sigma(3)}X^{\sigma(4)}\, ,
\eeq
and it can be thought of as a quantum Nambu bracket\cite{Nambu,CZ}. We could have placed such a bracket on the second term of (\ref{FS3}) if we had included the correct combinatorial factor. The equation is translation invariant under $X^i\rightarrow X^i+v^iI$, where $I$ is the identity.

$G_5$ and the scalars will belong to an algebra containing the fuzzy three-sphere. There are three main possibilities for what this algebra is. The first is \mnc, the algebra of $N \times N$ matrices, where $N$ is the dimension of the representation of $SO(4)$ in which the the fuzzy three-sphere we work with lies. For the fuzzy three-sphere only, this dimension coincides with the square of the radius in terms of the $G^i$ (i.e. $\sum_i G^iG^i=N$). $N$ is what we will identify with the number of membranes, and this is the $N$ that appears in the Basu-Harvey equation (\ref{BaH}) above\footnote{The original version of the Basu-Harvey equation did not contain this factor due to an error in calculating (\ref{FS3}). In an updated version of their paper they included a parameter $\lambda$ in the numerator rather than the $1/\sqrt{2N}$ factor we have included following \cite{Nogradi, bermancopland}. $\lambda$ was treated as a coupling with $\lambda^2 N$ required to be fixed in the large-$N$ limit for consistency.}. A second possibility for the algebra is that generated by the $G^i$, which is a sub-algebra of \mnc. The third option is the algebra which in the large-$N$ limit agrees with classical algebra of functions on $S^3$, namely the spherical harmonics. Though this third algebra is important in later sections, we will assume our fields are in \mncs at present. 

By analogy with the D1-D3 system the Basu-Harvey equation should appear as a Bogomol'nyi equation for minimising the energy, and should also follow from the vanishing of the supersymmetry variation of the Fermions on the membrane.

\section{The Membrane Fuzzy Funnel Solution}
Similar to the D1-D3 case we expect a static solution with scalars proportional to the fuzzy 3-sphere co-ordinates $G^i$ and only depending on $s$. An ansatz 
\beq
X^i(s)=f(s)G^i
\eeq
leads quickly to the solution
\beq\label{soln}
X^i(s)=\frac{i\sqrt{2\pi}}{M_{11}^\frac{3}{2}}\frac{1}{\sqrt{s}}G^i.
\eeq
The physical radius is given by
\beq\label{traces2}
R=\sqrt{\left|\frac{\mbox{Tr} \sum(X^i)^2}{\mbox{Tr} \openone}\right|}.
\eeq
Substitution and rearranging results in
\beq
s\sim \frac{N}{R^2} \, ,
\eeq
which is the self dual string behaviour we expect when we identify $N$ with the number of membranes (c.f (\ref{ridge})).

If this is the Bogomol'nyi equation we would expect that we could also minimise the energy by satisfying (\ref{BaH}) with a sign difference between the terms, leading to the opposite topological charge. In this case we can solve by removing the factor of $i$ from (\ref{soln}) or by multiplying each co-ordinate by $G_5$.

\section{Energy and Action for Multiple Membranes}\label{enact}

Given that the Basu-Harvey equation should arise as a Bogomol'nyi equation, as the Nahm equation did in the D1-D3 system (and as the generalised version does in (\ref{D1bog})), it is not unreasonable to define the energy of our static configuration with four non-zero scalars as in \cite{BasuH}
\bea \label{m2energy}
E = T_2 \int d^2 \sigma {\rm Tr} \bigg[ \left( \frac{d X^i}{ds} + 
\frac{M_{11}^3}{8\pi\sqrt{2N}}\frac{1}{4!} \epsilon_{ijkl} 
[G_5, X^j, X^k, X^l] \right)^2  \cr 
+ \left(1- \frac{M_{11}^3}{16\pi \sqrt{2N}} \frac{1}{4!}\epsilon_{ijkl} \left\{ \frac{d X^i}{ds} ,
[G_5, X^j, X^k, X^l] \right\} \right)^2 \bigg]^{1/2} .
\eea
The membrane tension $T_2$ is given by $T_2=M_{11}^3/(2\pi)^2$ and the integral is over the two spatial worldvolume directions $\sigma$ and $s$. In what follows we will consider the $X^i$ to obey $\{G_5,X^i\}=0$. In terms of the fuzzy three-sphere algebra this means restricting the $X^i$ to lie in $Hom(\cRp,\cRm)$ or $Hom(\cRm,\cRp)$. This is of course obeyed by the co-ordinates $G^i$. It means, for instance, that by multiplying out the squares in (\ref{m2energy}) $G_5$ can be eliminated. We can also remove the 4-bracket and combinatorial factor in (\ref{m2energy}) as the $\epsilon$ provides the anti-symmetrisation (though the bracket will be required in the action for translation invariance).

If the scalars obey the Basu-Harvey equation then Bogomol'nyi bound is satisfied and the energy density linearises:
\beq 
E = T_2 \int d^2 \sigma {\rm Tr} \left( 1- \frac{M_{11}^3}{8\pi \sqrt{2N}} \epsilon_{ijkl} 
\frac{d X^i}{ds} G_5 X^j X^k X^l\right).
\eeq    
The second term is a boundary term, and the boundary is the self dual string. Using the solution, the energy can be rewritten for large $N$ as
\beq\label{BHenergy}
 E = N T_2 L \int ds + T_5 L \int 2\pi^2 dR R^3,
\eeq
where we have used $T_5=M_{11}^6/(2\pi)^5$ and $L$ is the length of the string. These two terms have the energy densities you would expect for $N$ membranes and a single five-brane respectively.

As per the discussion of Section (\ref{validity}) we would expect this analysis to be only valid at the core ($s\rightarrow\infty$), though for large $N$ it agrees with the M5-brane picture, a description which should only be valid in the opposite limit. Examining (\ref{m2energy}) we can see that a Taylor expansion in terms of powers of $X^i$ is valid when $M_{11}^6R^6N^3\ll 1$, that is $R\ll \sqrt{N}M_{11}^{-1}$. Thus if $N$ is large, $R$ can be large as well.

Given an expression for the energy such as (\ref{m2energy}) we can expand and deduce terms in the associated action. On generalising away from a static solution with only four non-zero scalars dependent only on $\sigma$, we expect terms like
\bea \label{m2act} 
S &=& -T_2 \int d^3 \sigma {\rm Tr} \bigg[ 1 + 
\left( \p_a X^M \right)^2
-\frac{1}{2N.3!}[X^M,X^N,X^P][X^M,X^N,X^P]
\non\\
&+&\frac{1}{2N.4.3!} \left[ \p_a X^L ,[X^M,X^N,X^P] \right]  \times
\bigg( \left[\p^a X^L ,[X^M,X^N,X^P] \right] 
\non\\&+&\left[ \p^a X^M 
,[X^L,X^P,X^N] \right] + \left[ \p^a X^N ,[X^L,X^M,X^P] \right]\non\\ &+&\left[ 
\p^a X^P ,[X^L,X^N,X^M] \right] 
\bigg)  +\ldots \bigg]^{1/2} \eea
in a multiple membrane action. $L,M,\ldots$ labels the 8 transverse directions and $a,b,\dots$ the 3 worldvolume directions. The three-bracket used is defined analogously to the quantum Nambu 4-bracket (\ref{N4b}) as a sum over the six permutation of the entries, with sign.

\section{Fluctuations on the Funnel}

Basu and Harvey also performed a fluctuation analysis on the membrane fuzzy funnel similar to that performed on the D-string funnel in Section (\ref{furext}). The simplest fluctuations to consider are those in the four directions transverse to both the membrane and the five-brane. The analysis in the D-brane was linear, relying only on the dimensionally reduced super Yang-Mills action, which contained the kinetic term as well as a quartic potential. Basu and Harvey thus took a general kinetic term and the sextic coupling suggested in the last section as being sufficient for a linear analysis of the the membrane fluctuations.
In flat space and static gauge, the pull back of the metric is given by $P[G]_{ab}=\eta_{ab}+\partial_aX^M\partial_bX^M$, taking the determinant will lead to the first two terms of (\ref{m2act}). Therefore the action used for fluctuation analysis is
\beq\label{fluctact}
S=-T_2 \int d^3\sigma\mbox{Tr} \sqrt{-\det(P[G]_{ab})-\frac{1}{2N}\frac{1}{3!}[X^M,X^N,X^P][X^M,X^N,X^P]}\, .
\eeq
The fluctuations may depend on all three worldvolume co-ordinates and are proportional to the identity in the fuzzy sphere algebra, $\delta X^m(t,s,\sigma)=f^m(t,s,\sigma) \openone_N$. We keep terms up to quadratic order in the fluctuations, which gives
\beq
[X^M,X^N,X^P][X^M,X^N,X^P]=3(f^m)^2[X^i,X^j]^2\, ,
\eeq
where $M,N,P$ run over all indices, $m$ runs over the directions transverse to the both branes and $i,j$ run over the non-zero scalars of the solution. Evaluation of the commutator squared on the right-hand side proceeds via
\beq
[G^i, G^j]^2 = 2 G^i G^j G^i G^j -2 N^2\openone
\eeq
and
\beq\label{ijij}
G^i G^j G^i G^j\Prp=-(n+1)(n+3)\Prp=-2N\Prp.
\eeq
Finally this leads to 
\beq\label{ijsq}
[G^i, G^j]^2=-2N(N+2)\, .
\eeq
Note that (\ref{ijij}) and (\ref{ijsq}) correct an error in \cite{BasuH}. However, the leading order in $n$ behaviour of (\ref{ijsq}) remains unchanged and so the conclusions below remain the same.

Returning to the action (\ref{fluctact}) we now have
\beq
S = -T_2 \int d^3 \sigma {\rm Tr} \sqrt{ H -H (\p_t f^m)^2 +(\p_s f^m)^2 
+H (\p_\sigma f^m)^2 +\frac{N+2}{2s^2} (f^m)^2  },
\eeq
where 
\beq
H=1+\frac{\pi N}{2M_{11}^3s^3}.
\eeq
Here also the potential term differs from that of \cite{BasuH} by an additional factor of $1/2$. This is due to our use of the anti-symmetric 3-bracket squared $[X^M,X^N,X^P]^2$, as opposed to $Q^{MNP}H^{MNP}$, where $Q^{MNP}=\{[X^M,X^N],X^P\}$ and $H^{MNP}=Q^{MNP}+Q^{NPM}+Q^{PMN}$. With the correct factor these terms are the same for $M,N,P=i,j,k$, but differ when considering fluctuations in overall transverse directions.

The equation of motion for the linearised fluctuations becomes
\beq
(H \p_t^2 -\p_s^2 -H \p_\sigma^2 ) f^m (t,s,\sigma) 
+\frac{N+2}{2s^2} f^m (t,s,\sigma) =0.
\eeq
In the $s\rightarrow\infty$ limit (where we have a flat membrane) the equation of motion reduces to
\beq
(-\p_t^2 +\p_s^2 +\p_\sigma^2) f^m =0.
\eeq
The solutions to this equation are plane waves with $SO(2,1)$ symmetry in the worldvolume directions, as one would expect for a membrane.
Although in the opposite limit the analysis should not be valid, as per the earlier discussion we keep $N$ large and find agreement with what we would expect. As $s\rightarrow 0$, $H\sim s^{-3}$ and the equation of motion gives
\beq
(-\p_t^2 + \p_\sigma^2) f^m + R^{-3} \frac{\p }{\p R}
\left( R^3 \frac{\p f^m}{\p R} \right) =0.
\eeq 
This has exactly the $SO(2,1)\times SO(4)$ symmetry that we would expect for the M5 worldvolume with string soliton.

Analysis of more complicated fluctuations and in other directions has not been completed due to the increased complexity of the fuzzy three-sphere algebra compared with that of the fuzzy two-sphere.

\section{Issues with the Basu-Harvey Picture}

The main issue with the Basu-Harvey picture of M2-M5 intersections is that the membrane degrees of freedom for $N$ coincident membranes are represented by $N\times N$ matrices. This is also the case in the D1-D3 system, but here there is an understanding of how these degrees of freedom arise as light fundamental strings stretching between the branes. However in M-theory there are various indications \cite{KT,Kleb} that there should be $N^{3/2}$ degrees of freedom, but there is no picture of how this number (or $N^2$) degrees of freedom would arise.

One suggested resolution of the apparent contradiction\cite{BasuH} is the Basu-Harvey picture is an ultra-violet description which flows to an infra-red description in terms of $N^{3/2}$ degrees of freedom - this is the case for D2-branes (as discussed in Section 6.1 of \cite{AGMOO}). In later sections we will pursue the idea that we should use a different algebra than \mncs to describe the fuzzy three-sphere correctly in the large-$N$ limit.

Other weaknesses of the construction that needed to be investigated were a lack of dimensional reduction to the Nahm equation and D1-D3 system, and lack of a supersymmetry transformation which link satisfying the Basu-Harvey equation to preserving supersymmetry. Again, we will address these issues in later chapters.

\newpage

%% file: ch5.tex
\chapter{Calibrations and a Generalised Basu-Harvey Equation}\label{Ch:calib}

The previous section introduced the Basu-Harvey equation and the work of \cite{BasuH}. While this was successful in reproducing many of the properties desired for the M2-M5 system, it was not derived from an action principle and it satisfied many of the consistency checks by construction. Given this somewhat ad-hoc genesis, more complex checks are in order to try to establish its validity. In Section \ref{calib} we saw how the Nahm equation arose in the D1-D3 system and could be written in a more general form, which had solutions representing coincident D-strings ending on a calibrated intersection of D3-branes. Similar calibrated intersections of five-branes also exist, and so it is natural to ask whether the Basu-Harvey equation can be generalised to describe configurations of coincident membranes ending on one of these five-brane intersections. This would be essentially an M-theory version of \cite{CL}.

We find this extension is indeed possible, with additional complexities compared to the D1-D3 case. Firstly we will argue that as in \cite{CL} we can work with a `linearised' action to describe BPS solutions of the coincident membrane theory. We will then present the generalised Basu-Harvey equation, with conditions necessary for it to appear as a Bogomol'nyi equation. Then we will suggest a supersymmetry variation, something not discussed in \cite{BasuH}. With no Fermionic action we work by analogy with how other aspects of the Basu-Harvey picture have generalised the D1-D3 case. It will turn out that the most naive generalisation leads, via the introduction of projectors, exactly to the interpretation of the Basu-Harvey equation as describing coincident membranes ending on intersections of five-branes corresponding to the calibrations. We will then list these possible configurations and simple solutions. As in \cite{CL} we must solve the generalised form of the BPS equation and some algebraic conditions on the brackets. However, the increased complexity for three- and four-brackets leads to additional algebraic conditions whose two-bracket equivalents were identically solved in the D1-D3 case.

\section{A Linear Action for Coincident Membranes}

The generalised version of the Nahm equation was found as the Bogomol'nyi equation of the {\it linear} action in \cite{CL} (if the constraint (\ref{constraint}) holds). This linear action is that of dimensionally reduced super Yang-Mills. In Section \ref{sec:nonlin} we showed that for the states that we are interested in, the linear action was in fact equal to the full action (see also the discussion of \cite{Brecher}). We find something similar at work for the membrane theory. The starting point is the inferred energy for the coincident membranes ending on a single five-brane, given in \cite{BasuH} and reproduced in (\ref{m2energy}):
\bea \label{m2energy2}
E = T_2 \int d^2 \sigma {\rm Tr} \bigg[ \left( \frac{d X^i}{ds} + 
\frac{M_{11}^3}{8\pi\sqrt{2N}}\frac{1}{4!} \epsilon_{ijkl} 
[G_5, X^j, X^k, X^l] \right)^2  \cr 
+ \left(1- \frac{M_{11}^3}{16\pi \sqrt{2N}} \frac{1}{4!}\epsilon_{ijkl} \left\{ \frac{d X^i}{ds} ,
[G_5, X^j, X^k, X^l] \right\} \right)^2 \bigg]^{1/2} .
\eea
If the Basu-Harvey equation (\ref{BaH}) holds and $\{G_5,X^i\}=0$, then the remaining terms under the square-root  can be rewritten as the perfect square
\beq
E=T_2\int d^2\sigma \mbox{Tr}\left[\left(1+\frac{1}{2}(\partial_\sigma
X^i)^2-\frac{1}{2N}\frac{1}{2.3!}[X^i,X^j,X^k]^2\right)^2\right]^{1/2} .
\eeq
This is the energy is that one would get from an action
\beq\label{3action}
S=-T_2\int d^3\sigma
\mbox{Tr}\left(1+\frac{1}{2}(\partial_a
X^i)^2-\frac{1}{2N}\frac{1}{2.3!}[X^j,X^k,X^l]^2\right) \, ,
\eeq
the ``linearised" form of the membrane action (\ref{m2act}) for three non-zero scalars. It can also be compared with the action used for the fluctuation analysis (\ref{fluctact}). As per the discussion of Section \ref{sec:nonlin} we expect this ``linear'' action to be valid for more general flat BPS configurations. The action we will use when looking for a consistent generalised Basu-Harvey equation will therefore be (\ref{3action}) but with the indices $i,j,k$ running over all eight scalars. As we deal with static, $\sigma$-independent configurations we will work from corresponding expression for the energy instead. Although we do not have a coupling constant in which to expand, we can expand in powers of $X^i$ (as stated in Section \ref{enact}) and it is in this expansion that we are working to leading order.

\section{A Generalised Basu-Harvey Equation}

As explained in the previous section, our starting point is the action
 \beq
E=\frac{T_2}{2}\int d^2\sigma \mbox{Tr}\left( X^{i^ \prime}
X^{i^\prime }-\frac{1}{3!}[X^j,X^k,X^l][X^j,X^k,X^l]\right)
\eeq
(the constant piece has been subtracted). The indices $i,j,\dots$ run from 2 to 9 and $X^{10}$ is identified with $\sigma$. We have scaled out the factor of $1/(2N)$, as was previously done with the numerical factors to simplify the presentation. Reintroduction of this factor just involves inserting a factor of $1/\sqrt{2N}$ with each three- or four-bracket. We proceed as in Section \ref{sec:GNE} by using the usual Bogomol'nyi construction to write 
\beq\label{m2bog}
E=\frac{T_2}{2}\int d^2\sigma \left\{\mbox{Tr}\left(
X^{i^\prime}+g_{ijkl}\frac{1}{4!}[H^*,X^j,X^k,X^l]\right)^2+T\right\}\, ,
\eeq
where $T$ is a {\it{topological}} piece given by
\beq T=-T_2\int
d^2\sigma
\mbox{Tr}\left(g_{ijkl}X^{i^\prime}\frac{1}{4!}[H^*,X^j,X^k,X^l]\right)\, .
\eeq
When there are only four non-zero scalars ($X^2,\ldots,X^5$) and $g_{ijkl}=\epsilon_{ijkl}$ we can do this without any additional conditions. (Note that we now shift the earlier solution of Chapter \ref{BasuHarvey} from the 1 to 4 directions to the 2 to 5 directions, so that $X^i\sim G^{i-1}$ and $\epsilon_{2345}=1$ etc.) The topological piece gives the energy of the five-brane on which the membranes end. It is the second term of (\ref{BHenergy}).

If more than four scalars are non-zero, then we must impose
\bea\label{fullconst}
&&\frac{1}{3!}g_{ijkl}g_{ipqr}\mbox{Tr}\left([H^*,X^j,X^k,X^l][H^*,X^p,X^q,X^r]\right)\\
\nonumber &&=\mbox{Tr}\left([H^*,X^i,X^j,X^k][H^*,X^i,X^j,X^k]\right) \eea
in order to be able to rewrite the action as in (\ref{m2bog}). $H^*$ is a more general form of $G_5$, chosen to have the analogous properties $\{H^*,X^i\}=0$ and $(H^*)^2=1$. In fact using these properties (\ref{fullconst}) reduces to the simpler form
\beq\label{mconstraint}
\frac{1}{3!}g_{ijkl}g_{ipqr}\mbox{Tr}\left([X^j,X^k,X^l][X^p,X^q,X^r]\right)=\mbox{Tr}\left([X^i,X^j,X^k][X^i,X^j,X^k]\right)
\, ,   
\eeq 
which is the M-theory version of (\ref{constraint}).

Once we have written the energy in the form (\ref{m2bog}) using (\ref{mconstraint}) then we can clearly minimise it by imposing the generalisation of the Basu-Harvey equation 
\beq\label{myNahm}
\pd{X^i}{s}+\frac{M_{11}^3}{\sqrt{2N}8\pi}g_{ijkl}\frac{1}{4!}[H^*,X^j,X^k,X^l]=0.
\eeq
This equation has factors restored, and $g$ is a general anti-symmetric four-tensor. When $g_{ijkl}=\epsilon_{ijkl}$ we recover the Basu-Harvey equation and (\ref{mconstraint}) is an identity. 

\section{Equation of Motion}

The equation of motion following from the action (\ref{3action}) is given by 
\beq
X^{i^{\prime\prime}}=-\frac{1}{2}\lt X^j,X^k,[X^i,X^j,X^k]\rt
\eeq
where the three bracket $\lt A,B,C \rt$ is the sum of the six
permutations of the three entries, but with the sign of the permutation
determined only by the order of the first two entries; i.e. $ABC, ACB
\mbox{ and } CAB$ are the positive permutations.  By using the
Bogomol'nyi equation (\ref{myNahm}) twice on the left-hand side it is equivalent to: 
\beq
\label{3bconst} \frac{1}{3!} g_{ijkl} g_{jpqr}\lt
X^k,X^l,[X^p,X^q,X^r]\rt=-\lt X^j,X^k,[X^i,X^j,X^k]\rt \, \, .  
\eeq
After multiplying by $X^i$ and taking the trace, we recover the
constraint equation (\ref{mconstraint}). Thus in summary, the
solutions of the generalised Basu-Harvey equation (\ref{myNahm}) that
obey the algebraic equation of motion (\ref{3bconst}) are minimal energy solutions to the
equations of motion of the proposed membrane action (\ref{3action}).

\section{Supersymmetry}

In the D1-D3 system we saw that the Nahm equation could be derived {\it either} as the Bogomol'nyi equation for minimising the energy, {\it or} as a requirement for preserving half the supersymmetry (Section \ref{sec:d1susy}). We also saw how for D1-strings ending on calibrated intersections of D3-branes, preservation of some fraction of supersymmetry was again dependent on the generalised Nahm equation holding, and the imposition of projectors on the supersymmetry parameter. There was a different projector corresponding to each D3-brane making up the calibrated intersection. There also remained some algebraic conditions on the brackets (see (\ref{dc2})) and when these constraints were fulfilled the constraint (\ref{constraint}) was satisfied, implying so were the equations of motion. Thus requiring the preservation of some supersymmetry lead to a simpler way of expressing the conditions necessary to satisfy the constraint for each configuration.

For the D1-D3 system the linearised supersymmetry variation was that of dimensionally reduced super Yang-Mills. We do not have such information here. The linear multi-membrane theory may be related to a Yang-Mills-like theory based on a three-form rather than a two -form, but such a theory is unknown. We do not have a supersymmetry variation from which to start.

What we can do is to impose by fiat a simple generalisation of the linearised supersymmetry variation for the D1-strings and determine whether it leads to a consistent picture of membranes ending on five-branes. As for the D1-D3 case we find that if the generalised form of the Bogomol'nyi equation is satisfied then it leads to a simplified form of the supersymmetry variation, where the route to further simplification is the imposition of a set of projectors corresponding to the non-zero components of $g_{ijkl}$. Compatible sets of these projectors are in correspondence with the known calibrated five-brane intersections  and the imposition of these sets leads to the preservation of a certain fraction of supersymmetry, the fraction being that preserved by the corresponding five-brane intersection with a membrane attached. This is of course provided we satisfy the algebraic conditions on the brackets left over in the supersymmetry condition after imposition of the projectors. Similar to the case of D3-brane intersections, the total space of all the intersecting five-branes in the submanifold calibrated by $g=\frac{1}{4!}g_{ijkl}dx^i\wedge dx^j\wedge dx^k\wedge dx^l$.

It remains to check if the algebraic conditions on the brackets are enough to satisfy the constraint (\ref{mconstraint}) and thus the equations of motion. It turns out that it is not quite enough, there are additional algebraic conditions, a set of equations of similar form for all configurations, which must be satisfied to solve the constraint. For the D1-D3 case these had a simpler form and were satisfied identically. 

That the supposed supersymmetry variation reproduces the calibrated intersections and shows how to solve the constraint (\ref{mconstraint}) for these calibrated intersections  is evidence that it is in someway correct. At the very least it is a useful tool for finding configurations that solve the constraint. Whether a supersymmetric multi-membrane action with this supersymmetry variation can be determined remains an intriguing open question, and some progress in this direction has been made in \cite{baggerandlambert} where a model supersymmetric membrane theory with a supersymmetry variation of the same form as we use was found (however the supersymmetry algebra did not close, up to what seemed to be some kind of novel gauge transformation).

\subsection{The Supersymmetry Variation}

The most obvious suggestion for the supersymmetry variation is 
\beq
\delta\lambda=\left(\frac{1}{2}\partial_\mu
X^i\Gamma^{\mu
i}-\frac{1}{2.4!}\gtb{i}{j}{k}\Gamma^{ijk}\right)\epsilon. 
\eeq 
We substitute the generalised Basu-Harvey equation in the first term and
rearrange. The requirement that the supersymmetry variation
vanishes becomes that
\beq\label{susycond}
\sum_{i<j<k}\tb{i}{j}{k}\Gamma^{ijk}(1-g_{ijkl}\Gamma^{ijkl\#})\epsilon=0,
\eeq
where we have removed and overall factor of $H^*$ from the left-hand side
since, like $G_5$, it is the difference of projection operators onto
orthogonal sub-spaces and has trivial kernel. $\epsilon$ is the preserved supersymmetry on the membrane \wv and we
have $\Gamma^{01\#}\epsilon=\epsilon$, where the membrane's worldvolume
is in the 0, 1 and $10=\#$ directions. We can then solve the
supersymmetry condition (\ref{susycond}) by defining projectors
\beq
\label{proj} P_{ijkl}=\frac{1}{2}(1-g_{ijkl}\Gamma^{ijkl\#})\, ,
\eeq
where there is no sum over $i,j,k \mbox{ or }l$. We normalise
$g_{ijkl}=\pm1$ so they obey $P_{ijkl}P_{ijkl}=P_{ijkl}$. (Note, in all the
cases that we will consider, for each triplet $i,j,k$, $g_{ijkl}$ is only
non-zero for at most one value of $l$). We impose $P_{ijkl}\epsilon=0$
for each $i,j,k,l$ such that $g_{ijkl}\neq0$. Then by using
the membrane projection ($\Gamma^{01\#}\epsilon=\epsilon$) we can see
that each projector $P_{ijkl}$ corresponds to a five-brane in the
$0,1,i,j,k,l$ directions. To apply the projectors simultaneously, the matrices $\Gamma_{ijkl\#}$ need to commute with each other. $[\Gamma_{ijkl\#},\Gamma_{i'j'k'l'\#}]=0$ if and only if the
sets $\{i,j,k,l\}$ and  $\{i',j',k',l'\}$ have two or zero elements in
common, corresponding to five-branes intersecting over a three-brane
soliton or a string soliton.

Once we impose the set of mutually commuting projectors, the
supersymmetry transformation (\ref{susycond}) reduces to
\beq\label{susyids} \sum_{g_{ijkl}=
0}\tb{i}{j}{k}\Gamma^{ijk}\epsilon=0.  
\eeq
Here we sum over triplets $i,j,k,$ such that $g_{ijkl}=0$ for
all $l$. Using the projectors allows us to express these as a set of
conditions on the 3-brackets alone.

\section{Five-Brane Configurations}

We will now describe the specific equations that correspond 
to the various possible intersecting five-brane configurations. 

The five-branes must always have at least one spatial direction in common, 
corresponding to the direction in which the membrane intersects the five-branes. As discussed in Section \ref{subcalib} for D3-branes, these configurations of five-branes can also be thought
of as a single five-brane stretched over a calibrated manifold \cite{Harvey:1982xk}. These
five-brane intersections can be found in 
\cite{GP,Jerome,Acharya:1998en}. We list the
conditions following from the modified Basu-Harvey equation, those following 
from the supersymmetry conditions
(\ref{susyids}) (with $\nu$ the fraction of preserved supersymmetry) and then 
discuss any remaining conditions required to
satisfy the constraint (\ref{3bconst}). In the D1-string case
only the conditions on the brackets following from the  supersymmetry conditions were
required to satisfy the equivalent constraint.

\subsection{Configuration 1}

The first configuration corresponding to a single five-brane is that originally examined in \cite{BasuH}. We present it in a standard format so it can be compared with the more complex intersections.

\bea\label{c1} \m M5:&1&2&3&4&5&&&\cr
M2:&1&&&&&&&&\#\cr\em\nonumber\\ g_{2345}=1\quad\quad\quad
\nu=1/2\quad {} \eea \bea {X^2}' = -H^*\tb{3}{4}{5}&,& \quad {X^3}' =
H^*\tb{4}{5}{2}\ , \quad \nonumber\\ {X^4}' = -H^*\tb{5}{2}{3}&,&
\quad {X^5}' = H^*\tb{2}{3}{4}\ . \nonumber\\ \nonumber \eea

\subsection{Configuration 2}

For the next case we consider two five-branes intersecting over a three-brane corresponding to an SU(2) K\"{a}hler calibration of a two-surface embedded in four dimensions. In terms of the first five-brane's worldvolume theory the condition for preserved supersymmetry (and the intersection being a calibration) is just the Cauchy-Riemann equations: the complex scalar $Z=X^6+iX^7$ must be a holomorphic function of the complex worldvolume co-ordinate $z=x^4+ix^5$. The calibration form for this intersection contains the K\"{a}hler form. The activated scalars are $X^2$ to $X^7$.

\bea\label{c2} \m M5:&1&2&3&4&5\cr M5:&1&2&3&&&6&7\cr
M2:&1&&&&&&&&\#\cr\em\nonumber\\
g_{2345}=g_{2367}=1\quad\quad\quad\quad\qquad \nu=1/4\quad {} \eea
\bea {X^2}' =-H^*\tb{3}{4}{5} -H^*\tb{3}{6}{7} &,& \quad {X^3}' =
H^*\tb{4}{5}{2} +H^*\tb{6}{7}{2}\ , \quad\nonumber\\ {X^4}' =
-H^*\tb{5}{2}{3} &,& \quad {X^5}' = H^*\tb{2}{3}{4}\ , \quad
\nonumber\\ {X^6}' = -H^*\tb{7}{2}{3} &,& \quad {X^7}' =
H^*\tb{2}{3}{6}\ , \nonumber\\ \tb{2}{4}{6}= \tb{2}{5}{7} &,&  \quad
\tb{2}{5}{6}=-\tb{2}{4}{7}\ , \quad \nonumber\\ \tb{3}{4}{6}=
\tb{3}{5}{7} &,&  \quad \tb{3}{5}{6}=-\tb{3}{4}{7}, \quad \nonumber\\
\tb{4}{5}{6}=\tb{4}{5}{7}&=&\tb{4}{6}{7}=\tb{5}{6}{7}=0. \quad
\nonumber \eea

In order to satisfy the constraint we need the $X^i$'s to satisfy the
following equations:

$$ \mbox{Choose } m\in\{2,3\},\quad i,j,k,l\in\{4,5,6,7\}\, ,\ \mbox{then}\nonumber\\
$$\bea \epsilon_{ijk}\fbr{i}{m}{m}{j}{k}&=&0,\quad(\mbox{no sum over }
m)\nonumber\\ \epsilon_{ijkl}\fbr{i}{j}{m}{k}{l}&=&0.\label{jac1} \eea

In the string theory case there were no additional equations, as
apart from the Nahm like equations and algebraic conditions on the
brackets all that was needed to solve the constraint was the Jacobi
identity, $\epsilon_{ijk}[\Phi^i,[\Phi^j,\Phi^k]]$. If $X^m$ anti-commutes with $X^i,X^j,X^k$ then
the first equation of (\ref{jac1}) reduces to the Jacobi identity. Similarly if $X^m$
anti-commutes with $X^i,X^j,X^k,X^l$ the second equation reduces to
\beq
 \epsilon_{ijkl}X^iX^jX^kX^l=0.
\eeq
For all the following configurations the additional algebraic conditions take the same form as (\ref{jac1}), with only changes in the indices to be considered. Although equations of this form are not satisfied for general matrices in \mnc, we speculate that perhaps if the $X^i$ are restricted to a more refined algebra containing the fuzzy three-sphere (such as those considered in the next section) these equations could become identities.

\subsection{Configuration 3}

Three five-branes can intersect on a three-brane corresponding to an SU(3) K\"{a}hler
calibration of a two-surface embedded in six dimensions. The active scalars are $X^2$ to $X^9$.

\bea\label{c3} \m M5:&1&2&3&4&5\cr M5:&1&2&3&&&6&7\cr
M5:&1&2&3&&&&&8&9\cr M2:&1&&&&&&&\#\cr\em\nonumber\\
g_{2345}=g_{2367}=g_{2389}=1\quad\quad\quad\qquad \nu=1/8\quad {} \eea
\bea {X^2}' =-H^*\tb{3}{4}{5} &-&H^*\tb{3}{6}{7}-H^*\tb{3}{8}{9} ,
\quad\nonumber\\ {X^3}' = H^*\tb{4}{5}{2}
&+&H^*\tb{6}{7}{2}+H^*\tb{8}{9}{2}\ , \quad\nonumber\\ {X^4}' =
-H^*\tb{5}{2}{3} &,& \quad {X^5}' = H^*\tb{2}{3}{4}\ , \quad
\nonumber\\ {X^6}' = -H^*\tb{7}{2}{3} &,& \quad {X^7}' =
H^*\tb{2}{3}{6}\ , \nonumber\\ {X^8}' = -H^*\tb{9}{2}{3} &,& \quad
{X^9}' = H^*\tb{2}{3}{8}\ , \nonumber\\ \tb{2}{4}{6}= \tb{2}{5}{7} &,&
\quad \tb{2}{5}{6}=-\tb{2}{4}{7}, \quad \nonumber\\ \tb{2}{4}{8}=
\tb{2}{5}{9} &,&  \quad  \tb{2}{5}{8}=-\tb{2}{4}{9}\ , \quad
\nonumber\\ \tb{2}{6}{8}= \tb{2}{7}{9} &,&  \quad
\tb{2}{7}{8}=-\tb{2}{6}{9}, \quad \nonumber\\ \tb{3}{4}{6}=
\tb{3}{5}{7} &,& \quad \tb{3}{5}{6}=-\tb{3}{4}{7}, \quad \nonumber\\
\tb{3}{4}{8}= \tb{3}{5}{9} &,&  \quad  \tb{3}{5}{8}=-\tb{3}{4}{9}\ ,
\quad \nonumber\\ \tb{3}{6}{8}= \tb{3}{7}{9} &,&  \quad
\tb{3}{7}{8}=-\tb{3}{6}{9}, \quad \nonumber\\
\tb{4}{5}{6}+\tb{6}{8}{9}=0\ &,&\quad \tb{4}{5}{7}+\tb{7}{8}{9}=0\
,\quad \nonumber\\ \tb{4}{5}{8}+\tb{6}{7}{8}=0\ &,&\quad
\tb{4}{5}{9}+\tb{6}{7}{9}=0\ ,\quad \nonumber\\
\tb{4}{6}{7}+\tb{4}{8}{9}=0\ &,&\quad \tb{5}{6}{7}+\tb{5}{8}{9}=0\
,\quad \nonumber\\
\tb{4}{6}{8}=\tb{4}{7}{9}&+&\tb{5}{6}{9}+\tb{5}{7}{8}\ ,
\quad\nonumber\\
\tb{5}{7}{9}=\tb{5}{6}{8}&+&\tb{4}{7}{8}+\tb{4}{6}{9}\ .
\quad\nonumber \eea

In order to satisfy the constraint again we have additional algebraic
constraints for certain $X^i$'s; for this we define ``pairs'' as
$\{2,3\}$,$\{4,5\}$, $\{6,7\}$ and $\{8,9\}$.
$$ \mbox{Choose } m\in\{2,3\},\quad i,j,k,l\in\{4,5,6,7,8,9\}\mbox{
such that } \{i,j\},\{k,l\} \mbox{ are pairs. Then}\nonumber\\
$$\bea \epsilon_{ijk}\fbr{i}{m}{m}{j}{k}&=&0,\quad(\mbox{no sum over }
m)\nonumber\\ \epsilon_{ijkl}\fbr{i}{j}{m}{k}{l}&=&0. \label{jac3} \eea

\subsection{Configuration 4}

The next configuration has three five-branes intersecting over a string which corresponds to an SU(3) K\"{a}hler calibration of a four-surface in six dimensions.
There are only six activated scalars.

\bea\label{c4} \m M5:&1&2&3&4&5\cr M5:&1&2&3&&&6&7\cr
M5:&1&&&4&5&6&7\cr M2:&1&&&&&&&\#\cr\em\nonumber\\
g_{2345}=g_{2367}=g_{4567}=1\quad\quad\quad\qquad \nu=1/8\quad {} \eea
\bea {X^2}' &=&-H^*\tb{3}{4}{5} -H^*\tb{3}{6}{7}\ , \quad\nonumber\\
{X^3}' &=& H^*\tb{2}{4}{5} +H^*\tb{2}{6}{7}\ , \quad\nonumber\\ {X^4}'
&=&-H^*\tb{2}{3}{5} -H^*\tb{5}{6}{7}\ , \quad\nonumber\\ {X^5}' &=&
H^*\tb{2}{3}{4} +H^*\tb{4}{6}{7}\ , \quad\nonumber\\ {X^6}'
&=&-H^*\tb{2}{3}{7} -H^*\tb{4}{5}{7}\ , \quad\nonumber\\ {X^7}' &=&
H^*\tb{2}{3}{6} +H^*\tb{4}{5}{6}\ , \quad\nonumber\\
\tb{2}{4}{6}&=&\tb{2}{5}{7}+\tb{3}{5}{6}+\tb{3}{4}{7}\ ,
\quad\nonumber\\
\tb{3}{5}{7}&=&\tb{3}{4}{6}+\tb{2}{5}{6}+\tb{2}{4}{7}\ .
\quad\nonumber \eea

In order to satisfy the constraint again we have to satisfy additional
algebraic constraints for certain $X^i$'s; for this we define
``pairs'' as $\{2,3\}$,$\{4,5\}$ and $\{6,7\}$.
$$ \mbox{Choose } i,j,k,l,m\in\{2,3,4,5,6,7\}\mbox{ such that }
\{i,j\},\{k,l\} \mbox{ are pairs. Then}\nonumber\\
$$\bea \epsilon_{ijk}\fbr{i}{m}{m}{j}{k}&=&0,\quad(\mbox{no sum over }
m)\nonumber\\ \epsilon_{ijkl}\fbr{i}{j}{m}{k}{l}&=&0.\label{jac4}  \eea

\subsection{Configuration 5}

In the next configuration we are forced by supersymmetry to have an
additional anti-brane. Even though there are only three independent projectors this configuration has three five-branes and an anti-five-brane intersecting over a membrane. This corresponds to the SU(3) special Lagrangian calibration of a three-surface embedded in six dimensions.

\bea\label{c5} \m M5:&1&2&3&4&5\cr M5:&1&2&&4&&6&&8\cr
\bar{M5}:&1&2&3&&&6&7\cr M5:&1&2&&&5&&7&8\cr
M2:&1&&&&&&&\#\cr\em\nonumber\\
g_{2345}=g_{2468}=-g_{2367}=g_{2578}=1\quad\quad\qquad \nu=1/8\quad {}
\eea \bea {X^2}' =-H^*\tb{3}{4}{5} &-&H^*\tb{4}{6}{8} +H^*\tb{3}{6}{7}
-H^*\tb{5}{7}{8}\ , \nonumber\\ {X^3}' &=& H^*\tb{2}{4}{5}
-H^*\tb{2}{6}{7}\ , \quad\nonumber\\ {X^4}' &=&-H^*\tb{2}{3}{5}
-H^*\tb{2}{6}{8}\ , \quad\nonumber\\ {X^5}' &=& H^*\tb{2}{3}{4}
+H^*\tb{2}{7}{8}\ , \quad\nonumber\\ {X^6}' &=& H^*\tb{2}{3}{7}
-H^*\tb{2}{4}{8}\ , \quad\nonumber\\ {X^7}' &=&-H^*\tb{2}{3}{6}
-H^*\tb{2}{5}{8}\ , \quad\nonumber\\ {X^8}' &=& H^*\tb{2}{4}{6}
+H^*\tb{2}{5}{7}\ , \quad\nonumber\\ \tb{3}{4}{7}&=&\tb{3}{5}{6}\
,\quad \tb{4}{3}{8}=\tb{4}{5}{6}\ ,\quad\nonumber\\
\tb{5}{3}{8}&=&\tb{5}{7}{4}\ ,\quad \tb{6}{4}{7}=\tb{6}{8}{3}\
,\quad\nonumber\\ \tb{7}{3}{8}&=&\tb{7}{6}{5}\ ,\quad
\tb{8}{4}{7}=\tb{8}{6}{5}\ ,\quad\nonumber\\
\tb{2}{3}{8}&+&\tb{2}{4}{7}+\tb{2}{6}{5}=0\ , \quad\nonumber\\
\tb{6}{7}{8}&+&\tb{4}{5}{8}+\tb{3}{4}{6}+\tb{3}{5}{7}=0\ .
\quad\nonumber \eea

In order to satisfy the constraint once again we have similar
additional algebraic constraints for certain $X^i$'s; for this we
define ``pairs'' as $\{3,8\}$,$\{4,7\}$ and $\{5,6\}$.
$$ \mbox{Choose }m\in\{2,\dots,8\},\ i,j,k,l,\in\{3,4,5,6,7,8\}\mbox{
such that } \{i,j\},\{k,l\} \mbox{ are pairs. Then}\nonumber
$$ \bea \epsilon_{ijk}\fbr{i}{m}{m}{j}{k}&=&0,\quad(\mbox{no sum over
} m)\nonumber\\ \epsilon_{ijkl}\fbr{i}{j}{m}{k}{l}&=&0.\label{jac5}
\eea
We have described all of the configurations preserving $1/8$ of the membrane supersymmetry. There exist additional calibrations preserving less supersymmetry with more five-branes, which we expect can be treated in the same manner. This would involve tedious calculation leading to no new enlightenment and has not been attempted.

\section{Solutions}

Recall that the Basu-Harvey equation is solved by
\beq 
X^i(s)=\frac{i\sqrt{2\pi}}{M_{11}^{3/2}}\frac{1}{\sqrt{s}}G^i\, . 
\eeq
We can solve the cases of intersecting five-branes analogously to the intersecting three-branes of \cite{CL} by effectively using multiple copies of this solution. The first multi-five-brane case (\ref{c2}) is solved by setting
\beq 
X^i(s)=\frac{i\sqrt{2\pi}}{M_{11}^{3/2}}\frac{1}{\sqrt{s}}H^i, 
\eeq
where the $H^i$ are given by the block-diagonal $2N\times 2N$ matrices
\bea
H^2 &=& \mbox{diag}\, (G^1,G^1)\nonumber\\ H^3 &=& \mbox{diag}\,
(G^2,G^2)\nonumber\\ H^4 &=& \mbox{diag}\, (G^3,0)\nonumber\\ H^5 &=&
\mbox{diag}\, (G^4,0)\nonumber\\ H^6 &=& \mbox{diag}\,
(0,G^3)\nonumber\\ H^7 &=& \mbox{diag}\, (0,G^4)\nonumber\\ H^* &=&
\mbox{diag}\, (G^5,G^5)\, , 
\eea
which are such that
\beq 
H^i+\frac{1}{2(n+2)}g_{ijkl}\frac{1}{4!}[H^*,H^j,H^k,H^l]=0.  
\eeq
This makes sure that the conditions following
from the generalised Basu-Harvey equation vanish. The remaining
conditions in (\ref{c2}), that is those following from the
supersymmetry transformation, are satisfied trivially as all terms in the three
brackets involved vanish for this solution. The first additional algebraic equation of (\ref{jac1}) is satisfied for the solution as the indices must be chosen such that for each diagonal block at least one of the $X^i$'s appearing in the bracket has a zero there, thus the term with each permutation
vanishes independently. Again the second additional algebraic equation is trivially satisfied as there are no non-zero products of five different $X^i$'s.

The more complicated cases follow easily: configuration 3 is given by
the block-diagonal $3N\times 3N$ matrices

\bea H^2 &=& \mbox{diag}\, (G^1,G^1,G^1)\nonumber\\ H^3 &=&
\mbox{diag}\, (G^2,G^2,G^2)\nonumber\\ H^4 &=& \mbox{diag}\,
(G^3,0,0)\nonumber\\ H^5 &=& \mbox{diag}\, (G^4,0,0)\nonumber\\ H^6
&=& \mbox{diag}\, (0,G^3,0)\nonumber\\ H^7 &=& \mbox{diag}\,
(0,G^4,0)\nonumber\\ H^8 &=& \mbox{diag}\, (0,0,G^3)\nonumber\\ H^9
&=& \mbox{diag}\, (0,0,G^4)\nonumber\\ H^* &=& \mbox{diag}\,
(G^5,G^5,G^5)\, , \eea

and configuration 4 by

\bea H^2 &=& \mbox{diag}\, (G^1,G^1,0)\nonumber\\ H^3 &=&
\mbox{diag}\, (G^2,G^2,0)\nonumber\\ H^4 &=& \mbox{diag}\,
(G^3,0,G^1)\nonumber\\ H^5 &=& \mbox{diag}\, (G^4,0,G^2)\nonumber\\
H^6 &=& \mbox{diag}\, (0,G^3,G^3)\nonumber\\ H^7 &=& \mbox{diag}\,
(0,G^4,G^4)\nonumber\\ H^* &=& \mbox{diag}\, (G^5,G^5,G^5)\, . \eea
\newpage
Configuration 5 is

\bea H^2 &=& \mbox{diag}\, (G^1,G^1,G^1,G^1)\nonumber\\ H^3 &=&
\mbox{diag}\, (G^2,0,G^2,0)\nonumber\\ H^4 &=& \mbox{diag}\,
(G^3,G^2,0,0)\nonumber\\ H^5 &=& \mbox{diag}\,
(G^4,0,0,G^2)\nonumber\\ H^6 &=& \mbox{diag}\,
(0,G^3,G^4,0)\nonumber\\ H^7 &=& \mbox{diag}\,
(0,0,G^3,G^3)\nonumber\\ H^8 &=& \mbox{diag}\,
(0,G^4,0,G^4)\nonumber\\ H^* &=& \mbox{diag}\, (G^5,G^5,G^5,G^5) \, . \eea

As in the D1-D3 case there will exist many less trivial solutions  containing off-diagonal terms. These will describe configurations when the branes are no longer flat.
\newpage

%% file: ch6.tex
\chapter{Projection to Fuzzy Spherical Harmonics}\label{Ch:FSH}

Though so far throughout this dissertation we have taken the membrane fields to lie in \mnc, the algebra of complex $N\times N$ matrices, as stated earlier and in \cite{BasuH} this choice is not unique. One other possibility is to use the algebra generated by taking products of the fuzzy three-sphere co-ordinate matrices $G^i$, which is a subalgebra of \mnc. However, the choice that we shall be interested in is \anstp, the algebra that reduces to the classical algebra of functions on the sphere in the large-$N$ limit. We will describe this algebra in the next section, with additional details in Appendix \ref{app:pshydb}. It is a lot more complicated to work with than \mnc, stemming from the fact that it is not closed under multiplication. That means that we have to project back into the algebra after multiplying. This projection is then a source of non-associativity. This non-associativity, along with the non-commutativity, disappears in the large-$N$ limit. This is what we would expect if we are to reproduce the classical algebra of functions. 

To see how the projection works we have to break the full \mncs fuzzy three-sphere into a basis in terms of Young diagrams of $SO(4)$. The projection is a restriction on operators corresponding to allowing only completely symmetric diagrams, that is those with only one row. The intriguing thing about the algebra \anst is that the projection reduces the degrees of freedom. In fact we will see that it reduces the degrees of freedom to scale with $N$ like $N^{3/2}$, exactly the behaviour expected for the degrees of freedom of coincident membranes (the radius of the fuzzy sphere remains the same, namely $(G^i)^2=N)$. 

Classically there would be infinitely many degrees of freedom on the three-sphere, describable as spherical harmonics with arbitrarily high angular momentum. We saw earlier (Section \ref{furext}) how the D-string fluctuation analysis led to a cut-off on the angular momentum of modes propagating out along the string, solving a problem with the D3-brane fluctuation analysis. Again here the fuzziness of the sphere provides an ultraviolet cut-off which limits the number of degrees of freedom. The ratio of the cut-off and the radius is $N$-dependent such that the number of degrees of freedom scale like $N^{3/2}$. Classically one would expect the degrees of freedom on the surface of a three-sphere to scale like the surface's volume, i.e. like $R^3$. 

The appearance of a non-associative algebra on five-branes in the presence of background flux is described in \cite{Hofman:2001zt}. Here there is flux due to the end of the membrane and it would be natural for non-associativity to appear as the membranes open out to a five-brane. Non-associative fuzzy spheres can also occur for D-brane intersections\cite{AlekseevRS}. A non-associative algebra for coincident membranes is also suggested in \cite{baggerandlambert}, though it appears to take a different form.

After a brief description of the Young diagram basis of the fuzzy three-sphere we shall describe the projection to the `fuzzy spherical harmonics' and calculate the number of degrees of freedom surviving this projection. We will then show how non-associativity arises and then discuss the relationship with the Basu-Harvey equation.

\section{The Fuzzy Three-Sphere and its Young Diagram Basis}

Here we give a brief description of the fuzzy three-sphere and its Young diagram basis. This is intended to be sufficient to understand the results of this chapter and the next, and full details of the three-sphere construction are given in Appendix \ref{app:FS3} with some additional discussion on the Young diagram basis for the four-sphere given in Appendix \ref{app:YDB}.

The even sphere construction we use is based on $n$-fold symmetric tensor product representations of the spin cover of the sphere diffeomorphism group\cite{CLT,Ram1, Ram2, Ram3}. For the four sphere this means $n$-fold tensor products of $V$, the four-dimensional spinor representation, on which the $Spin(5)$
$\Gamma$-matrices act. The co-ordinates $\gh^\mu$ for the fuzzy four-sphere are given by 
\beq
\gh^\mu=(\Gamma^\mu\otimes1\otimes1\otimes\ldots\otimes1+1\otimes\Gamma^\mu\otimes1\otimes\ldots\otimes1+\ldots+1\otimes\ldots\otimes1\otimes\Gamma^\mu)_{sym}
\eeq
(from here onwards the indices $\mu,\nu,\ldots$ will run from 1 to 5, $i,j,\dots$
will run from 1 to 4 and $a,b,\dots$ will run from 1 to 3). Some intuition can be gained
from the $n=1$ case, where these are just the $Spin(5)$ gamma-matrices. $\rho_m(\Gamma^\mu)$ will
denote the
action of $\Gamma^\mu$ on the $m$th factor of the tensor product and
$e_i$, $i=1,\ldots 4$, is the basis of $V$,
\beq
\rho_m(\Gamma^\mu)(e_{i_1}\otimes e_{i_2}\otimes\dots\otimes
e_{i_m}\otimes\ldots\otimes e_{i_n})=(e_{i_1}\otimes
e_{i_2}\otimes\dots\otimes (\Gamma^\mu_{j_mi_m}e_{j_m})\otimes\ldots\otimes
e_{i_n})
\eeq
and $P_n$ denotes symmetrisation so that
\beq
\gh^\mu=P_n\sum_m\rho_m(\Gamma^\mu)P_n.
\eeq
In their non-Abelian algebra, the $\gh^\mu$ obey a version of the equation of a sphere, in that
$\sum_\mu \gh^\mu \gh^\mu=R^2\openone$, where $R$ is a function of
$n$ (this is calculated in the Appendix equations (\ref{FS4a}) and (\ref{FS4})). 

Each fuzzy odd-spheres is derived from the fuzzy even-sphere one dimension higher. To get the fuzzy three-sphere we use the projection $P_\pm=\frac{1}{2}(1\pm\Gamma^5)$ to
decompose $V$ into $V_+$ and $V_-$, the positive and negative
chirality two-dimensional spinor representations of $SO(4)$. The
co-ordinates act in a reducible representation $\cR=\cRp\oplus\cRm$. To
get $\cRp$ we take the symmetrised tensor product of $\frac{n+1}{2}$ factors of
$V_+$ and $\frac{n-1}{2}$ factors of $V_-$. Similarly, $\cRm$ is the
symmetrised tensor product of $\frac{n-1}{2}$ factors of
$V_+$ and $\frac{n+1}{2}$ of $V_-$. $\cRp$ is the irreducible representation of
$SO(4)$ with $(2j_L,2j_R)=(\frac{n+1}{2},\frac{n-1}{2})$ and
$\cRm$ is the irreducible representation with $(2j_L,2j_R)=(\frac{n-1}{2},\frac{n+1}{2})$ (this is described in more detail an Appendix \ref{app:FS3}).  The projector
$\cPpm=\left(P_+^{\otimes(n\pm1)/2}\otimes P_-^{\otimes(n\mp1)/2}\right)_{sym}$ projects
the fuzzy four-sphere onto $\cRpm$. $\Pr=\Prp+\Prm$ and the co-ordinates of the fuzzy
three-sphere are given by
\beq
G^i=\Pr\gh^i\Pr.
\eeq
The $G^i$ are linear maps acting on the the $N=\frac{(n+1)(n+3)}{2}$ dimensional representation $\cR$ and we represent them by $N\times N$ matrices. It can be shown that $\sum_i G^i G^i=\frac{(n+1)(n+3)}{2}\openone$ (see Appendix \ref{app:rad}). $\gh^5$ becomes a difference of projection operators; $G_5=\Pr \gh^5 \Pr= \Prp-\Prm$. 

The space of $N\times N$ matrices acting on the $N$-dimensional \rep $\cR$,
\mnc, can be decomposed into representations of $SO(4)$. This is a basis of operators corresponding to
Young diagrams\cite{Ram2}. Young diagrams for $SO(4)$ have
a maximum of two rows (as
$\Gamma^1\Gamma^2\Gamma^3\Gamma^4=\Gamma^5$, so all products of more
than two gamma matrices can be rewritten as products or two or less) and can be represented by the row lengths
$(r_1,r_2)$, where $r_2$ can be positive or negative. Each column is either
one or two boxes long and we represent these by factors of
$\rho_m(\Gamma)$ or $\rho_m(\Gamma\Gamma)$ respectively, each acting on
different factors of the tensor product. We suppress the indices on the
$\Gamma$'s. (They will be contracted with a traceless tensor of appropriate
symmetry). If $r_2$ is positive, the
$\rho_m(\Gamma\Gamma)$ all act on $V_+$ factors, if  negative
they all act on $V_-$ factors.  Diagrams with both $\rho_r(\Gamma\Gamma P_+)$ and $\rho_r(\Gamma\Gamma P_-)$ factors
vanish\cite{Ram2}. If $r_1-r_2$ is divisible by two then
we are in $End(\cRpm)$ and half of the $\rho_m(\Gamma)$'s come with
$P_+$ projector ($\rho_m(\Gamma P_+)$) and half come with a $P_-$ to
make sure that we stay within $\cRpm$. If
$r_1-r_2$ is not divisible by two then
we are in $Hom(\cRpm,\cRmp)$ and $(r_1-r_2\pm1)/2$ of the
$\rho_m(\Gamma)$'s come with a
$P_+$ and $(r_1-r_2\mp1)/2$ with a $P_-$. 
For example the diagram in $Hom(\cRp,\cRm)$ with row lengths $(4,1)$ is given by\\
\begin{equation}\nonumber
\includegraphics[clip=true, angle=-90]{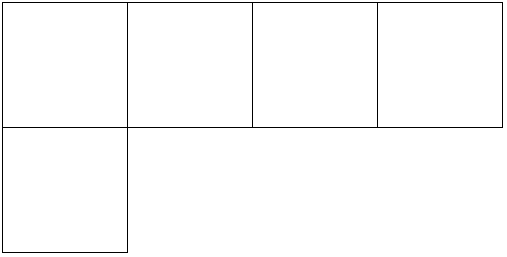}
\end{equation}\\
 and corresponds to operators of the form 
\beq
\sum_{\vec{r},\vec{s}}\rho_{r_1}(\Gamma\Gamma P_+)\rho_{s_1}(\Gamma
P_-)\rho_{s_2}(\Gamma P_+)\rho_{s_3}(\Gamma P_+)\, ,
\eeq
and the diagram in $End(\cRm)$ with row lengths $(5,-3)$ corresponds to the diagram
\begin{equation}\nonumber
\includegraphics[clip=true, angle=-90]{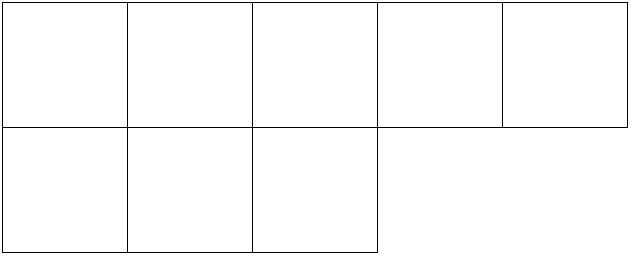}
\end{equation}\\
and to operators of the form
\beq
\sum_{\vec{r},\vec{s}}\rho_{r_1}(\Gamma\Gamma P_-)\rho_{r_2}(\Gamma\Gamma P_-)\rho_{r_3}(\Gamma\Gamma P_-)\rho_{s_1}(\Gamma
P_+)\rho_{s_2}(\Gamma P_-).
\eeq
The sum over $\vec{r},\vec{s}$ is such that $r_1 \neq r_2\neq\ldots
s_1\neq s_2\neq\ldots$ and all indices run from 1 to $n$. The vector index on the $\Gamma$'s is contracted
with a tensor of appropriate Young diagram symmetry. The product of two
$\Gamma$'s ensures that we have anti-symmetry down columns and the
symmetrisation of $\cR$ ensures we have the correct symmetry along
rows. 

\subsection{Dimension of Representations and the Young Diagram Basis}\label{sec:sumYDB}

In this section we describe how to obtain the the dimension of the Young diagram basis\cite{Ram2}. Representations of any even orthogonal group $SO(2k)$ can be labelled by the row lengths of a Young diagram, $r_1,r_2,\ldots,r_{2k}$\, , where $r_{2k}$ is allowed to be positive or negative (the only difference from $SO(2k+1)$). If $l_i=r_i+k-i$ and $m_i=k-i$ then the dimension of the representation with row lengths $r_i$ is given by 
\beq\label{dim}
D(l_i ) = \prod_{i \le j } \frac{ ( l_i^2 - l_j^2 )}{(m_i^2
 - m_j^2 ) }  \, .
\eeq
We would like to sum over a basis of all possible diagrams, the form of the operators for these is described in the previous section. We start by examining the `self-dual' operators (i.e. with $r_2\geq0$, so that all the products of two gamma matrices act on $P_+$ factors) in $End(\cRp)$ which will take the form
\bea
&&  \sum_{{\vec s} , {\vec t} }\rho_{s_1} ( \Gamma \Gamma P_+ ) \rho_{s_2}
 ( \Gamma \Gamma P_+ ) \cdots \rho_{s_{|r_2|}} ( \Gamma \Gamma P_+) 
 \rho_{t_1} ( \Gamma P_+) \rho_{t_2} ( \Gamma P_+ ) \times\cdots \non\\
&&\times\rho_{t_{k}  }
 ( \Gamma P_+ ) \rho_{t_{k+1} } ( \Gamma P_-) \rho_{t_{k+2} } 
( \Gamma P_- ) \cdots \rho_{t_{2k}}
 ( \Gamma P_- )\, ,
\eea
where $k=(r_1-|r_2|)/2$. $k$ can range from 0 to $(n-1)/2$ and $r_2$ can range from 0 to $(n+1)/2-k$. Thus the sum over self-dual diagrams gives
\beq\label{eppsum}
\sum_{k=0}^{\frac{(n-1)}{2} } \sum_{r_2=0}^{ {\frac{(n+1)}{2} } -
k} ( 2 p_1 + 2 k + 1 ) ( 2k + 1)  =
\frac{1}{96} ( n+1) ( 3n^3 + 41 n^2 + 97 n + 51 ).
\eeq
One can then repeat the calculation for the anti-self-dual diagrams, taking care not to count again the diagrams with $r_2=0$ in (\ref{eppsum}). These diagrams are completely symmetric and neither self-dual or anti-self-dual. They will be important in the next section.

Adding both sets of diagrams gives a total dimension for $End(\cRp)$ of
\beq
D=\frac{(n+1)^2(n+3)^2}{16}.
\eeq
This is $N^2/4$, exactly the dimension of matrices mapping from $\cRp$ to $\cRp$. The calculation for $End(\cRm)$ is identical, but with $P_+$ and $P_-$ interchanged. No new steps are needed to repeat the process for $Hom(\cRp,\cRm)$ and $Hom(\cRm,\cRp)$, and adding the four contributions gives a total dimension of $N^2$, showing that the Young diagram basis gives a complete basis of \mnc.

\section{Projection to Fuzzy Spherical Harmonics}\label{sec:shproj}

To properly describe the fuzzy three-sphere we want an algebra
which reproduces the classical algebra of functions on
the $S^3$ in the large-$N$ limit. For the fuzzy two-sphere there is no need for a projection as the matrix algebra has the correct behaviour. For the fuzzy three-sphere, \mnc\ does not give the correct large-$N$ limit (We use `large-$N$' and `large-$n$' limits interchangeably - recall $n$ is the number of tensor factors in the representation which the operators act on, and $N$ ($\sim n^2$ for the three-sphere) is the dimension of this representation, which coincides with the radius only for the fuzzy three sphere). To see this let
$x^i=G^i/n$, so that $x^ix^i\sim O(1)$. Then
$x^{[i}x^{j]}\sim O(1)$ as well, and the anti-symmetric part of the
product of co-ordinates persists in the large-$N$ limit. If we look at the commutator
\beq
[G^{i}, G^{j}]\cPpm=2\sum_{r}\rho_r(\Gamma^{ij}P_\pm)\cPpm +\sum_{r\neq
  s}2\rho_r(\Gamma^{[i} P_\mp)\rho_s(\Gamma^{j]} P_\pm)\cPpm\, ,
\eeq
the first term has eigenvalues of order $n$, but the second term has eigenvalues of order $n^2$ and it is this term that causes the problem.
For fuzzy even-spheres the symmetry
over all $n$ tensor factors causes such a term to vanish. Here we have a
division between $V_+$ and $V_-$ factors. 

The projection that does give the correct limit is
detailed in \cite{Ram2}. It is a  projection onto operators
corresponding to Young diagrams with only one row (i.e. completely
symmetric), the algebra of these operators is called ${\cal A}_n(S^3)$. We can extract the
terms corresponding to these operators
from the general sums given in Section \ref{sec:sumYDB} such as (\ref{eppsum}). Keeping only the $r_2=0$ terms in  (\ref{eppsum}) leads to 
\beq
\sum_{k=0}^{\frac{(n-1)}{2}}(2k+1)^2=n(n+1)(n +2)/6\, .
\eeq
The result for $End(\cRm)$ is the same, but the projection also requires the surviving operators to act identically on $\cRp$ and $\cRm$ under the exchange of $P_+$ and $P_-$, so we must combine one operator from $End(\cRp)$ with one from $End(\cRm)$ to make an a single operator in \anstp. Similarly we must do this for $Hom(\cRpm,\cRmp)$, noting that $G^i$ is already of this form, containing $\Prm\sum_r\rho_r(\Gamma^iP_+)\Prp$ in $Hom(\cRp,\cRm)$ and $\Prp\sum_r\rho_r(\Gamma^iP_-)\Prm$ in $Hom(\cRm,\cRp)$. There are $(n+1)(n+2)(n+3)/6$ of these operators giving a total number of degrees of freedom in the projected algebra \anst of
\beq
D=(n+1)(n+2)(2n+3)/6.
\eeq

As mentioned previously the algebra ${\cal A}_n(S^3)$ does not close under
multiplication, and we have to project back onto ${\cal A}_n(S^3)$
after multiplying. We denote the projected product by $A\bullet B$ or $(AB)_+$ for
$A,B\in {\cal A}_n(S^3)$. It is this projected product which is non-associative. (In the
large-$n$ limit one recovers associativity \cite{Ram2,Papageorgakis:2006ed}). 

\section{The Appearance of Non-Associativity}\label{sec:nonass}

The mechanism leading to non-associativity can be easily seen for the simple $n=1$ example,
where 
\beq
(\Gamma^1\bullet\Gamma^1)\bullet\Gamma^2=\openone\bullet\Gamma^2=\Gamma^2\neq0=\Gamma^1\bullet0=\Gamma^1\bullet(\Gamma^1\bullet\Gamma^2).
\eeq
Generalising to higher $n$ we must be careful to decompose into the
Young diagram basis and keep only the proscribed set. In particular we
should remember to keep traces of tensor operators that correspond to
smaller symmetric diagrams. Firstly, taking the projected product of
two of the co-ordinates we get
\beq
G^i\bullet G^j=\Prp\left[\sum_r\rho_r(\delta^{ij}P_+)+\sum_{r\neq
  s}\rho_r(\Gamma^{(i}P_-)\rho_s(\Gamma^{j)}P_+)\right]\Prp\;+\;
  (+\leftrightarrow-).
\eeq
We then use this to multiply a third co-ordinate giving
\bea
(G^i\bullet G^j)G^k&=&\Prp\left[\sum_{r\neq s\neq
    t}\rho_r(\Gamma^{(i}P_-)\rho_s(\Gamma^{j)}P_+)\rho_t(\Gamma^{k}P_-)+\sum_{r\neq s}\rho_r(\delta^{ij}P_+)\rho_s(\Gamma^kP_-)\right.\non\\
 &&+\sum_r\rho_r(\delta^{ij}\Gamma^kP_-)+\frac{1}{2}\sum_{r\neq s}\rho_r(\Gamma^{i}P_-)\rho_s(\Gamma^j\Gamma^kP_-)+\non\\
 &&\left.\frac{1}{2}\sum_{r\neq s}\rho_r(\Gamma^{j}P_-)\rho_s(\Gamma^i\Gamma^kP_-)\right]\Prm\;+
\;  (+\leftrightarrow-).
\eea
Now we must be careful to keep the traces of the last term when we
project. The traceless combination is
\bea
&&\frac{1}{2}\sum_{r\neq
  s}\left(\rho_r(\Gamma^{i}P_-)\rho_s(\Gamma^j\Gamma^kP_-)+\rho_r(\Gamma^{j}P_-)\rho_s(\Gamma^i\Gamma^kP_-)\right)\non\\&-&\frac{n-1}{6}\left(\delta^{ij}G^k+\delta^{ik}G^j+\delta^{jk}G^i\right),
\eea
as can be checked by contracting with a delta function. The trace parts
correspond to $(r_1,r_2)=(1,0)$ diagrams so they are retained.

In fact we have to be more careful with the first term. If
\beq
A_{(ij);k}=\sum_{r\neq s\neq t}\rho_r(\Gamma^{(i}P_-)\rho_s(\Gamma^{j)}P_+)\rho_t(\Gamma^{k}P_-)\, ,
\eeq
then the correct decomposition into Young diagrams is
\beq
 A_{(ij);k}=\frac{1}{3}\left(A_{(ij);k}+A_{(ik);j}+A_{(jk);i}\right)+\frac{1}{3}\left(2A_{(ij);k}-A_{(ik);j}-A_{(jk);i}\right)\, .
\eeq
The first bracket corresponds to the symmetric $(r_1,r_2)=(3,0)$ diagram, but the second is one of mixed symmetry. We keep the first and we must check the second for traces. It does possess traces and their sum is given by
\beq\label{nontrace}
\frac{(n-1)^2}{36}\left(2G^k-G^i-G^j\right)\, .
\eeq
Importantly this has eigenvalues of order $n^3$ and is going to lead to non-associativity in the large-$n$ limit. 

Combining the above means that projecting the second product gives
\bea
(G^i\bullet G^j)\bullet G^k&=&\Prp\left[\sum_{r\neq
  s\neq
  t}\rho_r(\Gamma^{(i}P_-)\rho_s(\Gamma^{j}P_+)\rho_t(\Gamma^{k)}P_-)+\frac{(n^2+10n+7)}{18}\delta^{ij}G^k\right.\non\\ &&\left.+\frac{(n^2-8n+7)}{36}\left(\delta^{ik}G^j+\delta^{jk}G^i\right)\right]\Prm\;+\;
  (+\leftrightarrow-).
\eea
A similar calculation can be done for $G^i\bullet (G^j\bullet G^k)$
and the non associativity can be expressed by
\beq
(G^i\bullet G^j)\bullet G^k-G^i\bullet(G^j\bullet G^k)=\frac{N+2}{6}(\delta^{ij}G^k-\delta^{jk}G^i).
\eeq
Contrary to what was said earlier the non-associativity shown above will persist in the large-$n$ limit; we can see if we replace $G^m$ by $\frac{G^m}{n}$ the right-hand side will be of order 1. This issue does not appear for even spheres and can be fixed by demanding the projection also removes the extra traces from terms when each gamma matrix acts on a different tensor factor, i.e. remove the contribution of (\ref{nontrace}). This is discussed in Appendix \ref{app:proj}.

Removing these additional contributions leads to 
\bea
(G^i\bullet G^j)\bullet G^k&=&\Prp\left[\sum_{r\neq
  s\neq
  t}\rho_r(\Gamma^{(i}P_-)\rho_s(\Gamma^{j}P_+)\rho_t(\Gamma^{k)}P_-)\right.\non\\ &&+\left.\frac{2n+1}{3}\delta^{ij}G^k+\frac{n-1}{6}\left(\delta^{ik}G^j+\delta^{jk}G^i\right)\right]\Prm\non\\ &&\;+\;
  (+\leftrightarrow-),
\eea
and the result for the associator
\beq
(G^i\bullet G^j)\bullet G^k-G^i\bullet(G^j\bullet G^k)=\frac{n+1}{2}(\delta^{ij}G^k-\delta^{jk}G^i).
\eeq
This contribution will vanish in the large-$n$ limit. Original versions of this calculation missed some trace terms and it was clarified in \cite{Papageorgakis:2006ed,BC2}

\section{Relationship of the Projection to the Basu-Harvey Picture}

One can ask if the Basu-Harvey equation still holds for this projection, and we
find that it does not. As one might expect, the anti-symmetric bracket
vanishes when we project onto symmetric representations. The co-ordinates
$G^i$ are contained in ${\cal A}_n(S^3)$. The commutator of two
of these co-ordinate matrices is given by
\bea\label{asdf}
[G^{i}, G^{j}]\cPpm&=&2\sum_{r}\rho_r(\Gamma^{ij}P_\pm)\cPpm +\sum_{r\neq
  s}2\rho_r(\Gamma^{[i} P_\mp)\rho_s(\Gamma^{j]} P_\pm)\cPpm\label{com}\\
&=&2\sum_{r}\rho_r(\Gamma^{ij}P_\pm)\cPpm-\frac{n+1}{2}\sum_{r}\rho_r(\Gamma^{ij}P_{\mp})\cPpm\non\\
&&+\frac{n-1}{2}\sum_{r}\rho_r(\Gamma^{ij}P_{\pm})\cPpm\, ,
\eea
where in the second line we have written everything in terms of the
Young diagram basis of \cite{Ram2}. We see that written in
this basis every term contains a product of two $\Gamma$'s acting on
the same tensor product factor. This corresponds to Young diagrams with $r_2\neq0$, so
that after applying the projection there are no surviving terms. 

Since the two-bracket vanishes, the anti-symmetric four-bracket
does also. It seems therefore the algebra ${\cal A}_n(S^3)$ is not compatible
with the Basu-Harvey equation. However the algebra ${\cal
  A}_n(S^3)$ has a tantalising property. Recall that the  solutions to
the Basu-Harvey equation are
\beq
X^i(\sigma)=\frac{i\sqrt{2\pi}}{M_{11}^\frac{3}{2}}\frac{1}{\sqrt{\sigma}}G^i,
\eeq
and recalling the physical radius
\beq\label{traces}
R=\sqrt{\left|\frac{Tr \sum(X^i)^2}{Tr \openone}\right|}\, ,
\eeq
we get
\beq
\sigma=\frac{2\pi N}{M_{11}^3R^2}\, .
\eeq
(Notice that in (\ref{traces}) we no longer have a matrix trace with
our modified algebra, however, because $\sum_i X^iX^i$ is proportional
to the identity in the algebra, and trace of the identity is precisely
what we divide by to obtain the physical radius, the form of the trace
is unimportant here). If we still identify $N$ with $Q$, the number of membranes, then we reproduce the profile expected from the self-dual string.

However, now the number of degrees of freedom in the co-ordinates is no longer $N^2$, but $D=(n+1)(n+2)(2n+3)/6\sim n^3$ so that for large $N$ (thus $n$) we have that
\beq
D\sim Q^\frac{3}{2},
\eeq
exactly as expected for $Q$ coincident membranes in the large $Q$
limit. This is interpreted as the result of the fuzzy three-sphere
being endowed with an ultraviolet cutoff and the scaling of the cutoff to the size
of sphere depending in the right way on $Q$ to give the correct number
of degrees of freedom. This means that one can interpret the $Q^{3/2}$
degrees of freedom corresponding to the non-Abelian membrane theory as coming
from modes on the fuzzy sphere.

This is encouraging but we must reconcile the non-associative
projection and the four-bracket. One possibility is that the projection does not act inside
the bracket, which is thought of as an operator $[G_5,\cdot,\cdot, \cdot]:({\cal
  A}_n(S^3))^3\rightarrow{\cal A}_n(S^3)$. In this case obviously
$X^i$ will still provide a solution. To understand this we should remember that the Basu-Harvey equation (or rather the fuzzy sphere equation (\ref{FS3}) on which it is based) is a quantised version of a higher Poisson bracket equation. If in this Poisson bracket equation we were to fix one of the co-ordinates, the derivatives acting on that co-ordinate would give zero and the bracket would not hold. However if we evaluate the derivatives  and then fix the co-ordinate, the bracket equation will still hold. Here the anti-symmetric bracket encodes the derivatives, giving an explanation of why we might evaluate the bracket before using any projections. This is what happens in the next section when we deal with the dimensional reduction, and is discussed in the Appendix \ref{app:dimred}.

\newpage

%% file: ch7.tex
\chapter{Fuzzy Sphere Reduction: Relating the Basu-Harvey and Nahm equations}\label{Ch:fuzred}

The M2-M5 brane system, on which we have been concentrating, is the M-theory analogue of the D1-D3 system. It should also be related to it by dimensional reduction and T-duality. This reduction was not present in \cite{BasuH} and its formulation is important for adding credibility to the Basu-Harvey picture of the M2-M5 system. Before we do this we must investigate the simpler task of finding a fuzzy two-sphere embedded in the fuzzy three-sphere that we can use for this reduction. 

In contrast the relationship between the fuzzy four-sphere and the fuzzy three-sphere is much clearer. Indeed the fuzzy three-sphere construction is derived from the starting point of the fuzzy four-sphere. This is true more generally; each fuzzy odd sphere is obtained from the fuzzy even sphere one dimension higher, with the relation to the fuzzy even sphere of one dimension lower unclear.  Compare this with the Clifford algebras (the $n=1$ fuzzy spheres!) associated to the spheres. For those related to $SO(2k)$ (and thus odd spheres) we can construct the additional matrix $\Gamma^{2k+1}=\Gamma^1\Gamma^2\ldots\Gamma^{2k}$ used in the chirality projection operator. This is the additional matrix needed to form the Clifford algebra for $SO(2k+1)$. However simply removing $\Gamma^{2k}$, for example, will not form the corresponding Clifford algebra one dimension less. Similarly the fuzzy three-sphere algebra contains the residual fifth co-ordinate $G_5$, but there is no place for it or a fourth co-ordinate in the fuzzy two-sphere algebra, which is well known. Reducing from odd to even spheres requires a different kind of reduction than going from even to odd.

For a classical sphere we could just fix one of the co-ordinates and via $(x^i)^2=R^2-x_o^2$ we would have a sphere of one dimension less. However doing so for the fuzzy sphere will violate (\ref{FS3}). We can see that putting $X^4$ proportional to the identity in the Basu-Harvey would not work either as when inserted in a four-bracket, the bracket would vanish, as we have required translation invariance under the addition of such terms proportional to the identity. 

What we need here is a projection which acts on states and operators of the fuzzy three-sphere leaving only those of the fuzzy two-sphere. We can look for fuzzy two-spheres, which are associated to $SO(3)\sim SU(2)$ inside the fuzzy three-sphere which is associated to $SO(4)\sim SU(2)\times SU(2)$. We can restrict to a sub-two-sphere, such as the equatorial one. There is also the well known Hopf fibration of the three-sphere as a circle fibred over a two sphere. However, we would like to obtain a two-sphere in such a way that the first three co-ordinates of the fuzzy three-sphere remain as the co-ordinates of the fuzzy two-sphere. We would then like to check if the restriction of the Basu-Harvey equation to a sub-algebra by this projection gives us the Nahm equation.

The projection from the fuzzy three-sphere to the two-sphere should be easily generalisable to any odd sphere. It also may be useful in other situations where fuzzy spheres of different dimensions occur (like those described at the beginning of Appendix \ref{app:FS}), and not just those which are embedded as the cross-section of a funnel which solves a Basu-Harvey/Nahm type equation.

\section{Reducing the Fuzzy Three-Sphere to the Fuzzy Two-Sphere}\label{sec:dimred}

Reducing the fuzzy three-sphere to the fuzzy two-sphere comes down to
finding a projection on the fuzzy three-sphere so that three of the fuzzy three-sphere
matrices obey the fuzzy two-sphere algebra.

We begin with n=1 and look for a projector $\bar{P}$ such that it
commutes with $\Gamma^a,\, a=1,2,3$ and
$[\Gamma^a,\Gamma^b]\Pb=2i\epsilon_{abc}\Gamma^c\Pb$. Looking ahead to
the Basu-Harvey equation, we will also require that $\Pb\Gamma^4\Pb=0$ to satisfy the fuzzy three-sphere equation. We use a
basis given by Appendix \ref{conv} and assume that the projector is
made of
$2\times 2$ blocks proportional to the identity, i.e. constructed
from ${\openone,\Gamma^4,\Gamma^5,\Gamma^{45}}$. The solution is given
by 

\beq
\bar{P}=\frac{1}{2}(1+i\Gamma^4\Gamma^5).
\eeq
One can then clearly see $\Pb^2=\Pb$, $\Pb$
commutes with $\Gamma^a$ and $\Pb\Gamma^4\Pb=\Pb\Gamma^5\Pb=0$.

Defining $\Gb^\mu=\Pb\Gamma^\mu\Pb$ we now have
\beq
[\Gb^a,\Gb^b]=2i\epsilon_{abc}\Gb^c.
\eeq
In other words, when restricted to the subspace onto which $\Pb$ projects,
the $\Gamma^a$ form an $SU(2)$ algebra. This is easy to see when we
choose $\{\eb_1=\frac{1}{2}(e_1+ie_3), \eb_2=\frac{1}{2}(e_2+ie_4)\}$
as a basis for $\Pb V$. Then
\beq\label{gsigma}
 \Gb^a(\eb_i)\equiv\sigma^a_{ji}\eb_j\, ,
\eeq
 where
$\sigma^a$ are the Pauli matrices.

We can generalise to any $n$ by
introducing $\cPb=\Pb^{\otimes n}$ which projects onto
$\cRb=(\Pb V)^{\otimes n}$. We denote the original fuzzy four-sphere
matrices, which act on $V^{\otimes n}$, by $\gh^\mu$. Then we
set
\beq
\gb^\mu=\cPb \gh^\mu\cPb= \cPb \sum_r\rho_r(\Gamma^\mu)\cPb\, ,
\eeq
which is just saying that we have restricted $\gh^\mu$ to $\cRb$. Now
because of (\ref{gsigma}) we will recover the construction of the fuzzy
two-sphere given in Appendix B of \cite{SJ2} and in Equation (\ref{2const}). Thus one can check that

\bea \label{projS2}
[\gb^a,\gb^b]=2i\epsilon_{abc}\gb^c, \\
\gb^a\gb^a=n(n+2)\openone.
\eea

Notice that acting with projectors onto $\cRb$ either side of  co-ordinate matrices of the fuzzy three-sphere (defined as four of the fuzzy four-sphere matrices between projectors onto $\cR$) does not give our fuzzy two-sphere matrices  (i.e if $G^{i}=\Pr \gh^{i} \Pr$ then $\gb^{a}\neq
 \cPb G^{a}\cPb$). However $\gb^{a}\propto\cPb G^{a}\cPb$ 
 because $\Pb P_\pm\Pb=\Pb/2$. Thus the constant of proportionality has a
 power of $2$. It also has combinatoric factor dependent on $n$ as there are
 ${(n-1) \choose (n-1)/2}$ ways of choosing the tensor
 product factors to be acted upon by the $ \frac{n-1}{2}$ $P_+$'s in $\Pr$ which
 do not act on the same factor as the $\Gamma$. In $\cPb$ we just have
 $(n-1)$ $\Pb$ factors so only one choice. Hence the constant of
 proportionality is given by
\beq\label{factor}
\gb^{a}={(n-1) \choose (n-1)/2}^{-1}2^{n-1}\cPb G^{a}\cPb.
\eeq
 The projectors are there to indicate that we project on to $\cR$ or $\cRb$, so this is not a
 problem; we can think of acting with the same original $\gh^\mu$
 of the fuzzy $S^4$ in both
 cases before we project back using $\Pr$ or $\cPb$ to the representation with which we are dealing. 

\section{Some Checks and Elucidation of the Projection}

To make sure that we can project from the fuzzy two-sphere to the fuzzy
 three-sphere, we should check that for any state,
$\Psi$,  in $\cRb$ we can find a state in $\cR$ such that $\cPb$ projects it onto $\Psi$. Indeed we can find many such
states. Similarly, for any operator on these states in the fuzzy
 two-sphere we can find operators in the full fuzzy three-sphere algebra,
 \mnc, that project on to it. In fact if we restrict ourselves to the
 non-associative algebra \anst there is a unique operator that
 projects onto each operator in the fuzzy two-sphere, up to addition of
 operators in the kernel of $\cPb$. A general operator of the
 form
\beq
\cPb\sum_{\overrightarrow{r}\neq\overrightarrow{s}\neq\overrightarrow{t}}\rho_{r_1}(\Gamma^1)\rho_{r_2}(\Gamma^1)\dots\rho_{r_i}(\Gamma^1)\rho_{s_1}(\Gamma^2)\dots\rho_{s_j}(\Gamma^2)\rho_{t_1}(\Gamma^3)\dots\rho_{t_k}(\Gamma^3)\cPb
\eeq
is proportional to that obtained by projecting
\bea
\Pr\sum_{\overrightarrow{r}\neq\overrightarrow{s}\neq\overrightarrow{t}}\rho_{r_1}(\Gamma^1P_\pm)\rho_{r_2}(\Gamma^1P_\pm)\dots\rho_{r_i}(\Gamma^1P_\pm)\rho_{s_1}(\Gamma^2P_\pm)\dots\rho_{s_j}(\Gamma^2P_\pm)\nonumber\\
\rho_{t_1}(\Gamma^3P_\pm)\dots\rho_{t_k}(\Gamma^3P_\pm)\Pr
\qquad +\qquad (+\leftrightarrow -),
\eea
where the signs of each $P_\pm$ are chosen to alternate from right to
 left. In both the fuzzy three-sphere and the fuzzy two-sphere we are still
 effectively using
 the co-ordinate matrices of the fuzzy four-sphere, but we are restricting
 to a much reduced set of states.

Notice also if we plug our projected fuzzy three-sphere matrices
straight back into the fuzzy three-sphere equation (\ref{FS3}) then it
is {\it not} satisfied due to the vanishing of $\gb^{4}$. This should
be compared with taking the classical version of the three-sphere
equation and reducing to a two-sphere (or any other sphere reduction) where the Nambu bracket is replaced by a Poisson
bracket. In the Poisson bracket if we reduce to a sub-sphere of lower degree, say by fixing one of the co-ordinates, then the equation will
not be satisfied. This occurs when the derivatives in the Poisson bracket act on the constant co-ordinate. However if we set the
co-ordinate to its fixed value \textit{after} evaluating the
derivatives then the higher sphere equation will still be satisfied. 

For the Nambu bracket there are no derivatives and the information is in the anti-commutation properties of the
matrices. Hence we should make our projection \textit{after}
evaluating the Nambu bracket in (\ref{FS3}). In our case we see that our
projected sphere satisfies the higher sphere equation trivially.
The main change is that both terms vanish for $i=4$, necessary to
obey the Basu-Harvey equation below.

We can also ask how the $SU(2)$ of the fuzzy two-sphere fits inside the
$SU(2)\times SU(2)=SO(4)$ of the fuzzy three-sphere. The $SO(4)$ has the six generators 
\beq
G^{ij}=\Pr\sum_r\rho_r(\Gamma^{ij})\Pr.
\eeq

From these we can construct two orthogonal $SU(2)$'s by 
\beq
\Sigma^{ij}_{L/R}=-i(G^{ij}\mp\epsilon_{ijkl}G^{kl})=-i\Pr\sum_r\rho_r(\Gamma^{ij}P_\pm)\Pr.
\eeq
The $\Sigma^{ij}_L$ and the $\Sigma^{ij}_R$ form $SU(2)$ algebras among themselves and commute with each other; we call these $SU(2)$'s $SU(2)_L$
and $SU(2)_R$ respectively. A general state in $\cRpm$ of the the fuzzy
$S^3$ is given by 
\beq
\left((e_1)^{\otimes p}\otimes (e_2)^{\otimes \frac{n\pm1}{2}-p}\otimes
(e_3)^{\otimes q}\otimes (e_4)^{\otimes \frac{n\mp1}{2}-q}\right)_{sym}\, ,
\eeq
where $e_i$ is the basis of $V$. If we label the generators of the two $SU(2)$'s by
$\sigma^a_{L/R}=1/2\epsilon_{abc}\Sigma^{bc}_{L/R}$ then these basis states are
eigenstates of $\sigma^3_{L/R}$, with eigenvalues $2m_{L/R}$. We can then
apply $\cPb$ to these states and examine the eigenvalues of $\gb^3$ on
these new $\cRb$ states,
$2\bar{m}$ say. Then we find that $\bar{m}=m_L+m_R$. Since adding the
size of the \reps of $SU(2)_L$ and $SU(2)_R$ in $\cRpm$ gives
$(n\pm1)/2+(n\mp1)/2=n$, which is the size of the \rep of the fuzzy two-sphere $SU(2)$. We see that we are effectively taking the sum of
$SU(2)_L$ and $SU(2)_R$.

\section{Disappearance of Non-associativity after Dimensional
  Reduction}

Given the non-associative nature of the fuzzy three-sphere algebra, it is
natural to ask how it gives rise to the associative algebra of the
fuzzy two-sphere. In the fuzzy two-sphere there exist only matrices
corresponding to symmetric Young diagrams. This is because the $SU(2)$ algebra
$\sigma^i \sigma^j =\delta^{ij}\openone+\epsilon^{ijk}\sigma^k$ implies
that an anti-symmetrised product of $\sigma$'s can be written as a single $\sigma$.

We can perform a check that the product in the fuzzy two-sphere is
associative,
\beq
\gb^a\gb^b\cPb=\left(\sum_r\rho_r(\delta^{ab}\openone+i\epsilon_{abd}\Gamma^d)+\sum_{r\neq
  s}\rho_r(\Gamma^a)\rho_s(\Gamma^b)\right)\cPb\, ,
\eeq
so that
\bea
(\gb^a\gb^b)\gb^c\cPb&=&\left(\sum_r\rho_r(\delta^{ab}\Gamma^c+i\epsilon_{abd}\Gamma^d\Gamma^c)+\sum_{r\neq
  s}\rho_r(\Gamma^a\Gamma^c)\rho_s(\Gamma^b)\right)\non\\
&&+\sum_{r\neq
  s}\rho_r(\Gamma^a)\rho_s(\Gamma^b\Gamma^c)+\sum_{r\neq
  s}\rho_r(\delta^{ab}+i\epsilon_{abd}\Gamma^d)\rho_s(\Gamma^c)\non\\&&\left.+\sum_{r\neq s\neq t}\rho_r(\Gamma^a)\rho_s(\Gamma^b)\rho_t(\Gamma^c)\right)\cPb\non\\
&=&\left(\sum_r\rho_r(\delta^{ab}\Gamma^c+i\epsilon_{abd}(\delta^{cd}\openone+i\epsilon_{dce}\Gamma^e))\right.\non\\
&&\left.+\sum_{r\neq
  s}\rho_r(\delta^{ac}\openone+i\epsilon_{acd}\Gamma^d)\rho_s(\Gamma^b)+\sum_{r\neq
  s}\rho_r(\Gamma^a)\rho_s(\delta^{bc}\openone+i\epsilon_{bcd}\Gamma^d)\right.\non\\
&&\left.+\sum_{r\neq
  s}\rho_r(\delta^{ab}+i\epsilon_{abd}\Gamma^d)\rho_s(\Gamma^c)+\sum_{r\neq s\neq
  t}\rho_r(\Gamma^a)\rho_s(\Gamma^b)\rho_t(\Gamma^c)\right)\cPb\non\\
&=&\Biggl( in\epsilon_{abc}+n\delta^{ab}\gb^c+(n-2)\delta^{ac}\gb^b+n\delta^{bc}\gb^a
\non\\&&\left.+i\epsilon_{acd}\sum_{r\neq s}\rho_r(\Gamma^d)\rho_s(\Gamma^b)
+i\epsilon_{bcd}\sum_{r\neq s}\rho_r(\Gamma^d)\rho_s(\Gamma^a)
\right.\non\\
&&\left.+i\epsilon_{abd}\sum_{r\neq s}\rho_r(\Gamma^d)\rho_s(\Gamma^c)+
\sum_{r\neq s\neq
  t}\rho_r(\Gamma^a)\rho_s(\Gamma^b)\rho_t(\Gamma^c)\right)\cPb\, .
\eea
Similarly, 
\bea
(\gb^a\gb^b)\gb^c\cPb&=&\Biggl( in\epsilon_{abc}+n\delta^{ab}\gb^c+(n-2)\delta^{ac}\gb^b+n\delta^{bc}\gb^a\non\\&&\left.+i\epsilon_{acd}\sum_{r\neq s}\rho_r(\Gamma^d)\rho_s(\Gamma^b)
+i\epsilon_{bcd}\sum_{r\neq s}\rho_r(\Gamma^d)\rho_s(\Gamma^a)
\right.\\
&&\left.+i\epsilon_{abd}\sum_{r\neq s}\rho_r(\Gamma^d)\rho_s(\Gamma^c)+
\sum_{r\neq s\neq
  t}\rho_r(\Gamma^a)\rho_s(\Gamma^b)\rho_t(\Gamma^c)\right)\cPb\, ,\non
\eea
so we have an associative product as expected. All terms can be written in terms
of symmetric operators as in the final line.

\section{Reducing the Basu-Harvey Equation to the Nahm Equation}

Consider the fuzzy three-sphere equation (\ref{FS3}) but replace the $G^i$  
by unknowns $\tilde{G}^i$ for which we must solve. Let us reduce to the  
fuzzy two-sphere so that the equation acts on $\cRb$,
\beq
\cPb \left(\tilde{G}^i  
+\frac{1}{2(n+2)}\epsilon_{ijkl}G_5\tilde{G}^j\tilde{G}^k\tilde{G}^l\right)\cPb =0.
\eeq
We let $\tilde{G}^4=G^4/c$, where $c$ is the same factor arising from  
projection we saw previously in equation (\ref{factor}), but fix the  
$\tilde{G^a}$ to be matrices in the algebra of the fuzzy two-sphere
($\tilde{G^a}=\cPb\tilde{G^a}\cPb$). Then, because $\cPb G^4\cPb=0$ and  
$\cPb G^5G^4\cPb=inc\cPb$, the $\tilde{G}^a$ must obey
\beq
\cPb \left(\tilde{G}^a +\frac{in}{2(n+2)}\epsilon_{abc}\tilde{G}^b\tilde{G}^c\right)\cPb =0.
\eeq
In  the large-$n$ limit this is the statement that the $\tilde{G}^a$  
must obey the fuzzy two-sphere $SU(2)$ algebra
\beq
[\tilde{G}^a,\tilde{G}^b]\cPb=2i\epsilon_{abc}\tilde{G}^c\cPb,
\eeq
which of course the $\gb^a$ obey (\ref{projS2}). Having to take the  
large-$n$ limit is expected because the fuzziness will make picking out  
a cross-section of the sphere difficult at small $n$.

We can now follow a similar procedure for the Basu-Harvey equation. We  
must take into account the additional anti-symmetrisation of the  
4-bracket and also use $R_{11}=M_{11}^{-3}\alpha'^{-1}$ (this can be deduced from the formulae at the end of Section \ref{sec:Mred} since $\alpha'=\ell_s^2$). We set  
$X^4=\frac{32\pi R_{11} G^4}{3c}$ and restrict the equation and the  
$X^a$ to the fuzzy two-sphere. Then we get
\beq
\left(\frac{dX^a}{d\sigma}+\frac{in}{\alpha'\sqrt{2N}}\epsilon_{abc}X^b 
X^c\right)\cPb=0.
\eeq
(We should also consider when the free index is $4$, in this case both terms  
vanish. This must happen as otherwise we could not choose $\sigma$  
dependence to solve the differential equation in both cases.) Rearranging, again in the large-$n$ limit, we get
\beq
\frac{dX^{a}}{d\sigma} +  
\frac{i}{2\alpha'}\epsilon_{abc}[X^{b},X^{c}]=0 \, ,
\eeq
which is the Nahm equation (\ref{Nahmeq}) but with the $X^a$ scaled by $\alpha'$ so that  
they have dimensions of length. It has solution
\beq
X^{a}=\frac{\alpha'\gb^{a}}{2\sigma} .
\eeq
The appearance of $R_{11}$ is expected as the length scale in  
11-dimensions is $1/M_{11}$ and in string theory it is  
$\sqrt{\alpha'}$. Changing from a three bracket for the Basu-Harvey  
equation to a two bracket for the Nahm produces a  factor of
$M_{11}^{-3}\alpha'^{-1}=R_{11}$ . Note, here we have considered either the  
case where the solutions before projection act in \mnc, or the case  
where they are in \anst but the projection does not act within the  
anti-symmetrised product.

%% file: ch8.tex
\chapter{The Doubled Formalism}\label{Ch:DF}

T-duality is an important symmetry of string theory. For superstrings it tells us that there are regimes in which two of the five consistent theories, for example type IIA and type IIB (see Section \ref{Sec:IIAB}), describe the same physics. A greater understanding of T-duality should lead to a greater understanding, not only of string theory, but of the underlying M-theory, which the five consistent string theories are believed to describe in a small region of its moduli space. 

In its simplest form, T-duality states that string theory with one of the co-ordinates, $X$ say, periodic with periodicity $R$ is equivalent to the theory obtained when the radius is instead $1/R$ (or rather $\alpha'/R$ with constants included). In identifying the two theories one interchanges momentum modes (which are quantised in the periodic dimension) with winding modes (a purely stringy effect not present in field theory). Momentum is dual to the co-ordinate $X$, and one can make the system more symmetric by introducing a co-ordinate $\xt$ dual to the winding number. This dual co-ordinate will have a radius $1/R$ and T-duality now interchanges $X$ and $\xt$. Of course in doing this we have doubled the degrees of freedom, but this can be remedied by introducing a constraint. When $R=1$ this constraint requires $X_L=X+\xt$ to be left-moving and $X_R=X-\xt$ to be right-moving, halving the number of degrees of freedom again to give the requisite amount.

T-duality can be extended to the case of many compact dimensions, and for a $T^d$ bundle the T-duality group is $O(d,d;\Z)$ (See \cite{Giveon:1994fu} for a review). This now includes integral shifts in the two-form gauge field $B$, re-orderings of the co-ordintates as well as generalisations of the $R\leftrightarrow1/R$ inversion described above. The doubling of the theory to include duals for all the periodic co-ordinates can be done\cite{Hull:2004in} and the doubled theory has an action and constraint on a $T^{2d}$ bundle which is invariant under $O(d,d;\Z)$. To relate it to the undoubled theory one must break this symmetry and choose a ``polarisation''; a $T^d$ sub-torus of the doubled $T^{2d}$ to be considered the physical torus. The $O(d,d;\Z)$ transformation of the doubled theory changes the torus picked out, and this change is equivalent to a change of the polarisation by the same $O(d,d;\Z)$ transformation. The undoubled theories obtained in each case are T-dual to each other, and the T-duality transformation between them is given by the same element of $O(d,d;\Z)$.

One reason why the formalism is of interest is that considering string theory on a torus bundle in the formalism can lead to `non-geometric' backgrounds. A standard string background has transition functions between patches which are diffeomorphisms and gauge transformations of $B$. This is a geometric background. T-duality acting on such a bundle could take it to one on which the transition functions were do longer diffeomorphisms, but could be any T-duality transformation. Hence a circle of radius $R$ in one patch could be patched to one of radius $1/R$ in another patch. Although this makes sense in string theory where both patches are described by the same CFT, this T-dualised version is no longer a geometric background. The transition functions can mix $G$ and $B$, and mix momentum with winding modes. These non-geometric backgrounds are named T-folds, by analogy with manifolds. A non-geometric T-fold is not a manifold, but  in the doubled theory it is described by a $T^{2d}$-bundle which is a manifold, and it can be analysed more clearly. Other symmetries such as U-duality for an M-theory compactification or mirror symmetry for a Calabi-Yau compactification could be used as transition functions, but T-duality has the advantage that it is a perturbative symmetry. Compactification of these non-geometric theories gives rise to consistent lower dimensional theories and T-folds (and other -folds) can be used as an additional source of 4-dimensional backgrounds from string theory\cite{tfold 1}.

In this chapter we will introduce the doubled formalism and show how the constraint is in effect a chirality condition on the doubled co-ordinates. We will also show how T-duality acts after choosing a polarisation. We will then show how Hull demonstrated quantum equivalence of the doubled and standard formalisms by gauging current whose vanishing is equivalent to the constraint\cite{Hull:2006va}, and how the T-duality transformation of the dilaton can be obtained from this approach. Finally, we will discuss briefly open strings, supersymmetry and other aspects of the formalism.

\section{Introducing the Formalism}
The doubled formalism was described by Hull in \cite{Hull:2004in,Hull:2006qs,Hull:2006va} and drew from previous duality covariant models\cite{Cremmer:1997ct,doublo}. Let $M$ be the target space of a sigma model which is locally a $T^d$ bundle over a base space $N$\footnote{T-duality is normally considered on torus bundles which have globally defined Killing vectors, so there are global isometries that can be gauged. In \cite{Hull:2006qs} T-duality was extended to general Torus fibrations where T-duality is done fibre-wise and we can consider such backgrounds here.}, this is expressed in the doubled formalism as a $T^{2d}$ bundle over $N$ with a constraint. Let $Y^m$ be the co-ordinates on (a given patch of) $N$ and let $\P^{I}=\P^{I}_{\alpha}d\sigma^\alpha$ be the $2d$ momenta on $T^{2d}$ (where $\alpha$ labels the worldsheet co-ordinates). Locally the momenta are exact and are given by
\beq
\P^{I}_{\alpha}=\partial_\alpha\X^I
\eeq
where $\X(\sigma)$ are the co-ordinates on a patch of the doubled torus, and thus have certain periodicity conditions. There are also connection 1-forms, functions of the base space co-ordinates given by $\A^I=\A^I_mdY^m$. The Lagrangian of the doubled formalism can be expressed in terms of the covariant fibre momenta
\beq
\Ph^{I}_{\alpha}=\P^{I}_{\alpha}+\A^{I}_{m}\partial_\alpha Y^m
\eeq
as
\beq\label{dlag}
\L_d=\frac{\pi}{2}\H_{IJ}\Ph^I\wedge*\Ph^J-\pi L_{IJ}\P^I\wedge\A^J+\L(Y).
\eeq
We have used the worldsheet Hodge dual $*$, $\H_{IJ}$ which is a positive definite metric on the $T^{2d}$ fibres and $L_{IJ}$ which is a constant $O(d,d)$ metric satisfying
\beq
L^{-1}\H L^{-1}\H=\openone,
\eeq
i.e. $I,J,\ldots$ indices are raised with $ L^{IJ}$. $\L(Y)$ is the Lagrangian for the sigma model on $N$. Note that we use different conventions to Hull, scaling our co-ordinates so that the periodicity is integral, rather than $2\pi$-integral, and absorbing a factor of $1/2\pi$ from the action into the Lagrangian. Note also that this leaves us a factor of $1/2$ less than what would appear to be the conventional sigma model normalisation, but this was the normalisation required in \cite{Hull:2006va} to prove quantum equivalence to the undoubled conventionally normalised sigma model (see Section \ref{Sec:quant}) by gauging currents, and also required by us when showing equivalence using holomorphic factorisation (see Chapter \ref{Ch:Holo}).

For both these methods it was also necessary to include the topological term 
\beq\label{ltop}
\L_t=\pi\Omega_{IJ}d\X^I\wedge d\X^J,
\eeq
for some anti-symmetric $\Omega_{IJ}$ to be determined. This term does not contribute to the classical equations of motion but affects the quantum analyses.

\subsection{The Constraint}

Restoring the correct number of degrees of freedom requires a covariant generalisation of the constraint touched upon earlier, which forces the $2d$ co-ordinates on the doubled torus to be, in a certain basis, either left-moving or right-moving. The generalised covariant form is 
\beq\label{gdconst}
\Ph=S*\Ph\, ,
\eeq
where $S^I_{\ K}=L^{IJ}\H_{JK}$ obeys $S^2=\openone$ and so defines an almost real structure (note that the formulae in this chapter are given for Lorentzian worldsheets unless stated). The $\X^I$ field equation can be rewritten
\beq
d*(S^I_{\ J}\Ph^J-*\Ph^I)=0\, ,
\eeq
so that the constraint (\ref{gdconst}) is a stronger condition.

\subsection{$O(d,d;\Z)$ Invariance}

The Lagrangian (\ref{dlag}) is invariant under rigid $GL(2d,{\mathbb R)}$ transformations acting as
\beq\label{htfm}
\H\rightarrow h^t\H h\, ,\qquad \P\rightarrow h^{-1}\P\, ,\qquad \A\rightarrow h^{-1}\A\, ,
\eeq
which would act on the co-ordinates via
\beq
\X\rightarrow h^{-1}\X\, .
\eeq
However, in order that the constraint continue to be satisfied, $h$ must preserve the $O(d,d)$ metric $L$, and so is restricted to be an element of $O(d,d)\subset GL(2d,{\mathbb R})$. Further, in order that the periodicity conditions of the co-ordinates be preserved, the symmetry group must be broken again to $O(d,d;\Z)$, which is of course the T-duality group of the undoubled bundle. Invariance of the topological term (\ref{ltop}) requires that $\Omega$ transform as $\Omega\rightarrow h^t \Omega h $.

\subsection{The Constraint and Chirality}\label{Sec:chicon}

One can introduce a vielbein $\V$ for the metric $\H$ which is an element of $O(d,d)$ invariant under the left action of $O(d)\times O(d)$. Thus $\H$ is a coset metric on $O(d,d)/O(d)\times O(d)$. Under a local  $O(d)\times O(d)$ transformation
\beq
\V\rightarrow k(Y)\V
\eeq
and $\H=\V^t\V$ is invariant. We can think of this as
\beq
\H_{IJ}=\left(\V^t\right)_I^{\ A}\delta_{AB}\V^B_{\ J}\, ,
\eeq
where $\delta$ is the $\H$ metric in the $O(d)\times O(d)$ frame. The indices $A,B=1,\dots,2d$ can be split into $a,b=1,\ldots,d$ and $a',b'=1,\ldots,d$ representing the two $O(d)$ factors. In this frame the $L$ metric is given by
\beq
L^{AB}=\left(\begin{array}{cc}\openone^{ab} & 0\\ 0&-\openone^{a'b'}\end{array}\right)\, .
\eeq
This specifies the form of $S$ in the frame and we can see that the constraint (\ref{gdconst}) can be written as
\bea\label{chiralPh}
\Ph^a&=&*\Ph^a\non\\
\Ph^{a'}&=&-*\Ph^{a'},
\eea
that is, as self-duality and anti-self-duality conditions. In terms of null co-ordinates on a flat worldsheet and components of $\Ph$ this is
\bea
\Ph^a_-&=&0\, ,\non\\
\Ph^{a'}_+&=&0.
\eea
For a flat fibre and no gauge field this is locally
\bea
\partial_-X^a&=&0\, ,\non\\
\partial_+X^{a'}&=&0,
\eea
meaning that half the co-ordinates are left-moving and the other half are right-moving. Note that this split is not the split into undoubled and dual co-ordinates and in general it is position dependent. For a Euclidean  toroidal worldsheet  with modulus $\tau$ the condition becomes one of (anti-) holomorphicity in terms of the complex co-ordinate $z=\sigma_1+\tau\sigma_0$, with the simplest case being
\bea\label{holoconst}
\bar{\partial}X^a&=&0\non\\
\partial X^{a'}&=&0.
\eea
This is what motivates us to use the holomorphic factorisation method to recover the undoubled partition function from the doubled formalism in Chapter \ref{Ch:Holo}.

\subsection{Polarisation}

To relate the doubled formalism to the conventional undoubled formalism one must pick a physical $T^d$ within the doubled $T^{2d}$, the co-ordinates of this $T^d$ will be the co-ordinates of the undoubled theory. However, by construction the doubled theory is T-duality invariant, whereas the undoubled is not: undoubled theories related by T-duality have the same doubled theory and which maximally null sub-torus $T^d\subset T^{2d}$ is picked out as physical determines which undoubled theory we recover. In the doubled theory this is called picking a polarisation. Group theoretically we are picking a $GL(d,{\mathbb R})$ subgroup of $O(d,d)$ so that the fundamental ${\bf 2d}$ of $O(d,d)$ splits into the fundamental ${\bf d}$ and its dual ${\bf d'}$ of $GL(d,{\mathbb R})$. This is facilitated by the use of a projector $\Pi$ onto the chosen $GL(d,{\mathbb R})$ basis, we choose a projection and then the projectors are constant. We label the $GL(d,{\mathbb R})$ indices by $i,j,\ldots=1,\ldots,d$ in superscript for the fundamental and in subscript for its dual. $\ti,\tj$ will indicate indices ranging over both. We use the projector 
\beq
\Phi^{\ti}_{\ I}=\left(\begin{array}{c}\Pi^{i}_{\ I}\\ \Pit_{i I}\end{array}\right)\, .
\eeq
This allows recovery of the physical and dual co-ordinates
\bea
X^i&=&X^i=\Pi^{i}_{\ I}\X^I, \non\\
\xt^i&=&X_i=\Pit_{i I}\X^I. 
\eea
We also let $P^i=\Pi^{i}_{\ I}\Ph^I$ and $Q_i=\Pit_{i I}\Ph^I$. In this basis the $O(d,d)$ metric is off-diagonal and given by
\beq
L_{\ti\tj}=\left(\begin{array}{cc}0&\openone\\ \openone&0\end{array}\right)\, .
\eeq
and we can introduce the dual of $\Phi^{\ti}_{\ I}$, $\Phih=L^{-1}\Phi L$ given by
\beq
\hat{\Phi}_{\ti}^{\ I}=\left(\begin{array}{c}\Pit_{i I}\\ \Pi_{i}^{\  I}\end{array}\right)\, .
\eeq
The metric in the $GL(d,{\mathbb R})$ basis is then given in terms of the metric, $G$, and two-form, $B$, of the undoubled torus by
\beq\label{HGB}
\Phih\H\Phih^t=\left(\begin{array}{cc}G-BG^{-1}B &BG^{-1}\\-G^{-1}B&G^{-1}\end{array}\right)\, .
\eeq
We can now recover the co-ordinates, metric and two form from the doubled set-up. Given a polarisation, (\ref{HGB}) tells us that under a transformation of the form (\ref{htfm}) restricted to $O(d,d;\Z)$ they transform as
\bea
G^{-1}=\Pi\H\Pi^t&\rightarrow& \Pi h^t\H h \Pi^t,\non\\
BG^{-1}=\Pit\H\Pi^t&\rightarrow& \Pit h^t\H h \Pi^t,\non\\
A=\Pi\A&\rightarrow& \Pi h^t\A.
\eea
Instead of the transformation on the doubled geometry (\ref{htfm}) we could have instead chosen a different projection using
\beq
\Pi\rightarrow\Pi h\, ,\qquad \Pit\rightarrow\Pit h\, .
\eeq
For given doubled geometry this spans the space of projections and this $O(d,d;\Z)$ symmetry of the doubled geometry, or equivalently $O(d,d;\Z)$ action on the polarisation, is the $O(d,d;\Z)$ symmetry of T-duality: the undoubled geometry obtained from the doubled geometry by a particular projection is related to that obtained using other choices of projection via an $O(d,d;\Z)$ T-duality symmetry.

\section{Quantum Equivalence via Conserved Currents}\label{Sec:quant}

In \cite{Hull:2004in} equivalence between the doubled formalism and the standard sigma model was shown by solving the constraint to express the $P^i$ in terms of the $Q_{i}$. This was then combined with the classical equations of motion for the doubled formalism to obtain the classical equations of motion for the standard undoubled sigma-model. 

However, demonstrating quantum equivalence is more complex. One way to do this is to first impose the constraint by gauging the symmetry associated to a current whose vanishing implies the constraint. The current
\beq
J_{I}=\H_{IJ}\Ph^J-L_{IJ}*\Ph^J,
\eeq
which is a worldsheet one-form, is conserved, being the sum of the Noether current generated by $\delta\X^I=\alpha^I$ and the topological current $j^I=L_{IJ}*\P^J$. Though these currents separately are not gauge invariant, the combination $J^I$ is. This current is of interest because its vanishing, $J^I=0$ is equivalent to the constraint (\ref{gdconst}). (The inclusion of the topological term (\ref{ltop}) does not change this as it adds only a trivially conserved topological piece to the Noether current). If we look in the $O(n)\times O(n)$ frame we see this clearly comparing (\ref{chiralPh}) and
\bea
J^a&=&\Ph^a-*\Ph^a,\non\\
J^{a'}&=&\Ph^{a'}+*\Ph^{a'}.
\eea
In terms of the chiral current components this is
\bea\label{chiralcomp}
J^a_+=0\,,\qquad &&J^a_-=\Ph^a_-\, ,\non\\
J^{a'}_+=\Ph^{a'}_+\,,\qquad&& J^{a'}_-=0\,.
\eea
We see that half of these vanish automatically, note that the conservation law for these currents is more complicated due to the vielbein not being constant.

In a given polarisation
\beq
J^i_+=\Pi^i_{\ a'}\Ph^{a'}_+\,,\qquad J^i_-=\Pi^i_{\ a}\Ph^a_-,
\eeq
where $\Pi^{\ti}_{\ A}=\Pi^{\ti}_{\ I}\V^I_{\ A}$ (with indices on $\V$ raised and lowered appropriately). This means that $J^i=0$ implies $J^I=0$ via non-degeneracy and (\ref{chiralcomp}), so that $J^i=0$ implies that the constraint holds.

For a trivial bundle 
\beq
J^i=d\xt_i+*dX^i.
\eeq
$d\xt_i$ is the Noether current from $\delta\xt_i=\alpha_i$ and $*dX^i$ is a trivially conserved topological current.

To gauge the symmetry associated to  $J^i$ we introduce a one-form gauge field $C_i$ and couple it minimally to $\P$ via 
\beq
\P^I\rightarrow \P^I+C_i\Pi^i_{\ J}L^{IJ}
\eeq
in the Lagrangian (\ref{dlag}). However, this gives a linear coupling of $C_i$ to only the Noether part of the current $J^i$ and we must introduce an additional term
\beq
\frac{\pi}{2}C_i\wedge P^i
\eeq
to couple $C$ to the topological current $*P$. (Note that had we attempted to gauge the whole current $J^I$ we would have met an obstruction to adding a topological piece in a gauge invariant way, gauging without coupling to this piece would lead to the elimination of all the $\X^I$, not just the dual ones). In effect we have added to the lagrangian
\beq
\L_g=\pi C_i\wedge*J^i+\frac{\pi}{2}\H^{ij}C_i\wedge *C_j\, ,
\eeq 
gauge invariance under large gauge transformations is only maintained if we include the topological term (\ref{ltop}) with $\Omega_{IJ}=\Pit_{i [I}\Pi^i_{\ j]}$.

The whole Lagrangian can be rewritten as
\beq
\L=\pi G_{ij}\ph^i\wedge*\ph^j+\pi G_{ij} B^i\wedge B^j -\pi\ph^iA_i+\frac{\pi}{2}G^{ij}D_i\wedge*D_j+2\pi A^i\wedge A_i +\L(Y),
\eeq
where $D_i=C_i+\Qh_i-G_{ij}*\ph^j-B_{ij}\ph^j$ is a non-dynamical auxiliary field. The first three terms give the standard undoubled Lagrangian (with the correct normalisation), hence showing classical equivalence. The final two terms depend only on $Y$.

\subsection{Quantisation}\label{Sec:qudil}

To quantise the model one would first like to integrate over the $\X$ to get a bundle of torus CFT's over $N$ with moduli $\tau(Y)$, as in the undoubled case. One can then integrate over $Y$ to complete the quantisation. However, one cannot proceed just by imposing canonical commutation relations on the $\X$ as there is also the constraint. We want to impose the constraint by gauging the current as above, but we have seen that to do this we must choose a polarisation, and in general there will not be a global one. The different CFT's in each patch will still fit together as the T-duality of the transition functions was a CFT symmetry. To quantise the gauged theory of the previous section in a given patch we gauge fix, but once we have included the topological term so that the theory is invariant under all gauge transformations (including large ones) we can fix the gauge so that $\xt$ is eliminated. The term 
\beq
\frac{\pi}{2}G^{ij}D_i D_j
\eeq
then does not effect the classical dynamics but is $Y$ dependent and integrating it out of the functional integral gives a determinant affecting the functional measure for $Y$. If we include the Fradkin-Tseytlin term in the Lagrangian
\beq
\L_{FT}=\sqrt{h}\phi R\, ,
\eeq
where $R$ is here the Ricci scalar for the world sheet metric $h$ and $\phi$ is a scalar field allowed to depend on $Y$, then at 1-loop the functional determinant above effectively replaces $\phi$ by
\beq
\phi\rightarrow \Phi=\phi-\frac{1}{2}\log \det(G_{ij}).
\eeq
Under a T-duality transformation taking $G\rightarrow G'$ this new scalar $\Phi$ shifts by
\beq
\Phi\rightarrow\Phi+\frac{1}{2} \log \frac{\det G'}{\det G}\, .
\eeq
This has the form of the Buscher rule for the T-duality transformation of the dilaton\cite{Buscher:1987sk} and in \cite{Hull:2006va} it was identified with the standard string theory dilaton. The field $\phi$ is T-duality invariant and in terms of $<e^{-\phi}>$ T-duality is a perturbative symmetry of the theory. This dilaton also occurs in string field theory.

Another approach to the quantum doubled theory is using Dirac brackets in a constrained Hamiltonian approach as in \cite{emily}, which uses results of \cite{walcher}. In a specific example, equivalence to conventional quantisation was found without picking a polarisation. In the following chapter we will detail how equivalence can also be found via a holomorphic factorisation argument.

\section{Other Aspects of the Formalism}

The doubled formalism can be extended to include open strings, D-branes and supersymmetry. There is also the relationship to other doubled theories, including those where all dimensions are doubled, to be considered. Finally we will mention the possible extension to U-folds for M-Theory compactifications.

For open strings each physical co-ordinate, $X^i(\sigma)$ will obey either Neumann or Dirichlet boundary conditions, $\partial_nX^i=0$ or $\partial_tX^i=0$ respectively. Each of these co-ordinates will have a dual co-ordinate in the doubled formalism that obeys the other choice of boundary condition, for example if $\partial_nX^i=0$ then $\partial_t\xt_i=0$. This means that for a doubled $T^{2d}$ there will always be $d$ Neumann and $d$ Dirichlet directions. There is a maximally Null $T^d$ within the $T^{2d}$ doubled fibre which has all Dirichlet directions and one can say there is a D-brane wrapping these directions. This means in the doubled formalism there is only one type of D-brane, which occupies exactly half the fibre directions. Depending on which polarisation is chosen, any number, from zero to all, of these D-brane directions are physical. Note that in the case of T-folds a certain type of D-brane in one patch can become a different type in another patch.

Extension of the formalism to a supersymmetric version is relatively straightforward and was performed in \cite{Hull:2006va,emily}. A superspace generalisation of the Lagrangian (\ref{dlag}) and the topological term (\ref{ltop}) were found, and the supersymmetric version of showing equivalence to the undoubled Lagrangian by gauging the current was also performed\cite{Hull:2006va}.

The doubled formalism of Hull is different to the doubled geometry of Hitchin\cite{Hitchin:2004ut}. In the latter only the tangent space is doubled, not the fibres. In both $O(d,d)$ has a central role, but only the doubled formalism involves the discrete $O(d,d;\Z)$ so that T-duality naturally appears. Doubled geometry, however, can be applied to any manifold, not just torus fibrations. In \cite{Hull:2006va} an extension of the doubled formalism in which all co-ordinates were doubled was introduced, but the utility of this for non-periodic spacetimes was unclear.

In \cite{Hull:2004in} it was suggested that a similar formalism could be applied to U-duality. When considering M-theory on a $d-$torus there should be extra dual co-ordinates introduced for each membrane winding mode, and each fivebrane winding mode if the torus has enough dimensions. For example, for M-theory on $T^7$, a $T^{56}$ bundle would be required. Transition functions would be in $E_7(\Z)$. Some form of constraint would be required to keep the correct degree of freedom counting but it is unclear how the theory could be formulated without a knowledge of the fundamental M-theory.

%% file: ch9.tex
\chapter{Holomorphic Factorisation of the Doubled Partition Function}\label{Ch:Holo}

In the previous section we introduced the doubled formalism of Hull and described how quantum equivalence of the doubled action to that of the standard undoubled string could be shown by gauging currents. To further demonstrate the consistency of the doubled formalism we will now show that it leads to the same partition function as the standard formulation when we use a holomorphic factorisation argument to apply the constraint.
 
We saw in Section \ref{Sec:chicon} how the constraint in a certain basis (the $O(d)\times O(d)$ frame) is a chirality constraint; the co-ordinates can be split into self-dual and anti-self dual ones. When we Wick rotate to a Euclidean worldsheet to do the path integral for the partition function this constraint becomes a constraint on the holomorphicity of the co-ordinates. Holomorphic factorisation (see for example\cite{ABMNV,Witten5,HNS,DVV,moore}) can be used to calculate the partition function for chiral Bosons and here we will proceed by interpreting the constraint as imposing that the co-ordinates are chiral Bosons and applying holomorphic factorisation techniques.

To use holomorphic factorisation to calculate the partition function for a self-dual worldsheet scalar (or a $2k$-form with $2k+1$-form field strength in $4k+2$ dimensions) one first proceeds by calculating the partition function for an ordinary scalar. Then the partition function is factorised into the product of a piece which is a holomorphic function of the worldsheet period matrix $\tau_{ij}$ (usually involving some sort of $\theta$-function) and its complex conjugate, this involves rewriting sums so that the sums in each factor are independent\footnote{There are more complications if one includes gauge fields, which we will not.}. If one wants a self-dual field then one takes only the holomorphic factor, while for an anti-self-dual field only the anti-holomorphic field is kept.  

We shall calculate the doubled partition function for a Bosonic co-ordinate compactified on a circle of radius $R$ (or equivalently radius $1/R$). Holomorphic factorisation is simplest at the free-Fermion radius ($R=1/\sqrt{2}$ in our conventions), and the sums can still be separated when the radius squared is a rational multiple of the free-Fermion radius. For an ordinary chiral Boson this would ensure that the partition function was a sum of finitely many products that could be factorised. At free-Fermion radius this sum is a sum over spin structures, and in general for the factorisation to be done properly many subtleties should be considered\cite{moore}. Here we find we can proceed without considering these issues, perhaps as we still wish to end up with an ordinary Boson, not a chiral one which would need a choice of spin structure or a generalisation thereof. We find we can factorise when our radius squared (which is 1 at the T-duality self-dual point) is a rational number.
 
We first write out the doubled action and calculate the constraint. As we have reiterated, in a certain basis the constraint is a chirality constraint, and we rewrite the action in terms of the fields of this basis.
These fields no longer have the simple periodicity conditions of the original co-ordinates, but we continue and evaluate the instanton sector of the partition function. We are then left with infinite sums over modes, and we Poisson re-sum and change summation variables until we can factorise the partition function into the contribution coming from the holomorphic co-ordinate and the one coming from the anti-holomorphic co-ordinate. We then take the appropriate holomorphicity square root of each contribution and recombine them. When we also include the contribution to the partition function from integrating over the fluctuations of the fields we obtain the standard string partition function for our original Bosonic co-ordinate (provided we include the topological term and use the normalisation that were both required to show equivalence using gauging of currents in Section \ref{Sec:quant}). We finish this chapter by calculating the partition function of an ordinary periodic Boson for comparison.

Although we have used the example of a non-fibred circle it is hoped the doubled formalism and the techniques of calculating the partition function used here will give more insight when applied to non-trivial T-folds.

\section{The Action and Constraint}

We begin with the doubled string action for a single toroidally
compactified Boson, so that we have the Boson $X$, 
which is associated to a circle of radius $R$, and its dual $\xt$ which is associated to a circle of radius $R^{-1}$. The action (\ref{dlag}) can be written as
\beq\label{Xact}
S=\frac{\pi}{2} R^2dX\wedge* dX+\frac{\pi}{2} R^{-2}d\xt\wedge* d\xt\, ,
\eeq
where the gauge fields are zero due to our simple background geometry and we are not concerned with the other co-ordinates. We have used the same unconventional normalisation as Hull in Section \ref{Sec:quant}, which is a factor of a half times the normal one. We too find this necessary to show equivalence. Similarly we will also need to include the topological interaction term (\ref{ltop}) that Hull required, this adds a term 
\beq
S_{top}=\pi dX\wedge d\xt\,
\eeq
to the action. Note that in our conventions $R=1$ corresponds to the self-dual radius, whereas in much of the literature on holomorphic factorisation $R=1$ is the free-Fermion radius (which would be $R=1/\sqrt{2}$ in our conventions).  

Recall from Section \ref{Sec:chicon} that the constraint in the $O(n)\times O(n)$ basis is the chirality constraint (\ref{chiralPh}), that is
\bea\label{Pconst}
\Ph^a&=&*\Ph^a\non\\
\Ph^{a'}&=&-*\Ph^{a'}.
\eea
Since we have a trivial bundle and calculate the partition function at one-loop, which requires a Euclidean worldsheet, we can rewrite this constraint as in (\ref{holoconst}) as
\bea
\bar{\partial}X^a&=&0\non\\
\partial X^{a'}&=&0\, .
\eea
The derivatives are with respect the complex worldsheet co-ordinates, $z=\sigma_1+\tau\sigma_0$ and $\zb$, where $\tau$ is the modular parameter of the worldsheet torus.

\subsection{Relating the Bases}

Choosing $X$ as the original co-ordinate means we are working with a specific polarisation, that is we have chosen a basis where the co-ordinates separate into the fundamental and the dual representations of $GL(n,{\mathbb R})$ ($n=1$ here), labelled by $\ti=(^i, \,_i)$. Recall the projectors 
\beq
\Phi^{\ti}_{\ I}=\left(\begin{array}{c}\Pi^{i}_{\ I}\\ \Pi_{i\,I}\end{array}\right)
\eeq
take us to this basis and
\bea
X&=&X^i=\Pi^{i}_{\ I}\X^I, \non\\
\xt&=&X_{i}=\Pi_{i\,  I}\X^I. 
\eea
Also in this basis we have
\beq
\H_{\ti\tj}=\left(\begin{array}{cc}R^2 & 0\\0&R^{-2}\end{array}\right)
\eeq
and
\beq
L_{\ti\tj}=\left(\begin{array}{cc}0&1\\1&0\end{array}\right)\, .
\eeq
We introduce $K^A_{\ \ti}$ via $ K=\V\Phi^t$, which will allow us to relate indices in the $O(n,n)$ basis (where the constraint is the simple chirality constraint we want) to the $GL(n,{\mathbb R})$ basis (where our original Boson can be seen). As we know $L$ and $\H$ in both bases we can determine that $K$ is given by\footnote{We have made some sign choices which do not affect the constraint.}
\beq
K^A_{\ \ti}=\frac{1}{\sqrt{2}}\left(\begin{array}{cc}R & R^{-1}\\ R&-R^{-1}\end{array}\right)
\eeq
This means that $\P^a$ appearing in the constraint (\ref{Pconst}) is related to $\P^i=dX$ and $\P_i=d\xt$ via $\P^a=K^a_{\ \ti}\P^{\ti}$, giving
\beq\label{Pa}
\P^a=\frac{1}{\sqrt{2}}\left(RdX+R^{-1}d\xt\right)\, ,\P^{a'}=\frac{1}{\sqrt{2}}\left(RdX-R^{-1}d\xt\right)\, .
\eeq
We will work in terms of $P$ and $Q$ where $\sqrt{2}\P^a=dP$ and $\sqrt{2}\P^{a'}=dQ$. In terms of $X$ and $\xt$ they are given by the linear combinations
\beq\label{PQ}
P=RX+R^{-1}\xt\, ,\qquad Q=RX-R^{-1}\xt.
\eeq
As a consequence of (\ref{Pa}) the constraint they obey is
\bea
\bar{\partial} P&=&0\, ,\\
\partial Q &=&0\, ,
\eea
which is the motivation for holomorphic factorisation. We can rewrite the action in terms of $P$ and $Q$, as follows
\beq\label{PQact}
S=\frac{\pi}{4} dP\wedge* dP+\frac{\pi}{4} dQ\wedge* dQ.
\eeq
The topological term can be rewritten as 
\beq\label{PQtop}
S_{top}=\frac{\pi}{4} dQ\wedge dP-\frac{\pi}{4} dP\wedge dQ.
\eeq
Wick rotation will introduce a relative factor of $i$ to this term. 

\section{Factorising the Instanton Sector}

The standard way of obtaining a partition function for a chiral Boson
is to factorise the partition function for an ordinary Boson into
holomorphic and anti-holomorphic parts and keep only the factor with
the correct holomorphic dependence \cite{moore}. Here we would like to do
this for $P$ and $Q$. However, crucially $P$ and $Q$ do not have standard
periodicity properties, and are linked to the periodicities of $X$ and $\xt$ through (\ref{PQ}). This is why
we cannot directly identify these fields as left and right movers on a
circle as the periodicity conditions would not match.

\subsection{The Cohomology Basis} 

To evaluate the instanton sector of the partition function we need
to examine contributions which from the field 
theory point of view will be in the cohomological sector.
That is, when the action is written as in (\ref{Xact}) what is really
meant is that 
$dX$ should be replaced by $L$, where $L=dX+\omega$, with $\omega\in
H^1(\Sigma,\Z)$. 
We can express the cohomological part in terms of the standard cohomology basis, which
for a toroidal worldsheet consists of just one $\alpha$ cycle and one $\beta$ cycle. 
\beq
L=dX+n\alpha+m\beta\, ,\qquad \Lt=d\xt+\nt\alpha+\mt\beta,
\eeq
where $n,\nt,m,\mt\in \Z$, which in turn means that we should replace $dP$ and $dQ$ by
\bea
M&=&dP+(Rn+R^{-1}\nt)\alpha+(Rm+R^{-1}\mt)\beta\, ,\non\\ N&=&dQ+(Rn-R^{-1}\nt)\alpha+(Rm-R^{-1}\mt)\beta\,.
\eea
The classical or {\it{instanton}} sector of the partition function is
the sum over all field 
configurations of $\exp[-S]$ and can be written as
\bea\label{Rpf}
Z&=&\sum_{m,n,\mt,\nt} \exp \left[ -(Rm+R^{-1}\nt)^2\frac{
\pi|\tau|^2}{4\tau_2}+(Rn+R^{-1}\nt)(Rm+R^{-1}\mt)\frac{\pi\tau_1}{2\tau_2}\right.\non\\&&\left.\qquad\qquad\quad-(Rm+R^{-1}\mt)^2\frac{\pi}{4\tau_2}\right]\nonumber \\ 
&&\times \exp \left[ -(Rm-R^{-1}\nt)^2\frac{\pi|\tau|^2}{4\tau_2}+(Rn-R^{-1}\nt)(Rm-R^{-1}\mt)\frac{\pi\tau_1}{2\tau_2}\right.\non\\&&\left.\qquad\quad\,-(Rm-R^{-1}\mt)^2\frac{\pi}{4\tau_2}\right]
\eea
where the first factor corresponds to $P$ and the second to $Q$. 
To holomorphically factorise the partition function we must Poisson
re-sum, but first we separate the sums so that the $P$ 
and the $Q$ parts of the partition function sum over different 
independent variables. The contribution from 
the topological term within the sum, is
\beq
\exp[i\pi (n\mt -m\nt)].
\eeq
This therefore only contributes a sign to the terms in the partition
sum though it is crucial in showing the equivalence to the usual partition sum.

\subsection{Rewriting the Sums}

To be able to separate the sums we assume $R^2=\frac{p}{q}$, with $p,q$ coprime, and let $k=pq$. Then we have
\beq
Rn\pm R^{-1}\nt=\sqrt{k}\left(\frac{n}{q}\pm\frac{\nt}{p}\right).
\eeq
Making the substitutions $n=cq+q\gamma_q$ and $\nt=\ct p+p\gamma_p$ (where $c,\ct\in\Z$ and $\gamma_q\in \{0,\frac{1}{q},\ldots,\frac{q-1}{q}\}$) we can further say
\beq
\sqrt{k}\left(\frac{n}{q}\pm\frac{\nt}{p}\right)=\sqrt{k}(c\pm\ct +\gamma_q\pm \gamma_p).
\eeq
Then we let $h=c+\ct$ and $l=c-\ct$. We have rewritten the sum over
$n$ and $\nt$ as a sum over $c,\ct,\gamma_q$ and $\gamma_p$, and then 
rewritten the $c$ and $\ct$ sums as a sum over $h$ and $l\in\Z$, 
but $h-l=2\ct$ so we must restrict to even values of $h-l$ by inserting a factor of 
\beq
\sum_{\phi\in\{0,\frac{1}{2}\}}\frac{1}{2}\exp[2\pi i \phi(h-l)]
\eeq
in the partition function. Repeating the process to split the $m,\mt$ sum, the partition function becomes
\bea\label{pqpf}
Z=&&\sum_{\phi,\theta,\gamma_q,\gamma_p,\gamma'_q,\gamma'_p} \sum_{h,
i,j,l}\frac{1}{4}\exp\left[-\frac{k\pi}{4}\left\{(h+\gamma_q+\gamma_p)^2\frac{|\tau|^2}{\tau_2}\right.\right.\nonumber \\ 
&&\left.-2(h+\gamma_q+\gamma_p)(i+\gamma'_q+\gamma'_p)\frac{\tau_1}{\tau_2}+(i+\gamma'_q+\gamma'_p)^2\frac{1}{\tau_2}\right\}\non\\
&&-\frac{k\pi}{4}\left\{(l+\gamma_q-\gamma_p)^2\frac{|\tau|^2}{\tau_2}-2(l+\gamma_q-\gamma_p)(j+\gamma'_q-\gamma'_p)\frac{\tau_1}{\tau_2}+(j+\gamma'_q-\gamma'_p)^2\frac{1}{\tau_2}\right\}\nonumber \\ 
&&+2\pi i\left\{\phi(h-l)+\theta(i-j)\right\}+\left.\frac{i\pi k}{2} \left((l+\gamma_q-\gamma_p)(i+\gamma'_q+\gamma'_p)\right.\right.\non\\
&&\left.\left.-(h+\gamma_q+\gamma_p)(j+\gamma'_q-\gamma'_p)\right)\right].
\eea
Using the notation $\gamma_\pm=\gamma_q\pm\gamma_p$ we can rewrite the partition function again as
\bea\label{PQpf}
Z=&&\sum_{\phi,\theta,\gamma_q,\gamma_p,\gamma'_q,\gamma'_p}\sum_{h,
i,j,l} \left(\frac{1}{2}\exp \left[-\frac{k\pi}{4}\left\{(h+\gamma_+)^2\frac{|\tau|^2}{\tau_2}-2(h+\gamma_+)(i+\gamma'_+)\frac{\tau_1}{\tau_2}\right.\right.\right. \nonumber\\
&&\left.+(i+\gamma'_+)^2\frac{1}{\tau_2}\right\}+2\pi i\left\{\phi h+\theta i\right\}\Biggr]\nonumber \\ 
&&\times\frac{1}{2}\exp\left[-\frac{k\pi}{4}\left\{(l+\gamma_-)^2\frac{|\tau|^2}{\tau_2}-2(l+\gamma_-)(j+\gamma'_-)\frac{\tau_1}{\tau_2}\right.\right. \nonumber\\
&&\left.+(j+\gamma'_-)^2\frac{1}{\tau_2}\right\}-2\pi i\left\{\phi l+\theta j\right\}\Biggr]\nonumber\\
&&\times\exp\left.\left[\frac{i\pi k}{2} \left((l+\gamma_-)(i+\gamma'_+)-(h+\gamma_+)(j+\gamma'_-)\right)\right]\right)
\eea
where we have split the terms in the sum into three factors, the piece coming from the $P$
kinetic term (which depends on $h$ and $i$), the piece coming from the
$Q$ kinetic term (which depends on $l$ and $j$) and the topological
piece, a cross term depending on all four integer indices.

\subsection{Poisson Resummation}

We now wish to perform Poisson resummation on $i$ and $j$. 
Let us focus on the $i$ resummation, replacing it with a sum over
$r$. Here we rewrite the $P$ part of the partition function and the first term of the topological piece:
\bea\label{Ppf}
Z_P&=&\sum_{\phi,\theta,\gamma_q,\gamma_p,\gamma'_q,\gamma'_p}\sum_{h,l,
r} \frac{1}{2}\sqrt{\frac{4\tau_2}{k}}\exp \Biggl[ -\frac{k\pi}{4}\left\{(h+\gamma_+)^2\frac{|\tau|^2}{\tau_2}-2\gamma'_+(h+\gamma_+)\frac{\tau_1}{\tau_2} \right.\nonumber\\
&&\left.\qquad\qquad\quad+(\gamma'_+)^2\frac{1}{\tau_2}\right\}+2\pi i\phi h+\frac{i\pi k}{2}(l+\gamma_-)\gamma'_+\non\\&&
\qquad-\frac{4\pi\tau_2}{k}\left(r-\theta+ik\frac{(h+\gamma_+)}{4}\frac{\tau_1}{\tau_2}-\frac{ik\gamma'_+}{4\tau_2}-\frac{k}{4}(l+\gamma_-)\right)^2\Biggr]\\
&=&\sum_{\phi,\theta,\gamma_q,\gamma_p,\gamma'_q,\gamma'_p}\sum_{h,l,r}\sqrt{\frac{\tau_2}{k}}\exp \Biggl[ \tau_2\Biggl(-\frac{k\pi}{4}(h+\gamma_+)^2\non\\
&&\qquad\qquad\qquad\qquad\qquad-4\pi k\left(\frac{r-\theta}{k}-\frac{1}{4}(l+\gamma_-)\right)^2\Biggr)\nonumber\\
&&-2\pi i \tau_1\left(h+\gamma_+\right)(r-\theta-\frac{k}{4}(l+\gamma_-))+2\pi i\phi h+2\pi i \left(r-\theta\right)\gamma'_+\Biggr].
\eea

The appearance of the squares with $\tau_2$ and the cross term with
$\tau_1$ is standard and Poisson resumming to replace $j$ with $s$ in
the $Q$ part of the partition function we can rewrite the whole partition function as
\bea
Z&=&\sum_{\phi,\theta,\gamma_q,\gamma_p,\gamma'_q,\gamma'_p,h,l,r,s}\Biggl( \sqrt{\frac{\tau_2}{2k}}\exp \left[i\pi k\tau \frac{p_L^2}{2}-i\pi k\twb \frac{p_R^2}{2}+2\pi i  \left(\phi h+(r-\theta)\gamma'_+\right)\right]\nonumber \\ 
&&\qquad\quad\times\sqrt{\frac{\tau_2}{2k}}\exp \left[i\pi k\tau \frac{q_L^2}{2}-i\pi k\twb \frac{q_R^2}{2}+2\pi i  \left(-\phi l+(s+\theta)\gamma'_-\right)\right]\Biggr)
\eea
where
\bea\label{pqlr}
p_L&=&\frac{1}{2}\left(h+\gamma_+\right)-2\left(\frac{r-\theta}{k}-\frac{1}{4}(l+\gamma_-)\right)\, ,\non\\
p_R &=&\frac{1}{2}\left(h+\gamma_+\right)+2\left(\frac{r-\theta}{k}-\frac{1}{4}(l+\gamma_-)\right)\, , \non\\
q_L&=&\frac{1}{2}\left(l+\gamma_-\right)-2\left(\frac{s+\theta}{k}+\frac{1}{4}(h+\gamma_+)\right)\, ,\non\\
q_R &=&\frac{1}{2}\left(l+\gamma_-\right)+2\left(\frac{s+\theta}{k}+\frac{1}{4}(h+\gamma_+)\right)\, .
\eea
We can now clearly see the holomorphic and anti-holomorphic
parts of the partition function for both $P$ and $Q$, as well as
additional pieces which restrict the sum over ``momenta'',
$p_L,p_R,q_L,q_R$. 
However, because the sums are linked we cannot remove the extra
pieces. So we rewrite the sums again, reconstructing them back to a sum over just four integers.

\subsection{Recombining the Sums}
The $h,l$ terms in the momenta can easily be recombined just by undoing the substitutions we made earlier to write 
\beq 
h+\gamma_+=\frac{n}{q}+\frac{\nt}{p}\, , \qquad l+\gamma_-=\frac{n}{q}-\frac{\nt}{p}\, ,
\eeq
where we have replaced the sums over $h,l,\gamma_+,\gamma_-$ and
$\phi$ with a sum over integers $n$ and $\nt$, we also remove 
one of the factors of $\frac{1}{2}$ we inserted outside the partition function.

The other sum has been Poisson resummed so the recombination is more complicated. We use the identity
\beq\label{rootsof1}
\sum_{k=0}^{n-1}\left(\exp \left(\frac{2\pi i k}{n}\right)\right)^j=\sum_{\gamma_n}\exp (2\pi i \gamma_n j)=\left\{\begin{array}{l}
n\ \mbox{if}\ j \equiv 0 \mod n\\ 0\ \mbox{otherwise}
\end{array}\right. .
\eeq
With that in mind we see the only occurrence of $\gamma'_q$ and $\gamma'_p$ is in the factor
\bea
&&\sum_{\gamma'_q,\gamma'_p}\exp \left[2\pi i\left(\frac{(p_L-p_R)}{2}\gamma'_++\frac{(q_L-q_R)}{2}\gamma'_-\right)\right]\non\\
&=&\sum_{\gamma'_q,\gamma'_p}\exp\left[2\pi i (r+s)\gamma'_q+2\pi i(r-s-2\theta)\gamma'_p\right]
\eea
which has the effect of enforcing
\bea
r+s&\equiv& 0 \mod q\, ,\nonumber\\
r-s-2\theta&\equiv& 0 \mod p\, ,
\eea
in the rest of the partition sum. We see that these requirements are fulfilled exactly by putting
\beq
\frac{r-\theta}{k}=\frac{1}{2}\left(\frac{w}{p}+\frac{\wt}{q}\right)\, , \qquad \frac{s+\theta}{k}=\frac{1}{2}\left(\frac{w}{p}-\frac{\wt}{q}\right)\, .
\eeq
We have replaced the sums over $r,s,\theta,\gamma'_p$ and $\gamma'_q$
by a sum over integers $w,\wt\in\Z$. 
Note that importantly it is the term with $q$ in the denominator which
changes sign between the two combinations. 
Also, due to (\ref{rootsof1}) we get a factor of $k=pq$ outside the
exponential, which cancels the factor of $1/k$ we got from the Poisson resummation. 
We can now rewrite (\ref{pqlr}) as
\bea\label{pqlr2}
p_L=\frac{n}{q}-\left(\frac{w}{p}+\frac{\wt}{q}\right)\, ,\qquad p_R &=&\frac{\nt}{p}+\left(\frac{w}{p}+\frac{\wt}{q}\right)\, , \nonumber\\
q_L=-\frac{\nt}{p}-\left(\frac{w}{p}-\frac{\wt}{q}\right)\, ,\qquad q_R &=&\frac{n}{q}+\left(\frac{w}{p}-\frac{\wt}{q}\right)\, .
\eea

The instanton contribution doubled partition function is now in the simple form
\beq\label{Zd}
Z_d=\sum_{p_L,
p_R} \sqrt{2\tau_2}\exp\left[i\pi k\tau \frac{p_L^2}{4}-i\pi k\twb \frac{p_{R}^{2}}{4}\right] \sum_{q_L,q_R} \sqrt{2\tau_2}\exp\left[i\pi k\tau \frac{q_L^2}{4}-i\pi k\twb \frac{q_R^2}{4}\right].
\eeq

Now we can make the following final substitution
\bea
u=n-\wt\, ,\qquad v&=&-w\, ,\nonumber\\
\ut=\wt\, ,\qquad \vt&=&\nt+w\, ,
\eea
leading to  
\bea\label{pqlr3}
p_L=\frac{u}{q}+\frac{v}{p}\, ,\qquad p_R &=&\frac{\ut}{q}+\frac{\vt}{p}\, , \nonumber\\
q_L=\frac{\ut}{q}-\frac{\vt}{p}\, ,\qquad q_R &=&\frac{u}{q}-\frac{v}{p}\, .
\eea
It is the shift in the momenta caused by the topological term that
allows us to rewrite $n$ and $\wt$ 
in terms of {\it independent} summation variables $u$ and $\ut$ etc. 

\subsection{Holomorphic Factorisation}

We see that now the pieces of the right holomorphicity that we wish to
keep are summed over the same indices and are not linked to the pieces
which we wish to remove. We may therefore remove the anti-holomorphic part 
of the partition function coming from $P$ (the $p_R$ piece) and the 
holomorphic part of the partition function coming from $Q$ (the $q_L$
piece). This leaves us with
\beq\label{Zf}
Z_f=\sum_{p_L,
p_R} \sqrt{2\tau_2}\exp\left[i\pi \tau \frac{p_L^2}{4}-i\pi\twb \frac{q_{R}^{2}}{4}\right]
\eeq
where
\beq
p_L=uR+\frac{v}{R}\, , \qquad q_R=uR-\frac{v}{R}\, .
\eeq
Alternatively we can see that the doubled partition function is the final form of the instanton part of the partition function, $Z_f$, times its complex conjugate:
\beq
Z_d=Z_f\times \bar{Z_f}.
\eeq
$Z_f$ is the partition function for a standard Boson with radius $R$
(or $R^{-1}$ by relabelling the sums) up to a factor outside the exponential, in the next section we will find that when we considered the doubled form of the rest of the partition function, the inverse of this factor arises. To recap: the approach taken here has been to treat the Bosons in the doubled formalism 
as chiral Bosons when trying to quantise. There is a key difference however, 
for chiral Bosons one must pick a spin structure \cite{Witten5} and we have not 
done so here. If one were to do so then there would not be enough degrees of 
freedom to reconstruct the usual non-chiral Boson. 
Thus, when one is holomorphically factorising here, one is effectively 
keeping a sum of chiral Bosons with all spin structures. This prescription is an essential part of the quantum prescription of the theory.

\section{Factorising The Oscillator Sector}

So far we have only included the sum over solutions to the classical equations of motion, to complete the quantum path integral we must include the fluctuations around these classical solutions\cite{polchinski}. We show the the same contribution to the partition function is obtained whether one works in terms of $X$ and $\xt$ or in terms of $P$ and $Q$.

\subsection{Oscillators in terms of $X$ and $\xt$}

For a Boson $X$ with action
\beq
S=-\frac{\pi R^2}{2} dX\wedge *dX\, ,
\eeq
we must do the Gaussian integral
\beq
\int\D X e^{-\int\frac{\pi R^2}{2}X \bx X},
\eeq
where $\bx$ is the Laplacian. The $\D X$ integration is split into the zero-mode piece and the integral over $\D X'$, orthogonal to the zero-mode. As $X$ has period 1 in our conventions, the zero-mode contribution is only a factor of 1. To normalise the measure we insert a factor of
\bea\label{norm}
&&\left(\int dx\, e^{-\frac{\pi R^2}{2}\int x^2}\right)^{-1}\non\\
&=&\left(\frac{\pi}{\frac{\pi R^2}{2}\int 1}\right)^{-1/2}\non\\
&=&\frac{R}{\sqrt{2}},
\eea
where we have used the fact that with our conventions $\int 1 $ over the torus is 1. 
This means
\beq
Z_{osc}=\frac{R}{\sqrt{2}}\frac{1}{\det' \bx}.
\eeq
We evaluate the determinant of $\bx=-4\tau_2\partial\bar{\partial}$ as a regularised product of eigenvalues, where the $'$ indicates this does not include zero-modes. We use a basis of eigenfunctions
\beq\label{efns}
\psi_{nm}=\exp \left[\frac{2\pi i}{2i\tau_2}\left(n(z-\zb)+m(\tau \zb-\twb z)\right)\right],
\eeq
which is single valued under $z\rightarrow z+1$ and $z\rightarrow z+\tau$, where $z=\sigma_1+\tau\sigma_0$. The regularised determinant is then the product of eigenvalues
\beq
\mbox{det} ' \bx=\prod_{\{m,n\}\neq\{0,0\}}\frac{4\pi^2}{\tau_2}(n-\tau m)(n-\twb m)\,.
\eeq
This can be evaluated using $\zeta$-function regularisation (see for example \cite{ginsparg}) as
\beq
\mbox{det}' \bx=\tau_2\eta^2(\tau)\bar{\eta}^2(\twb)\, ,
\eeq
where $\eta(\tau)$ is the Dedekind $\eta$-function
\beq
\eta(\tau)=e^{i\pi \tau/12}\prod_{n>1}(1-e^{2\pi i n \tau}).
\eeq
We now have that the oscillator part of the partition function is given by
\beq\label{dZoscX}
Z_{osc}=\frac{R}{\sqrt{2\tau_2}|\eta|^2}.
\eeq
The contribution due to the $\xt$ functional integral is an identical factor with $R$ replaced by $1/R$, so the square root of the doubled oscillator contribution, which we expect to be (and is) the same as the constrained contribution, is given by
\beq\label{dZosc}
Z_{osc}=\frac{1}{\sqrt{2\tau_2}|\eta|^2}.
\eeq

\subsection{Oscillators in terms of $P$ and $Q$}

To factorise the classical part of the partition function we worked in
terms of $P$ and $Q$ and used holomorphic factorisation, and we can
check that we get the same answer if we do that here. The substitution
(\ref{PQ}) introduces a Jacobian factor of $1/2$. Once the
substitution is made we do the path integral for the two Bosons, $P$ and $Q$,
just like the path integral for $X$ and $\xt$, except for a factor of
$1/2$ in the action and the more complex target space boundary
conditions that $P$ and $Q$ inherit as a result of their definition
(\ref{PQ}) in terms of $X$ and $\xt$. As the eigenfunctions of $\bx$,
(\ref{efns}), do not depend on these boundary conditions (unlike the
instanton pieces) the determinants for $P$ and $Q$ are the same as
those for  $X$ and $\xt$. However, the zero-mode integral does depend on the boundary conditions: although $P$ and $Q$ can take any value, the periodicity condition means we should only integrate over one fundamental region, we choose the one inherited from the fundamental region for $X$ and $\xt$, the region where $X$ and $\xt$ range from 0 to 1. The volume of this region is given by an integral over the values $P$ can take, of the range of values $Q$ can take for that value of $P$, that is
\beq
\int_{P=0}^{R^{-1}}2PdP+\int_{P=R^{-1}}^{R}2R^{-1}dP+\int_{P=R}^{R+R^{-1}}2((R+R^{-1})-P)dP=2\, ,
\eeq
cancelling the factor from the Jacobian.

The normalisation factor (\ref{norm}) remains the same as the additional factor of $1/2$ on the action is cancelled by the Jacobian which should also be included in this integral (or rather the root of the Jacobian as there is one Jacobian to be split between this and the $Q$ normalisation integrals). The $P$ oscillator contribution is then 
\beq\label{Zoscp}
Z_{osc;P}=\frac{1}{\sqrt{2\tau_2}|\eta|^2}.
\eeq
The $Q$ contribution is identical, and again one can take the $\tau$ dependent holomorphic square root of the $P$ factor and the $\twb$ dependent anti-holomorphic square root of the $Q$ factor and multiply them together to again get (\ref{Zoscp}). Taking this together with (\ref{Zf}) we find that the partition function for a Boson of radius $R$ in the doubled formalism is 
\beq\label{Z}
Z=\sum_{p_L,
p_R} \frac{1}{|\eta|^2}\exp\left[i\pi \tau \frac{p_L^2}{4}-i\pi\twb \frac{q_{R}^{2}}{4}\right]
\eeq
with
\beq
p_L=mR+\frac{n}{R}\, , \qquad q_R=mR-\frac{n}{R}\, .
\eeq
This is exactly what one obtains for the same Boson using the undoubled formalism, as we will now calculate with our conventions.

\section{Partition Function of the Ordinary Boson}

In order to aid comparison with the result of the doubled formalism, 
we describe below the partition function of the ordinary Boson at one 
loop using appropriate conventions so as to compare results. We proceed 
in the same way as above. 

\subsection{The Instanton Sector}

The action should be written
\beq
S=-\pi R^2 L\wedge *L\, ,
\eeq
with $L=dX+n\alpha+m\beta$, $m,n\in\Z$. We can then write the instanton sum part of the partition function as
\beq
Z_{inst}=\sum_{m,n} \exp \left[ -\pi R^2\left(n^2\frac{|\tau|^2}{\tau_2}-2mn\frac{\tau_1}{\tau_2}+\frac{m^2}{\tau_2}\right)\right]\, .
\eeq
Poisson resummation on $m$ gives
\bea\label{Bpf}
Z_{inst}&=&\sum_{n,w} \sqrt{\frac{\tau_2}{R^2}}\exp \left[ -\pi R^2\frac{n^2|\tau|^2}{\tau_2}-\frac{\tau_2\pi}{R^2}\left(w-\frac{in\tau_1R^2}{2\tau_2}\right)^2\right]\nonumber\\
&=&\sum_{n,w} \sqrt{\frac{\tau_2}{R^2}}\exp \left[ -\pi\tau_2\left(R^2n^2+\frac{w^2}{R^2}\right)+2\pi i n w \tau_1\right]\nonumber\\
&=&\sum_{n,w} \sqrt{\frac{\tau_2}{R^2}}\exp \left[ i\pi\tau \frac{p_L^2}{2}-i\pi\twb \frac{p_R^2}{2}\right]\, ,
\eea
where
\beq
p_L=Rn+\frac{w}{R}\, ,\qquad p_R=Rn-\frac{w}{R}\, .
\eeq
Performing Poisson resummation on $n$, rather than $m$, leads to a the same
result up to a modular transformation 
taking $\tau\rightarrow-\frac{1}{\tau}$. 

\subsection{Combining with the Oscillator Sector}

Evaluation of the oscillator part of the partition function proceeds
much as the previous section leading up to (\ref{dZoscX}). The only
difference is that there is no factor of $1/\sqrt{2}$ due to the standard normalisation of the action, giving
\beq
Z_{osc}=\frac{R}{\sqrt{\tau_2}|\eta|^2},
\eeq
leading to the full partition function
\beq
Z=\sum_{m,n} \frac{1}{|\eta|^2}\exp \left[ i\pi\tau \frac{p_L^2}{2}-i\pi\twb \frac{p_R^2}{2}\right]\, ,
\eeq
where
\beq
p_L=Rn+\frac{w}{R}\, ,\qquad p_R=Rn-\frac{w}{R}\, .
\eeq
The partition function is now invariant for $R\rightarrow 1/R$, after
the relabelling of sums. In general there will be an $R$-dependent
factor outside the exponential which is absorbed into the dilaton
shift, but in the case of the torus there is no shift due to the vanishing Euler characteristic. In our doubled calculation there was also no $R$ dependence in
the partition function, but both the instanton and oscillator pieces
were separately independent of $R$. For higher genus we expect
both pieces of the doubled partition function to remain $R$
independent, but for the ordinary Boson the instanton part will give
higher powers of $R$ where as the $R$-dependence of the oscillator
part will remain the same (this contribution effectively comes from
the volume of the zero-mode but we have scaled it into the target space
metric). This R-dependence will give the dilaton shift which is not
present in the doubled formalism, as here perturbation theory is in
terms of a differently defined, T-duality invariant, dilaton\cite{Hull:2006va}. Hull, however did find a shift coming from a functional determinant in the quantisation process (Section \ref{Sec:qudil}) but it is unclear how this generalises to different loops.  Although we have used the example of a non-fibred circle it is hoped the doubled formalism and the techniques of calculating the partition function used here will give more insight when applied to non-trivial T-folds.

%% file: conc.tex
\chapter{Conclusions and Further Work}

In this dissertation we started by reviewing some aspects of M-theory. We also reviewed how the D1-D3 system could be examined either from the D3-brane's or the D1-strings' perspective, and how these pictures agree for large $N$. We then saw how the M2-M5 system could be described similarly from both worldvolume viewpoints: the self-dual string picture from the M5-brane point of view and Basu and Harvey's recent proposal in terms of a fuzzy funnel for the M2 brane picture. 

We then showed how this Basu-Harvey picture could be generalised to M2-branes ending on a calibrated intersection of M5-branes, a process which required an effective linearised action for the coincident membranes as well as a hypothetical supersymmetry variation. It was also detailed how a proper description of the fuzzy three-sphere requires a projection on the allowed degrees of freedom, resulting in the number of degrees of freedom scaling like $N^{3/2}$, the result expected from scattering calculations. We proceeded to demonstrate how the Basu-Harvey picture of the M2-M5 system could be related to the fuzzy funnel picture of the D1-D3 system via a reduction from the fuzzy three-sphere to the fuzzy two-sphere.

We began the second part of the dissertation by describing Hull's doubled formalism, where the number of fibre co-ordinates for a torus bundle are doubled leading to a more geometric picture of how T-duality acts on the system. Hull demonstrated quantum equivalence to the standard formulation by gauging currents, here we have demonstrated quantum equivalence by calculating the partition function using holomorphic factorisation techniques to apply the constraint.

There are numerous avenues of future research using the M2-M5 system to investigate properties of the M2- or M5-brane theory. There has already been a recent attempt to model a supersymmetric membrane theory with non-associative fields\cite{baggerandlambert}. This was motivated by analogies to the D2 supersymmetry algebra and used a non-associative algebra and supersymmetry transform similar to ours, but different. An obvious next step would be to try to establish a connection between the two. The supersymmetry algebra of \cite{baggerandlambert} did not close up to a contribution that was taken to be some kind of novel gauge transform (again by analogy with the D2-brane case). The nature of this transform could also be investigated.

In the dissertation we also discussed how the Nahm transform can relate the fuzzy funnel solutions of the D1-string worldvolume theory to monopole solutions on the D3-brane worldvolume via the Nahm equation (Section \ref{sec:Nahm}). A similar transform should exist from solutions of the Basu-Harvey equation to more complicated self-dual string style solitons on the five-brane. The transform would need a formulation of the higher gauge theory of the two-form $B_{\mu\nu}$, possibly involving gerbes\cite{gerbe}.

The recent work of Bergshoeff, Gibbons and Townsend\cite{BGT} describing open M5-branes ending on M9-branes (M9-branes being the `end-of-the-universe' branes of \cite{HW}) leads to a natural generalisation of this work to fuzzy four-spheres. The M5-branes open up in a fuzzy four sphere to give the M9. The existence of a Yang-type monopole in the M9-brane as observed in \cite{BGT} is consistent with the appearance of a fuzzy four-sphere in the dual five-brane worldvolume description and hence the M5-brane theory could be investigated using a fuzzy sphere description. 

There are also plenty of possibilities for further work in the doubled formalism. We have considered the one-loop partition function for a single periodic direction with constant radius, so there are generalisations to higher genus and fibre dimension, as well as to more complex geometry with non-trivial B-field. It would also be interesting to see whether the partition function could be obtained without picking a polarisation, which should then demonstrate its invariance under choice of polarisation. More importantly, one could consider non-trivial T-folds without a global choice of polarisation and calculate the partition function there.

The relation to M-theory and how the T-duality invariant dilaton of the doubled formalism arises could also be investigated. While we found no dilaton shift, Hull \cite{Hull:2006va} did find one at one-loop that should be accounted for. In \cite{Hull:2004in} it was proposed how one might construct a U-duality invariant formulation of M-theory torus compactifications, with the theory compactified on $T^7$ (for example) needing to be reformulated on $T^{56}$. This could be investigated further and the form of the constraint needed established.

This dissertation has been a small step towards understanding the interactions of the fundamental branes of M-theory and a small step towards understanding the nature of an important symmetry special to string theory. We hope this will help us, and others, to take more steps (including those suggested in this section), steps that will lead eventually to a full understanding of string and M-theory, and what may lie beyond. What this is we do not know, but what is likely is that the true nature of space and time is stranger than we do, or can, imagine now.

%% file: app1.tex
\chapter{Conventions}\label{conv}

We use the following basis for the $Spin(4)$ $\Gamma$ matrices:
\begin{equation}\label{gamma-convention}
\Gamma^{i} = \left(\begin{matrix} 0 && \sigma^i \\
\bar\sigma^i && 0 \end{matrix}\right),\ \Gamma^5 =
\left(\begin{matrix} \openone_{2\times2} && 0 \\ 0 &&
-\openone_{2\times2} \end{matrix}\right).
\end{equation}
where
\begin{equation}\label{sigma-sigma-bar}
\sigma^i = (-i\vec{\sigma}_{Pauli}, \openone_{2\times2}),\
\bar\sigma^i = ( i\vec{\sigma}_{Pauli}, \openone_{2\times2})\, ,
\end{equation}
with  $\vec{\sigma}_{Pauli}$ being the standard Pauli sigma
matrices:
\begin{equation}
\sigma^1 = \left(\begin{matrix}  0 && 1 \\ 1 && 0
\end{matrix}\right),\ \sigma^2 = \left(\begin{matrix}  0 && -i \\ i &&
0 \end{matrix}\right),\ \sigma^3 = \left(\begin{matrix}  1 && 0 \\
0 && -1 \end{matrix}\right) .
\end{equation}
Thus
$\Gamma^5=\Gamma^1\Gamma^2\Gamma^3\Gamma^4=\frac{1}{4!}\epsilon_{ijkl}
\Gamma^i\Gamma^j\Gamma^k\Gamma^l$.

We will also use the notation
\beq
\Gamma^{ij}=\frac{1}{2}[\Gamma^i,\Gamma^j]
\eeq
and $\Gamma^{ij...p}$ for the generalisation to higher antisymmetrised products.
\newpage

%% file: app2.tex
\chapter{Fuzzy Spheres}\label{app:FS}

Since the fuzzy sphere was first defined in \cite{Madore}, fuzzy spheres of various dimensions have had many applications. The fuzzy sphere has appeared before in membrane theory in \cite{dWHN} in connection with its quantisation. Fuzzy spheres have been proposed as models of non-commutative space underlying the stringy exclusion principle in \cite{Maldacena:1998bw,Jevicki:1998rr}. They have also found application in Matrix Theory\cite{CLT,Ho}, as well as Tiny Graviton Matrix Theory\cite{SJ1,SJ2}. The fuzzy two-sphere has been used to describe polarised D0-branes polarised in a background field strength\cite{Myers}. Of course they occur in the fuzzy funnel description of branes ending on other branes as we have described; there are more recent extensions of this work too, for example, curved backgrounds and fluctuating funnels\cite{Bha2,Bhattacharyya:2005cd,Bha3,Thomas:2006ac}. Time-dependent collapsing fuzzy spheres have also been studied\cite{MPRS}. Recently \cite{SJN} commented on the inherent holography of fuzzy spheres and a possible application to the cosmological constant problem.

\section{Motivation and Desired Properties}

When the fuzzy sphere was first defined\cite{Madore} the motivation was to do away with the concept of a point in spacetime. In quantum field theory ultraviolet divergences appear when field oscillations are measures at a precise point in space time. Various ways are sought around this problem. String theory uses strings to replace point particles to remove such divergences. Loop quantum gravity modifies the the structure of space-time so that the notion of a point is lost.

For the fuzzy sphere the microscopic structure of space-time is modified such that at a significantly small length scale the co-ordinates of a point are non-commuting operators\cite{Connes}. One can no longer simultaneously diagonalise the co-ordinates, so the particle's exact position cannot be modified and hence the idea of a point loses its meaning. However, this intrinsic fuzziness in the location of a particle should not be observable on large length scales, and should not be greater than the quantum uncertainty in a particle's position.

The algebra of functions on a manifold defines its differential geometry: the co-ordinates generate the algebra and the vector fields act as derivations. To define the fuzzy sphere we would like to replace the classical algebra of functions with a non-commuting version, losing the idea of localisation and well-defined points. This algebra will be finite. A standard way of replacing the algebra with a finite version to remove divergences is to use a lattice, but this is not the non-commutative algebra we seek here. The fuzzy sphere should look like a classical sphere at large length scales. In effect we truncate the infinite classical algebra and introduce a non-commutative product that maintains the symmetry of the sphere. The truncation can be made at different orders and as the dimension of the truncation increases the approximation to the classical sphere gets better.

\section{The Fuzzy Two-sphere}\label{app:FS2}

The equation of a sphere embedded in $\mathbb{R}^3$ is 
\beq\label{S2}
\delta_{ab}x^ax^b=r^2.
\eeq
Any complex valued function on the sphere can be expanded as a polynomial in the co-ordinates
\beq
f(x^a)=f_o+f_ax^a+f_{ab}x^ax^b+\ldots
\eeq
and an approximation to the sphere can be found by truncating this expansion. 

Trivially, we could keep only the constant term and the algebra of functions would become $\mathbb{C}$, the same as the algebra of functions of a point.

If we allow the linear coefficients to be non-zero then we have a four dimensional vector space spanned by $f_o$ and the $f_a$. We would like to make this into an algebra. One possible manner of doing this gives a direct sum of four copies of $\mathbb{C}$ - this would be the lattice approximation. Instead we make the choice
\beq
x^a=\kappa \sigma^a
\eeq
where $\sigma^a$ are the Pauli matrices and $\kappa$ a parameter which determines the minimum length scale. This gives us the algebra of $2\times 2$ matrices and the co-ordinates obey
\beq\label{apfs2}
[x^a,x^b]=2i\kappa\epsilon_{abc}x^c\, . 
\eeq
To obey (\ref{S2}) we require $r^2=3\kappa^2$. We end up with a very poor approximation to a sphere, only the north and south poles can be distinguished.

The next degree of approximation is to keep the quadratic terms in the expansion as well. This gives a 9-dimensional vector space, and this can be made into the algebra of $3\times 3 $ matrices by replacing the co-ordinates $x^a$ with matrices
\beq
x^a=\kappa J^a,
\eeq
where $J^a$ is a 3 dimensional representation of the $SU(2)$ Lie algebra: $[J^a,J^b]=2i\epsilon_{abc}J^c$ and $r^2=8\kappa^2$.

More generally we can approximate the algebra of functions on the two-sphere by \mc{N_2}by replacing $J^a$ with the $N_2$ dimensional irreducible representation of $SU(2)$ and using the general relation $r^2=(N_2^2-1)\kappa^2$. Then (\ref{S2}) and (\ref{apfs2}) will still hold. As $N_2$ becomes large $\kappa\sim r/N_2$, and so the scale of fuzziness is small compared to the radius and more and more points can be distinguished.

Every element of \mc{N_2}can be expanded in the matrix versions of the co-ordinates using traceless symmetric tensors $f_{a_1a_2\ldots a_l}$ as
\beq\label{sh}
f=\sum_{l=0}^{N_2-1}\frac{1}{l!}f_{a_1a_2\ldots a_l}x^{a_1}x^{a_2}\ldots x^{a_{l}}.
\eeq
Replacing the matrix co-ordinates with the original continuous sphere co-ordinates gives a map from \mc{N_2}to the algebra of functions on the two-sphere. This is a very good approximation of an algebra morphism for large $N_2$, for functions whose expansion in polynomials is of degree significantly less than $N_2$. Bounded functions on the sphere are approximated by nearly-diagonal matrices, those which commute to order $\kappa$.

As the co-ordinates do not commute we cannot define a position for a particle, but in quantum mechanics the particle is defined by its state vector and its co-ordinates are observables. For \mc{N}the state vectors, $\psi$, are vectors of $\mathbb{C}^N$ with $\psi^*\psi=1$. These state vectors have $2N-4$ more degrees of freedom than just position.

The equivalent of general co-ordinate transforms are changes of the co-ordinate generators of the matrix algebra. Diffeomorphisms of the two-sphere are automorphisms of its algebra of smooth functions, so we look for automorphisms of \mcp{N}. Since \mc{N}is simple, these must be of the form $f\rightarrow f^\prime=g^{-1}fg$ for a fixed $g$ in \mcp{N}. Automorphisms corresponding to diffeomorphisms of the two-sphere will respect the operation of complex conjugation. If we are to have $f^{\prime*}=f^{*\prime}$ then we need $g^*=g^{-1}$ so that $g\in SU(N)$. General co-ordinate transforms of the fuzzy sphere are given by $x^{a\prime}=g^{-1}x^ag$ on the matrix co-ordinates for $g\in SU(N)$. If $g$ is not in $SU(N)$ then the transform will take us to a different differential or topological structure, for example the fuzzy torus, whose algebra of functions can also be represented by \mcp{N}.

\section{Poisson and Nambu Brackets}\label{app:PN}

As we have seen, the generators of the fuzzy two-sphere are taken to obey $[x^a,x^b]=2i\kappa\epsilon_{abc}x^c$. This is like a quantised Poisson bracket, with the fuzziness parameter $\kappa$ playing the role of Planck's constant. Indeed the classical two-sphere obeys a classical version of this equation\cite{SJ1}. We use the standard embedding of the two-sphere in $\mathbb{R}^3$
\bea\label{spherepol}
x^1&=&r\sin\theta\cos\phi,\nonumber\\
x^2&=&r\sin\theta\sin\phi,\nonumber\\
x^3&=&r\cos\theta.
\eea
We then calculate the Poisson bracket
\beq\label{poisson}
\{x^1,x^2\}=\partial_\theta x^1\partial_\phi x^2-\partial_\theta x^2\partial_\phi x^1=r^2\sin\theta\cos\theta=vrx^3,
\eeq
where $v=\sin\theta$ is the volume form on the unit two-sphere. More generally
\beq
\{x^a,x^b\}=\epsilon_{abc}vrx^c,
\eeq
giving us a classical form of (\ref{apfs2}). Note that along with (\ref{S2}) 
we have equations involving the co-ordinates and both invariant tensors of $SO(3)$.

To examine higher dimensional spheres we introduce the Nambu $p$-bracket
\beq\label{nambu}
\left\{f_1,f_2,\ldots,f_p\right\}=\epsilon_{r_1\ldots r_p}\frac{\partial f_{r_1}}{\partial\sigma_1}\frac{\partial f_{r_2}}{\partial\sigma_2}\ldots \frac{\partial f_{r_p}}{\partial\sigma_p},
\eeq
where the $f_i$ are functions on a $p$-dimensional space parameterised by $\sigma_1,\ldots,\sigma_p$. For $p=2$ this is the Poisson bracket. Higher $p$-spheres obey a classical equation
\beq\label{genN}
\{x^i,x^j,\ldots,x^p\}=\epsilon_{ij\ldots p}vr^{p-1}x^p,
\eeq
where $v$ is the unit volume form of the $p$-sphere. When constructing higher fuzzy spheres we need to satisfy a quantum version of this equation. As we will see this is much simpler for higher even spheres than for odd ones.

\subsection{Dimensional Reduction of Nambu Brackets}\label{app:dimred}

As we would like to relate the Nahm equation (\ref{Nahmeq}) to the Basu-Harvey equation (\ref{BaH}) we must first, in Section \ref{sec:dimred}, find how to embed the fuzzy two-sphere in the fuzzy three-sphere. We would especially like to know what happens to the fuzzy three-sphere equation (\ref{FS3}) and how it is obeyed by the two-sphere. Since these are quantum versions of the sphere Nambu bracket equations (\ref{genN}) we can look at what happens when we embed a lower sphere in these classical Nambu bracket equations. For simplicity we consider embedding a circle in a two-sphere. 

We know that the co-ordinates (\ref{spherepol}) obey (\ref{poisson}). Suppose we try to restrict to a circle of constant $x^3$. Then putting this value into the Poisson bracket will cause it to vanish, as the derivatives will annihilate the constant. However the remaining co-ordinate on the other side of the equation will not vanish in general for co-ordinates restricted to the circle. So though the co-ordinates obey the two sphere equation $x^ax^a=r^2$, the Poisson bracket is not solved.

The correct procedure is to fix the value of the co-ordinate {\it after} acting with the derivatives in the Poisson bracket. For example, fixing $x^3$ is the same as fixing $\theta$. From (\ref{spherepol})
we should use the fact that
\bea
\partial_\theta x^1&=& \frac{\cos\theta}{\sin\theta}x^1,\nonumber\\
\partial_\theta x^2&=& \frac{\cos\theta}{\sin\theta}x^2,\nonumber\\
\partial_\theta x^3&=& -\frac{\sin\theta}{\cos\theta}x^3=-r\sin\theta
\eea
and $\p_\phi x^3=0$ to rewrite the three Poisson bracket equations as
\bea\label{pois}
\cos\theta (x^1\p_\phi x^2-x^2\p_\phi x^1)&=&r\sin^2\theta x^3,\nonumber\\
\p_\phi x^2&=&x^1,\nonumber\\
-\p_\phi x^1&=&x^2.
\eea
The second and third equations are the one-dimensional Nambu bracket relation $\p_\phi x^a=\epsilon_{ab}x^b$, where because of our choice of co-ordinates (\ref{spherepol}) $\epsilon_{12}=-1$. With these relations, once we have fixed $x^3$ and thus $\theta$, we can rewrite the first equation of (\ref{pois}) as
\beq
(x^1)^2+(x^2)^2=\sin^2\theta_o r^2=r'^2.
\eeq
This equation gives us the radius of the new sphere in terms of the old sphere. In higher sphere cases, where $r$ appears in the Nambu bracket equation, it is this modified radius that should appear.

\section{The Fuzzy Four-sphere}\label{app:FS4}

There are different ways to construct fuzzy spheres, and we will follow the group-theoretical approach utilised in \cite{CLT} (the fuzzy four-sphere was first discussed in \cite{GKP}). Here the fuzzy four sphere was first constructed using symmetric tensor product representations. The basic equations that the fuzzy four-sphere co-ordinates, $\gh$, must satisfy are of the form 
\beq\label{FS4a}
\gh^\mu \gh^\mu=R^2
\eeq
and
\beq\label{FS4}
\epsilon_{\mu\nu\sigma\tau\rho}\gh^\mu \gh^\nu \gh^\sigma \gh^\tau =\alpha \gh^\rho\, .
\eeq
The $\epsilon$ equation can alternatively be expressed in terms of a quantum Nambu bracket,
\beq
[\gh^\mu,\gh^\nu, \gh^\sigma, \gh^\tau] =\alpha\epsilon_{\mu\nu\sigma\tau\rho}\gh^\rho\, .
\eeq
The quantum Nambu bracket is just a sum over the $4!$ signed permutations of the entries 
\beq
[A^1,A^2, A^3, A^4] =\epsilon_{i_1i_2i_3i_4}A^{i_1}A^{i_2}A^{i_3}A^{i_4}.
\eeq

As with the fuzzy two-sphere a good place to start is what we will call $n=1$, the minimal non-trivial case. Putting $\gh^\mu=\Gamma^\mu$ (where $\Gamma^\mu$ are the five $4\times 4$ Euclidean gamma matrices, when necessary we will use the basis given in Appendix \ref{conv}) solves the fuzzy four-sphere equations with $R=2$ and $\alpha=4!$. Here we are approximating the algebra of functions on the sphere by \mcp{4}.

The key step is to generalise this gamma matrix representation to higher $N$, and to this end we introduce the $n$th symmetric tensor representation of the gamma matrices,
\beq
\gh^\mu=(\Gamma^\mu\otimes1\otimes1\otimes\ldots\otimes1+1\otimes\Gamma^\mu\otimes1\otimes\ldots\otimes1+\ldots+1\otimes\ldots\otimes1\otimes\Gamma^\mu)_{sym}.
\eeq
The $\Gamma^\mu$ act on the spinor representation $V$ which has a basis $e_a$, $a=1,\dots,4$, via $\Gamma^\mu(e_a)=(\Gamma^\mu)_{ba}e_b$. Thus the $\gh^\mu$ act on the $n$-fold tensor product of $V$, $\left(V^{\otimes N}\right)_{sym}$. Basis states of this will be $n$-fold symmetrised products of the $e_a$,
\beq\label{state}
\left(e_{a_1}\otimes e_{a_2}\otimes\cdots\otimes e_{a_n}\right)_{sym}\, ,
\eeq
and enumerating all such possible states gives the dimension of this basis, 
\beq\label{N_4}
N_4=\frac{(n+1)(n+2)(n+3)}{6}\, .
\eeq
Following the discussion of Section (\ref{app:FS2}) the state vectors of the fuzzy four sphere are vectors of $\mathbb{C}^{N_4}$ and the $\gh^\mu$ are in \mcp{N_4}. We will refer to objects of the form (\ref{state}) and their linear combinations as ``states'', and we will refer to elements of the algebra of functions on the fuzzy sphere, here \mcp{N_4}, as ``operators''. We will often refer to both as being composed of $n$ tensor factors.

We now check that the $\gh^\mu$ satisfy the fuzzy four-sphere equation (\ref{FS4a}). We can write $\gh^\mu$ as $P_n\sum_r\rho_r(\Gamma^\mu)P_n$ where $P_n$ is the symmetrisation operator and 
\beq
\rho_r(A)=\openone\otimes\openone\otimes\ldots\otimes \underbrace{A}_{r^{th} \mbox{factor}} \otimes \ldots \otimes \openone
\eeq
with $A$ acting on the $r$th tensor factor: for example
\beq
\rho_m(\Gamma^\mu)(e_{i_1}\otimes e_{i_2}\otimes\dots\otimes
e_{i_m}\otimes\ldots\otimes e_{i_n})=(e_{i_1}\otimes
e_{i_2}\otimes\dots\otimes (\Gamma^\mu_{j_mi_m}e_{j_m})\otimes\ldots\otimes
e_{i_n})\, .
\eeq
Using this notation we can see $\gh^\mu\gh^\mu$ splits into two kinds of term
\bea\label{RS4a}
\gh^\mu\gh^\mu&=&\sum_{r, s}\rho_r(\Gamma^\mu)\rho_s(\Gamma^\mu)\nonumber\\
&=&\sum_r\rho_r(\Gamma^\mu\Gamma^\mu)+\sum_{r\neq s}\rho_r(\Gamma^\mu)\rho_s(\Gamma^\mu)\, .
\eea
Since $\Gamma^\mu\Gamma^\mu=5\openone$ the first term is equal to $5n\openone_{N_4}$. To evaluate the second equation we must use $\Gamma^\mu\otimes\Gamma^\mu=1$. To deduce this requires showing that $\Gamma^\mu\otimes\Gamma^\mu$ tensored with $(n-2)$ identities commutes with the generators of $SO(5)$, which are given by 
\beq
\gh^{\mu\nu}=\frac{[\gh^\mu,\gh^\nu]}{2}=\sum_r\rho_r(\Gamma^{\mu\nu}).
\eeq
Once this is established, Schur's lemma tells us that $\Gamma^\mu\otimes\Gamma^\mu$ is proportional to the identity. Acting on any basis state gives the constant of proportionality. The number of distinct pairs of tensor factors upon which the two $\Gamma^\mu$'s can act is given by $n(n-1)$. Putting all this together we get
\beq\label{RS4}
R^2=5n+n(n-1)=n(n+4).
\eeq
A slightly more complicated calculation using similar techniques establishes that (\ref{FS4}) is also satisfied, with $\alpha=8(n+2)$.

Notice that this construction gives fuzzy four-spheres only for $N=(n+1)(n+2)(n+3)/6$ with $n$ a positive integer. \cite{CLT} demonstrated that a spin representation of $SO(5)$ could not be constructed for $N=5$ as the $\gh^i$ would mix between representations.

This construction is generalisable to higher fuzzy even spheres using the $Spin(2k+1)$ gamma matrices. In general $R^2=n(n+2k)$. We could also use this construction for the fuzzy two-sphere using
\beq\label{2const}
\Sigma^a=(\sigma^a\otimes1\otimes1\otimes\ldots\otimes1+1\otimes\sigma^a\otimes1\otimes\ldots\otimes1+\ldots+1\otimes\ldots\otimes1\otimes\sigma^a)_{sym}\, ,
\eeq
where $\sigma^a$ are the Pauli matrices. This has $N=n+1$. One could think of this as a spin $n/2$ representation with the spin given by the eigenvalue of $\Sigma^3/2$.

\section{The Fuzzy Three-sphere}\label{app:FS3}

The fuzzy three-sphere was first constructed in \cite{Ram1} and it was developed in \cite{Ram2,Ram3}. We look for a fuzzy three-sphere embedded in the fuzzy four-sphere by looking at how the symmetric tensor representations of $Spin(5)$ decompose into representations of $Spin(4)$. As $Spin(4)=SU(2)\times SU(2)$, its irreducible representations can be labelled by the pair of spins $(j_L,j_R)$. Using the projectors $P_\pm=\frac{1}{2}(\openone\pm\Gamma^5)$ we can decompose the fundamental spinor representation of $Spin(5)$, $V$, into $P_+V$ and $P_-V$. These are the $(\frac{1}{2},0)$ and $(0,\frac{1}{2})$ representations of $Spin(4)$ respectively. The symmetrised tensor \rep $V^{\otimes n}$ can be decomposed similarly into the direct sum of irreducible representations of $Spin(4)$ containing $k$ factors of $P_+V$ tensored with $(n-k)$ factors of $P_-V$ and then symmetrised. This can be seen from
\beq  ( V^{\otimes n } )_{sym} = P_n V^{\otimes n } 
 = P_n ( P_+ + P_- )^{\otimes n}  V^{\otimes n } 
 = P_n \sum_{k} ({P_+}^{\otimes k} {P_-}^{\otimes n-k})_{sym} 
V^{\otimes n} .
\eeq
It can be easily checked that summing the dimensions of these $SU(2)\otimes SU(2)$ \reps reproduces the dimension of the $n$-fold symmetric tensor (given in (\ref{N_4})).

The fact that $\Gamma^iP_\pm=P_\mp\Gamma^i$ tells us that the $G^i$ are going to map between different \irreps and the fuzzy three-sphere will require a reducible sum of \irrepsp. Evaluation of the radius $G^iG^i$ will be different on each \irrep and will also depend on whether adjacent \irreps (with $k$ one less or greater) are included. Obviously the sphere should have fixed radius, and requiring the $G^i$ to act non-trivially (they map an \irrep to an adjacent one) fixes which \irreps are part of the fuzzy three-sphere: we restrict to those with $k=\frac
{(n\pm1)}{2}$ when $n$ is odd. We call these $\cRp$ and $\cRm$ and in terms of $SU(2)\times SU(2)$ they are given by
\beq
\cR=\cRp+\cRm=\left(\frac{n+1}{4},\frac{n-1}{4}\right)\oplus\left(\frac{n-1}{4},\frac{n+1}{4}\right)\, .
\eeq
The dimension can be found by enumerating the states or simply from the $SU(2)$ spins,
\beq\label{N_3}
N=2\times\left(\frac{n-1}{2}+1\right)\left(\frac{n+1}{2}+1\right)=\frac{(n+1)(n+3)}{2}\, .
\eeq
As for the fuzzy four-sphere this construction gives solutions only for certain values of $N$.

\subsection{Projectors for the Fuzzy Three-sphere}

When calculating with $G^i$ we should make sure we do not pass through states in \irreps outside the fuzzy three-sphere. We do this by explicitly realising the restriction to the fuzzy three-sphere by introducing projectors $\Prp$ and $\Prm$ onto $\cRp$ and $\cRm$. These can be constructed from $\frac{n\pm1}{2}$ tensor factors of $P_+$ and $\frac{n\mp1}{2}$ tensor factors of $P_-$, for example 
\beq
\Prp=\left(P_+^{\otimes\frac{n+1}{2}}\otimes P_-^{\otimes\frac{n-1}{2}}\right)_{sym}\, .
\eeq
For the first non-trivial case, $n=3$, the explicit form of $\Prp$ is
\beq
\Prp=P_+\otimes P_+ \otimes P_-+P_+\otimes P_- \otimes P_++P_-\otimes P_+ \otimes P_+\, .
\eeq
We can write $G^i$  in terms of the first four fuzzy four-sphere matrices, $\gh^i$, as
\beq
G^i=\Prm \gh^i \Prp+\Prp \gh^i\Prm=\Prm \sum_r\rho_r(\Gamma^iP_+) \Prp+\Prp \sum_r\rho_r(\Gamma^iP_-)\Prm\, .
\eeq
This also shows that $G^i$ contains a piece in $Hom(\cRp,\cRm)$ and a piece in $Hom(\cRm,\cRp)$. Writing both the $P_\pm$ and $\cPpm$ projectors is not necessary but is useful to avoid errors on longer calculations. For example when the gamma matrix factor acts on a $ P_- $ factor in $\cRp$ it will map to state in $\left(\frac{n+3}{2},\frac{n-3}{2}\right)$ and so will be projected out by $\cPr=\Prp+\Prm$.

Since $\Gamma^5P_\pm=\pm P_\pm$, acting with the fifth co-ordinate matrix of the fuzzy four-sphere, $G^5$, will count the difference between the number of $P_+$ and $P_-$ factors. Thus $G_5$ gives $+1$ on $\cRp$ and $-1$ on $\cRm$ (we write it as $G_5$ with the index down as it is different from the four fuzzy three-sphere co-ordinates). It can be expressed in various ways:
\beq
G_5=\cPr \gh^5 \cPr=\cPr\sum_r\rho_r(\Gamma^5)\cPr=\cPr\sum_r\rho_r(P_+-P_-)\cPr=\Prp-\Prm\, .
\eeq

\subsection{Calculating the Radius}\label{app:rad}

We have chosen the \irreps so that the radius is fixed,  and we will calculate the radius but first we show a useful identity. For the fuzzy four-sphere we had that
\beq
\Gamma^\mu\otimes\Gamma^\mu=\openone\otimes\openone\, ,
\eeq
where both sides act on $V\otimes V$. Because of the symmetrisation it follows that
\beq
\left(\Gamma^\mu\otimes\Gamma^\mu\right)\left(P_+\otimes P_-\right)=\Gamma^\mu P_+\otimes \Gamma^\mu P_-=P_-\otimes P_+=\left(\openone\otimes\openone\right)\left(P_+\otimes P_-\right)\, .
\eeq
Then we can separate off the first four gamma matrices, labelled by $i,j,\ldots=1,\ldots,4$, to get
\bea\label{gammaP}
\left(\Gamma^i\otimes\Gamma^i\right)\left(P_+\otimes P_-\right)&=&\left(\Gamma^\mu\otimes\Gamma^\mu\right)\left(P_+\otimes P_-\right)-\left(\Gamma^5\otimes\Gamma^5\right)\left(P_+\otimes P_-\right)\non\\&=&2\left(\openone\otimes\openone\right)\left(P_+\otimes P_-\right)\, .
\eea
A similar calculation shows $\left(\Gamma^i\otimes\Gamma^i\right)\left(P_\pm\otimes P_\pm\right)=0$. Alternatively, (\ref{gammaP}) can be demonstrated by showing that $\Gamma^i\otimes\Gamma^i$ (tensored with $(n-2)$ $\openone$'s) commutes with the generators of $SO(4)$. Schur's lemma then implies that it is proportional to the identity in both \irrepsp, and each \irrep has the same proportionality constant due to symmetry under exchanging $+$ and $-$. The constant can then be found by acting on any basis state.

We can now proceed by calculating
\bea
G^iG^i\Prp&=&\Prp\sum_{r}\rho_r(\Gamma^iP_-)\Prm\sum_{s}\rho_s(\Gamma^iP_+)\Prp\nonumber\\
&=&\Prp\rho_r(\Gamma^i\Gamma^i)\Prp+\Prp\sum_{r\neq s}\rho_r(\Gamma^iP_-)\rho_s(\Gamma^iP_+)
\Prp\, .
\eea
In the first term there are $\frac{n+1}{2}$ factors on which the $P_+$ could act. Similarly in the second term there are $\left(\frac{n+1}{2}\right)\left(\frac{n-1}{2}\right)$ possible choices of $r$ and $s$. Using (\ref{gammaP}) we get
\beq\label{FS4rad}
G^iG^i\Prp=4\left(\frac{n+1}{2}\right)\Prp +2\left(\frac{n+1}{2}\right)\left(\frac{n-1}{2}\right)\Prp=\frac{(n+1)(n+3)}{2}\Prp\, .
\eeq
The calculation is the same when $G^iG^i$ acts on $\Prm$, except that one just interchanges $+$ and $-$ on all projectors. Notice that the radius is the same as the dimension $N$ given in equation (\ref{N_3}), that these coincide is unique to the fuzzy three-sphere. This construction and calculation can be generalised to higher odd spheres: for a fuzzy $2k-1$ sphere the radius\footnote{In the literature the term `radius' is often applied to the dimensionful number, $r$, which appears in the scaled co-ordinates $x^i=\frac{r}{n}G^i$. Here we use radius to refer to that of the `bare' fuzzy sphere.}  is given by
\beq
R^2=\frac{(n+1)(n+2k-1)}{2}\, .
\eeq

\subsection{The Nambu Bracket Equation}\label{app:G4b}

For the four sphere we found equation (\ref{apfs2}), which was the analogue of the $SU(2)$ algebra equation (\ref{FS4}) obeyed by the fuzzy two-sphere. The four-sphere could be expressed in terms of a quantum Nambu four-bracket and the two-sphere equation contained a commutator, a quantum Nambu two-bracket. These equations are quantum analogues of the classical Nambu bracket equation (\ref{nambu}) obeyed by the classical co-ordinates. As discussed in \cite{SJ1,SJ2} quantising odd Nambu brackets is a much more difficult task. Even quantum Nambu brackets preserve many of the properties of the original classical Nambu brackets, but an anti-symmetric odd-bracket has fewer of the desired properties, for example the classical `trace' property
\beq
\int d^p\sigma\{f_1,f_2 ,\ldots,f_p\}=0
\eeq
is obeyed in a quantised form by the even quantum Nambu four bracket 
\beq
\mbox{Tr}[A^1,A^2,\ldots, A^{2k}]=0\, ,
\eeq
but not for the odd brackets. This property is important for the existence of conserved quantities. 

As the $n=1$ equations are solved by the Clifford algebra of the spin cover of the sphere symmetry group we can look at the equivalent equation for the $Spin(4)$ gamma matrices. We have that $[\Gamma^i,\Gamma^j,\Gamma^k]=3!\epsilon_{ijkl}\Gamma^5\Gamma^l$, or in terms of a four-bracket
\beq\label{gamma4b}
\epsilon_{ijkl}[\Gamma^5,\Gamma^i,\Gamma^j,\Gamma^k]=3!4!\Gamma^l\, .
\eeq 
$\Gamma^5$ is not a generator of the Clifford algebra, but it necessarily appears here. Recall that $\Gamma^5=\Gamma^1\Gamma^2\Gamma^3\Gamma^4$ and it appears in the chirality projector.

Hence we look for a general form of (\ref{gamma4b}) involving $G_5$ which still appears as a difference of projection operators for the $n>1$ cases. The calculation is quite long but we have repeated it here as it is prototypical of most fuzzy three-sphere calculations. Also, when originally performed in \cite{BasuH} some terms were missed leading to different large-$n$ behaviour. After our discovery of this it was corrected in later versions of the paper. 

It turns out that it will be easier to exploit the symmetry of the epsilon if we first calculate $\epsilon_{ijkl} G_5 G^i G^j G^k G^l$ which can be shown to be proportional to the identity by the same Schur's lemma and $+$/$-$ symmetry argument used earlier. Acting on $\Prp$ we have
\beq \label{5G}
\epsilon_{ijkl} G_5 G^i G^j G^k G^l \Prp = \epsilon_{ijkl} \sum_{r,s,t,u=1}^n
\rho_r (\Gamma^i P_-) \rho_s (\Gamma^j P_+) \rho_t (\Gamma^k P_-) 
\rho_u (\Gamma^l P_+) \Prp. \eeq
We separate into terms depending on which, if any, $G^m$'s act on the same tensor factor. The anti-symmetry provided by the epsilon means that any terms with two $G^m$'s acting individually on tensor factors of the same chirality will vanish, as they would also have to be symmetric. To evaluate the remaining terms we use
\beq\label{gammas}
\Gamma^i\Gamma^j\Gamma^k=\epsilon_{ijkl}\Gamma^5\Gamma^l, \qquad \Gamma^i\Gamma^j=-\frac{1}{2}\epsilon_{ijkl}\Gamma^5\Gamma^k\Gamma^l
\eeq
to eliminate the $\epsilon$ from the remaining terms. We can then use (\ref{gammaP}) and the similar relations 
\bea
\sum_i (\Gamma^i \otimes \Gamma^i) (P_+ \otimes P_-)_{\rm{sym}} =
2 (P_- \otimes P_+)_{\rm{sym}},\cr
\sum_{ij} (\Gamma^{ij} \otimes \Gamma^{ji}) (P_+ \otimes P_+)_{\rm{sym}} =
4 (P_+ \otimes P_+)_{\rm{sym}},\cr
\sum_{ij} (\Gamma^i \otimes \Gamma_{ij} \otimes \Gamma^j) (P_- \otimes P_+ \otimes P_+)_{\rm{sym}} =
-2 (P_+ \otimes P_+ \otimes P_-)_{\rm{sym}},\cr
\sum_{ij} (\Gamma^i \otimes \Gamma_{ij} \otimes \Gamma^j) (P_- \otimes P_- \otimes P_+)_{\rm{sym}} =
2 (P_+ \otimes P_- \otimes P_-)_{\rm{sym}}
\eea
to eliminate the remaining gamma matrices and all terms  are proportional to the identity. That just leaves us to enumerate the possible choices of which tensor factors are acted upon for each term, and we see that in doing this calculation we are just repeating the steps of evaluating $G^iG^i$:
\bea \label{enum} \sum_r \rho_r (P_+) \Prp = \frac{(n+1)}{2} \Prp ,
\qquad \quad \cr
\sum_{r \neq s} \rho_r (P_-) \rho_s (P_+) \Prp = \frac{(n+1)(n-1)}{4}
\Prp ,\cr
\sum_{r \neq s} \rho_r (P_+) \rho_s (P_+) \Prp = \frac{(n+1)(n-1)}{4}
\Prp ,\cr
\sum_{r \neq s \neq t} \rho_r (P_-) \rho_s (P_+) \rho_t (P_+) \Prp = \frac{(n+1)(n-1)^2}{8}
\Prp ,\cr
\sum_{r \neq s \neq t} \rho_r (P_-) \rho_s (P_-) \rho_t (P_+) \Prp = \frac{(n+1)(n-1)(n-3)}{8}
\Prp .\cr
\eea
This finally allows us to evaluate the non trivial terms in (\ref{5G})
\bea
\epsilon_{ijkl} \sum_r \rho_r (\Gamma^i \Gamma^j \Gamma^k \Gamma^l P_+) \Prp &=&
12 (n+1) \Prp, \non\\
\epsilon_{ijkl} \sum_{r \neq s} \rho_r (\Gamma^i \Gamma^j \Gamma^k P_-)
\rho_s (\Gamma^l P_+) \Prp &=& 3(n+1)(n-1) \Prp,\non\\
\epsilon_{ijkl} \sum_{r \neq s} \rho_r (\Gamma^i P_-) \rho_s (\Gamma^j \Gamma^k
\Gamma^l P_+) \Prp &= &3(n+1)(n-1) \Prp,\non\\
\epsilon_{ijkl} \sum_{r \neq s} \rho_r (\Gamma^{ij} P_+) \rho_s (\Gamma^{kl} 
P_+) \Prp &=& 2 (n+1)(n-1) \Prp, \non\eea
\bea&&\epsilon_{ijkl} \sum_{r \neq s \neq t} \rho_r (\Gamma^i P_-) \rho_s (\Gamma^j \Gamma^k P_- )\rho_t (\Gamma^l P_+)\Prp\non\\&&\qquad\qquad\qquad\qquad\qquad\qquad=\frac{(n+1)(n-1)(n-3)}{2} \Prp, \non\\
&&\epsilon_{ijkl} \sum_{r \neq s \neq t} \rho_r (\Gamma^i P_-) \rho_s (\Gamma^j P_+) \rho_t (\Gamma^k \Gamma^l P_+)]\Prp\non\\&&\qquad\qquad\qquad\qquad\qquad\qquad=\frac{(n+1)(n-1)(n-1)}{2} \Prp .\label{terms} 
\eea
Note the it was the terms in the final two lines that were originally missed. In the first of these, for example, the anti-symmetry in $k$ and $i$ does not lead to vanishing as the tensor factors are not symmetric due to the presence of an additional $\Gamma^j$ multiplying one of them.

Summing up all the terms and repeating the identical calculation for $\Prm$ we get
\beq \label{5Gn}
\epsilon_{ijkl} G_5 G^i G^j G^k G^l =(n+1)(n+2)(n+3)\openone.
\eeq
To get the fuzzy three sphere equation we would now like to calculate $\epsilon_{ijkl} G_5 G^j G^k G^l$. From the index structure of the operator it must be of the form
\beq 
\epsilon_{ijkl} G_5 G^j G^k G^l = f(n) G^i + g(n) G_5 G^i.
\eeq
We then multiply both sides by $G^i$ and use (\ref{5Gn}) and (\ref{FS4rad}), which fix the unknown functions of $n$ to
\beq
f(n) = - 2(n+2), \qquad g(n) =0,
\eeq
leading to the equation 
\beq
G^i +\frac{1}{2(n+2)} \epsilon_{ijkl} G_5 G^j G^k G^l =0,
\eeq
which can be written with a four bracket as
\beq
G^i +\frac{1}{2(n+2)} \epsilon_{ijkl}\frac{1}{4!}[ G_5, G^j, G^k ,G^l] =0,
\eeq
giving us a realisation of a fuzzy version of the three-sphere equation involving a quantum Nambu 4-bracket. Both this and (\ref{FS4rad}) are invariant under $SO(4)$ rotations which act on the $G^i$ as conjugation by unitary matrices belonging to $SU(N)$. 

If the $G^i$ are to be interpreted as the co-ordinates of the sphere, we can ask what is the interpretation of $G_5$. In \cite{SJ2} $G_5$ keeps the interpretation as the fifth co-ordinate on a fuzzy four-sphere, set as close to zero as the fuzzy algebra will allow. It is sufficiently small that it vanishes in the large-$n$ limit. This implies that the fuzzy three-sphere constructed in this way is actually a finite band around the equator of the four-sphere, and retains the topology of the four-sphere.  In the specific context of Tiny Graviton Matrix Theory it is given the interpretation of a hidden eleventh dimension appearing in the quantisation of IIB string theory. In contrast in \cite{Ram1} it plays the role of the tachyon for unstable $D0$-branes in a background field strength.

\section{Projection to Fuzzy Spherical Harmonics via Young Diagram Basis}\label{app:pshydb}

We have constructed finite algebras of matrices which reproduce the symmetries of the four-sphere and three-sphere. However, if we recall the two sphere case, we wish to reproduce the algebra of functions on the sphere as we take the large-$n$ limit. This algebra is given by the spherical harmonics, which can be represented by {\it symmetric traceless} tensors. We will now expand the fuzzy four-sphere algebra in terms of a Young diagram basis, before describing the projection which leaves only the fuzzy spherical harmonics. This was described in \cite{Ram2}.

\subsection{The Young Diagram Basis}\label{app:YDB}

To make a general operator on our $n$-fold symmetric tensor product space we can act on the tensor-factors symmetrically with different products of gamma matrices. Since $\Gamma^1\Gamma^2\Gamma^3\Gamma^4\Gamma^5=\openone$, the maximum length of a product of gammas is two. Therefore our operators are made up from single gamma factors, $\rho_r(\Gamma^i)$, and double  gamma factors, $\rho_r(\Gamma^{ij})$. The operators will exactly correspond to Young diagrams of $SO(5)$, made up of boxes in two rows with lengths $r_1\geq r_2$. For example the $(r_1,r_2)=(4,1)$ diagram is\\
\begin{equation}\nonumber
\includegraphics[clip=true, angle=-90]{yd41.pdf}
\end{equation}\\
Each box corresponds to an index on a gamma matrix and they are symmetric across the rows and anti-symmetric down each column. Each column corresponds to a $\rho_r$ acting on a tensor factor, and the length of the column to how many gamma matrices it contains. The above Young diagram corresponds to the operator
\beq
\sum_{\vec{s}}\rho_{s_1}(\Gamma^i\Gamma^j)\rho_{s_2}(\Gamma^k)\rho_{s_3}(\Gamma^l)\rho_{s_4}(\Gamma^m).
\eeq
The vector indices on the gamma matrices are contracted with a tensor of the appropriate symmetry. This is forced as the gamma matrix products are anti-symmetric and the symmetry across tensor factors ensures symmetrisation across rows. Importantly these tensors must also be {\it traceless}; that is they must vanish if any two indices are contracted together. This ensures we do not over-count diagrams with fewer boxes when enumerating the dimension of the space of independent operators. The dimension of space of traceless tensors corresponding to a diagram can be calculated using the row lengths via a group theoretical formula. For the fuzzy odd spheres this is given in the main text in equation (\ref{dim}) of Section \ref{sec:sumYDB}. Summing over both row lengths up to $n$ we get that all operators of this form span an
\beq
\frac{(n+1)^2(n+2)^2(n+3)^2}{3!^2}
\eeq
dimensional space - exactly the dimension of \mcp{N_4}. A more explicit calculation of this for the fuzzy three-sphere is done in Section \ref{sec:sumYDB}

\subsection{The Projection}\label{app:proj}

The projection to spherical harmonics is simply a restriction to those diagrams with $r_2=0$; i.e. completely symmetric diagrams. We denote the space of these operators by \ansfp, but this space is not closed under multiplication. We can project back into \ansf after multiplication but this projection leads to non-associativity. The projection and how the non-associativity arises in the more complicated fuzzy-three sphere case is described in Sections \ref{sec:shproj} and \ref{sec:nonass}.
For the fuzzy four-sphere this non-associativity disappears in the large-$n$ limit without the further projection, needed for fuzzy odd-spheres, which was described in Section \ref{sec:nonass}.

For the fuzzy odd-spheres, this projection is needed, not only to get the correct algebra of functions in the large-$n$ limit, but also to restore commutativity in that limit. This persistence of non-commutativity for the fuzzy odd-spheres is due to the loss of symmetry between the different chirality factors. For example, in the fuzzy four-sphere the commutator of two co-ordinates is given by
\beq
[\gh^\mu,\gh^\nu]=\sum_r\rho_r(2\Gamma^{\mu\nu})+\sum_{r\neq s}\rho_r(\Gamma^{[\mu})\rho_s(\Gamma^{\nu]}).
\eeq
The second term vanishes because of the symmetry between the tensor factors. The `physical' co-ordinates on the fuzzy sphere $x^\mu=\frac{r}{n}\gh^\mu$ are defined so that the dimensionful parameter $r$ gives the radius for large-$n$:
\beq
x^\mu x^\mu=r^2+O\left(\frac{1}{n}\right)\, .
\eeq
Operators with $l$ $\rho$'s acting on different tensor factors in general have eigenvalues of order $n^l$. Thus 
\beq
[x^\mu,x^\mu]=O\left(\frac{1}{n}\right)\, ,
\eeq
and at large-$n$ commutativity is restored. However, for the fuzzy three-sphere the same commutator gives
\beq\label{comm}
[G^{i}, G^{j}]\cPpm=2\sum_{r}\rho_r(\Gamma^{ij}P_\pm)\cPpm +\sum_{r\neq
  s}2\rho_r(\Gamma^{[i} P_\mp)\rho_s(\Gamma^{j]} P_\pm)\cPpm\, .
\eeq
As there is no symmetry between the different chirality tensor factors the second term does not vanish and we have
\beq
[x^\mu,x^\mu]=r^2\times O(1).
\eeq
The non-commutativity persists in the large-$n$ limit if we do not impose the projection. By allowing only symmetric diagrams the projection removes the second term of (\ref{comm}), which can be written in terms of diagrams with $(r_1,r_2)=(1,1)$ in $End(\cPpm)$ (equation (\ref{asdf})). It is the same lack of symmetry that leads to the need for the extra projection of Section \ref{sec:nonass}.

It is not obvious how to realise this projection as an operator. It can be done by using the Casimirs of $SO(5)$ (for the four-sphere). These will be functions of $r_1$ and $r_2$ and this relation can be inverted to write $r_2$ as an operator in terms of the Casimirs. The projection will be to the kernel of this operator. A way of realising the projected product would be to trace with the spherical harmonics $Y^*_{r_1,r_2}$ for $r_2=0$. The projected product is then given by
\beq
A\bullet B=\sum_{r_1}\mbox{Tr}(ABY^*_{r_1,0})Y_{r_1,0}\, .
\eeq

%% file: thesis.bbl
\begin{thebibliography}{100}

\bibitem{BasuH}
A.~Basu and J.~A.~Harvey,
``The M2-M5 brane system and a generalized Nahm's equation,''
Nucl.\ Phys.\ B {\bf 713} (2005) 136
[arXiv:hep-th/0412310].

\bibitem{Hull:2004in}
  C.~M.~Hull,
  ``A geometry for non-geometric string backgrounds,''
  JHEP {\bf 0510}, 065 (2005)
  [arXiv:hep-th/0406102].
  
\bibitem{Hull:2006va}
  C.~M.~Hull,
  ``Doubled geometry and T-folds,''
  arXiv:hep-th/0605149.

\bibitem{CMT}
N.~R.~Constable, R.~C.~Myers and O.~Tafjord,
``The noncommutative bion core,''
Phys.\ Rev.\ D {\bf 61} (2000) 106009
[arXiv:hep-th/9911136].

\bibitem{CL}
N.~R.~Constable and N.~D.~Lambert,
``Calibrations, monopoles and fuzzy funnels,''
Phys.\ Rev.\ D {\bf 66}, (2002) 065016
[arXiv:hep-th/0206243].

\bibitem{bermancopland}
D.~S.~Berman and N.~B.~Copland,
``Five-brane calibrations and fuzzy funnels,''
Nucl.\ Phys.\ B {\bf 723} (2005) 117
[arXiv:hep-th/0504044].

\bibitem{BC2}
  D.~S.~Berman and N.~B.~Copland,
  ``A note on the M2-M5 brane system and fuzzy spheres,''
  Phys.\ Lett.\ B {\bf 639} (2006) 553
  [arXiv:hep-th/0605086].
  

\bibitem{BC3}
  D.~S.~Berman and N.~B.~Copland,
  ``The string partition function in Hull's doubled formalism,''
  arXiv:hep-th/0701080.

\bibitem{CJS}
  E.~Cremmer, B.~Julia and J.~Scherk,
  ``Supergravity Theory In 11 Dimensions,''
  Phys.\ Lett.\ B {\bf 76}, 409 (1978).
  
\bibitem{Polchinski:1995mt}
  J.~Polchinski,
  ``Dirichlet-Branes and Ramond-Ramond Charges,''
  Phys.\ Rev.\ Lett.\  {\bf 75} (1995) 4724
  [arXiv:hep-th/9510017].

\bibitem{HuqN}
  M.~Huq and M.~A.~Namazie,
  ``Kaluza-Klein Supergravity In Ten-Dimensions,''
  Class.\ Quant.\ Grav.\  {\bf 2} (1985) 293
  [Erratum-ibid.\  {\bf 2} (1985) 597].

\bibitem{GianiP}
  F.~Giani and M.~Pernici,
  ``N=2 Supergravity In Ten-Dimensions,''
  Phys.\ Rev.\ D {\bf 30} (1984) 325.

\bibitem{Dirac}
  P.~A.~M.~Dirac,
  ``An Extensible Model Of The Electron,''
  Proc.\ Roy.\ Soc.\ Lond.\ A {\bf 268} (1962) 57.

\bibitem{Duff:1990xz}
  M.~J.~Duff and K.~S.~Stelle,
  ``Multi-Membrane Solutions Of D = 11 Supergravity,''
  Phys.\ Lett.\ B {\bf 253} (1991) 113.


\bibitem{Bergshoeff:1987cm}
  E.~Bergshoeff, E.~Sezgin and P.~K.~Townsend,
  ``Supermembranes And Eleven-Dimensional Supergravity,''
  Phys.\ Lett.\ B {\bf 189} (1987) 75.

\bibitem{Brink:1980a}
  L.~Brink and P.~S.~Howe,
  ``Eleven-Dimensional Supergravity On The Mass - Shell In Superspace,''
  Phys.\ Lett.\ B {\bf 91}, 384 (1980).

\bibitem{Cremmer:1980ru}
  E.~Cremmer and S.~Ferrara,
  ``Formulation Of Eleven-Dimensional Supergravity In Superspace,''
  Phys.\ Lett.\ B {\bf 91}, 61 (1980).

\bibitem{Duff:1987bx}
  M.~J.~Duff, P.~S.~Howe, T.~Inami and K.~S.~Stelle,
  ``Superstrings In D = 10 From Supermembranes In D = 11,''
  Phys.\ Lett.\ B {\bf 191} (1987) 70.

\bibitem{Townsend:1995af}
  P.~K.~Townsend,
  ``D-branes from M-branes,''
  Phys.\ Lett.\ B {\bf 373} (1996) 68
  [arXiv:hep-th/9512062].

\bibitem{Gueven}
  R.~Gueven,
  ``Black p-brane solutions of D = 11 supergravity theory,''
  Phys.\ Lett.\ B {\bf 276} (1992) 49.

\bibitem{PST}
  P.~Pasti, D.~P.~Sorokin and M.~Tonin,
  ``Covariant action for a D = 11 five-brane with the chiral field,''
  Phys.\ Lett.\ B {\bf 398} (1997) 41
  [arXiv:hep-th/9701037].

\bibitem{BLNPST}
  I.~A.~Bandos, K.~Lechner, A.~Nurmagambetov, P.~Pasti, D.~P.~Sorokin and M.~Tonin,
  ``Covariant action for the super-five-brane of M-theory,''
  Phys.\ Rev.\ Lett.\  {\bf 78} (1997) 4332
  [arXiv:hep-th/9701149].

\bibitem{PerryS}
  M.~Perry and J.~H.~Schwarz,
  ``Interacting chiral gauge fields in six dimensions and Born-Infeld
  theory,''
  Nucl.\ Phys.\ B {\bf 489} (1997) 47
  [arXiv:hep-th/9611065].

\bibitem{APPS1}
  M.~Aganagic, J.~Park, C.~Popescu and J.~H.~Schwarz,
  ``World-volume action of the M-theory five-brane,''
  Nucl.\ Phys.\ B {\bf 496} (1997) 191
  [arXiv:hep-th/9701166].

\bibitem{d11p5}
  P.~S.~Howe and E.~Sezgin,
  ``D = 11, p = 5,''
  Phys.\ Lett.\ B {\bf 394} (1997) 62
  [arXiv:hep-th/9611008].

\bibitem{cov5eom}
  P.~S.~Howe, E.~Sezgin and P.~C.~West,
  ``Covariant field equations of the M-theory five-brane,''
  Phys.\ Lett.\ B {\bf 399} (1997) 49
  [arXiv:hep-th/9702008].

\bibitem{APPS2}
  M.~Aganagic, J.~Park, C.~Popescu and J.~H.~Schwarz,
  ``Dual D-brane actions,''
  Nucl.\ Phys.\ B {\bf 496} (1997) 215
  [arXiv:hep-th/9702133].

\bibitem{Berman53}
  D.~Berman,
  ``M5 on a torus and the three brane,''
  Nucl.\ Phys.\ B {\bf 533} (1998) 317
  [arXiv:hep-th/9804115].

\bibitem{WAG}
  L.~Alvarez-Gaume and E.~Witten,
  ``Gravitational Anomalies,''
  Nucl.\ Phys.\ B {\bf 234}, 269 (1984).

\bibitem{VafaW}
  C.~Vafa and E.~Witten,
  ``A One loop test of string duality,''
  Nucl.\ Phys.\ B {\bf 447}, 261 (1995)
  [arXiv:hep-th/9505053].
  
\bibitem{Duff:1995wd}
  M.~J.~Duff, J.~T.~Liu and R.~Minasian,
  ``Eleven-dimensional origin of string / string duality: A one-loop test,''
  Nucl.\ Phys.\ B {\bf 452} (1995) 261
  [arXiv:hep-th/9506126].

\bibitem{Witten5}
  E.~Witten,
  ``Five-brane effective action in M-theory,''
  J.\ Geom.\ Phys.\  {\bf 22}, 103 (1997)
  [arXiv:hep-th/9610234].

\bibitem{FHMM}
  D.~Freed, J.~A.~Harvey, R.~Minasian and G.~W.~Moore,
  ``Gravitational anomaly cancellation for M-theory fivebranes,''
  Adv.\ Theor.\ Math.\ Phys.\  {\bf 2}, 601 (1998)
  [arXiv:hep-th/9803205].

\bibitem{Strominger:1995ac}
  A.~Strominger,
  ``Open p-branes,''
  Phys.\ Lett.\ B {\bf 383} (1996) 44
  [arXiv:hep-th/9512059].

\bibitem{HLW}
P.~S.~Howe, N.~D.~Lambert and P.~C.~West,
``The self-dual string soliton,''
Nucl.\ Phys.\ B {\bf 515} (1998) 203
[arXiv:hep-th/9709014].

\bibitem{3brane}
  P.~S.~Howe, N.~D.~Lambert and P.~C.~West,
  ``The threebrane soliton of the M-fivebrane,''
  Phys.\ Lett.\ B {\bf 419} (1998) 79
  [arXiv:hep-th/9710033].

\bibitem{revisited}
  P.~K.~Townsend,
  ``The eleven-dimensional supermembrane revisited,''
  Phys.\ Lett.\ B {\bf 350} (1995) 184
  [arXiv:hep-th/9501068].
  
\bibitem{Giveon:1994fu}
  A.~Giveon, M.~Porrati and E.~Rabinovici,
  ``Target space duality in string theory,''
  Phys.\ Rept.\  {\bf 244} (1994) 77
  [arXiv:hep-th/9401139].

\bibitem{T4lecs}
  P.~K.~Townsend,
  ``Four lectures on M-theory,''
  arXiv:hep-th/9612121.

\bibitem{BHO}
  E.~Bergshoeff, C.~M.~Hull and T.~Ortin,
  ``Duality in the type II superstring effective action,''
  Nucl.\ Phys.\ B {\bf 451}, 547 (1995)
  [arXiv:hep-th/9504081].

  P.~S.~Aspinwall,
  ``Some relationships between dualities in string theory,''
  Nucl.\ Phys.\ Proc.\ Suppl.\  {\bf 46} (1996) 30
  [arXiv:hep-th/9508154].

  J.~H.~Schwarz,
  ``The power of M theory,''
  Phys.\ Lett.\ B {\bf 367} (1996) 97
  [arXiv:hep-th/9510086].

\bibitem{K3}
  P.~K.~Townsend,
  ``String - membrane duality in seven-dimensions,''
  Phys.\ Lett.\ B {\bf 354}, 247 (1995)
  [arXiv:hep-th/9504095].

\bibitem{HW}
  P.~Horava and E.~Witten,
  ``Heterotic and type I string dynamics from eleven dimensions,''
  Nucl.\ Phys.\ B {\bf 460}, 506 (1996)
  [arXiv:hep-th/9510209].

\bibitem{Duff96}
  M.~J.~Duff, J.~T.~Liu and J.~Rahmfeld,
  ``Four-dimensional string-string-string triality,''
  Nucl.\ Phys.\ B {\bf 459}, 125 (1996)
  [arXiv:hep-th/9508094].

  M.~J.~Duff, R.~Minasian and E.~Witten,
  ``Evidence for Heterotic/Heterotic Duality,''
  Nucl.\ Phys.\ B {\bf 465} (1996) 413
  [arXiv:hep-th/9601036].

\bibitem{HullT}
  C.~M.~Hull and P.~K.~Townsend,
  ``Unity of superstring dualities,''
  Nucl.\ Phys.\ B {\bf 438}, 109 (1995)
  [arXiv:hep-th/9410167].

\bibitem{Duffsw}
  M.~J.~Duff,
  ``Strong / weak coupling duality from the dual string,''
  Nucl.\ Phys.\ B {\bf 442}, 47 (1995)
  [arXiv:hep-th/9501030].

\bibitem{Witten:1995ex}
  E.~Witten,
  ``String theory dynamics in various dimensions,''
  Nucl.\ Phys.\ B {\bf 443}, 85 (1995)
  [arXiv:hep-th/9503124].

\bibitem{Schwarz:1996bh}
  J.~H.~Schwarz,
  ``Lectures on superstring and M theory dualities,''
  Nucl.\ Phys.\ Proc.\ Suppl.\  {\bf 55B} (1997) 1
  [arXiv:hep-th/9607201].

\bibitem{DuffMth}
  M.~J.~Duff,
  ``M theory (the theory formerly known as strings),''
  Int.\ J.\ Mod.\ Phys.\ A {\bf 11} (1996) 5623
  [arXiv:hep-th/9608117].

\bibitem{BFSS}
  T.~Banks, W.~Fischler, S.~H.~Shenker and L.~Susskind,
  ``M theory as a matrix model: A conjecture,''
  Phys.\ Rev.\ D {\bf 55} (1997) 5112
  [arXiv:hep-th/9610043].

\bibitem{Maldacena}
  J.~M.~Maldacena,
  ``The large N limit of superconformal field theories and supergravity,''
  Adv.\ Theor.\ Math.\ Phys.\  {\bf 2} (1998) 231
  [Int.\ J.\ Theor.\ Phys.\  {\bf 38} (1999) 1113]
  [arXiv:hep-th/9711200].

\bibitem{blackholes}
  I.~R.~Klebanov and A.~A.~Tseytlin,
  ``Intersecting M-branes as four-dimensional black holes,''
  Nucl.\ Phys.\ B {\bf 475}, 179 (1996)
  [arXiv:hep-th/9604166].

\bibitem{Nahm}
W.~Nahm,
``A Simple Formalism For The Bps Monopole,''
Phys.\ Lett.\ B {\bf 90}, 413 (1980).

\bibitem{Callan}
  C.~G.~.~Callan and J.~M.~Maldacena,
  ``Brane dynamics from the Born-Infeld action,''
  Nucl.\ Phys.\ B {\bf 513}, 198 (1998)
  [arXiv:hep-th/9708147].

\bibitem{Gibbons}
G.~W.~Gibbons,
``Born-Infeld particles and Dirichlet p-branes,''
Nucl.\ Phys.\ B {\bf 514} (1998) 603
[arXiv:hep-th/9709027].

\bibitem{Myers}
  R.~C.~Myers,
  ``Dielectric-branes,''
  JHEP {\bf 9912}, 022 (1999)
  [arXiv:hep-th/9910053].

\bibitem{Kastor}
  D.~Kastor and J.~H.~Traschen,
  ``Dynamics of the DBI spike soliton,''
  Phys.\ Rev.\ D {\bf 61} (2000) 024034
  [arXiv:hep-th/9906237].

\bibitem{Lee}
  S.~M.~Lee, A.~W.~Peet and L.~Thorlacius,
  ``Brane-waves and strings,''
  Nucl.\ Phys.\ B {\bf 514} (1998) 161
  [arXiv:hep-th/9710097].

\bibitem{Fairlieetal}
E.~Corrigan, C.~Devchand, D.~B.~Fairlie and J.~Nuyts,
``First Order Equations For Gauge Fields In Spaces Of Dimension Greater Than
Four,''
Nucl.\ Phys.\ B {\bf 214}, 452 (1983).

\bibitem{Harvey:1982xk}
R.~Harvey and H.~B.~Lawson,
``Calibrated Geometries,''
Acta Math.\  {\bf 148}, 47 (1982).

\bibitem{GP}
G.~W.~Gibbons and G.~Papadopoulos,
``Calibrations and intersecting branes,''
Commun.\ Math.\ Phys.\  {\bf 202}, 593 (1999)
[arXiv:hep-th/9803163].

\bibitem{Jerome}
J.~P.~Gauntlett, N.~D.~Lambert and P.~C.~West,
``Branes and calibrated geometries,''
Commun.\ Math.\ Phys.\  {\bf 202} (1999) 571
[arXiv:hep-th/9803216].


\bibitem{Diaconescu}
D.~E.~Diaconescu,
``D-branes, monopoles and Nahm equations,''
Nucl.\ Phys.\ B {\bf 503}, 220 (1997)
[arXiv:hep-th/9608163].

\bibitem{Tsimpis}
D.~Tsimpis,
``Nahm equations and boundary conditions,''
Phys.\ Lett.\ B {\bf 433}, 287 (1998)
[arXiv:hep-th/9804081].

\bibitem{Manton}
  N.~S.~Manton and P.~Sutcliffe,
  ``Topological solitons,'' Cambridge University Press (2004) 493 p

\bibitem{Hashimoto}
  K.~Hashimoto and S.~Terashima,
  ``Stringy derivation of Nahm construction of monopoles,''
  JHEP {\bf 0509} (2005) 055
  [arXiv:hep-th/0507078].

\bibitem{Tseytlin}
  A.~A.~Tseytlin,
  ``On non-abelian generalisation of the Born-Infeld action in string
  theory,''
  Nucl.\ Phys.\ B {\bf 501}, 41 (1997)
  [arXiv:hep-th/9701125].

\bibitem{BrecherP}
  D.~Brecher and M.~J.~Perry,
  ``Bound States Of D-Branes And The Non-Abelian Born-Infeld Action,''
  Nucl.\ Phys.\ B {\bf 527} (1998) 121
  [arXiv:hep-th/9801127].

\bibitem{Brecher}
D.~Brecher,
``BPS states of the non-Abelian Born-Infeld action,''
Phys.\ Lett.\ B {\bf 442}, 117 (1998)
[arXiv:hep-th/9804180].

\bibitem{Lambert}
N.~D.~Lambert,
``Moduli and brane intersections,''
Phys.\ Rev.\ D {\bf 67}, 026006 (2003)
[arXiv:hep-th/0209141].

\bibitem{Helling}
J.~Erdmenger, Z.~Guralnik, R.~Helling and I.~Kirsch,
``A world-volume perspective on the recombination of intersecting branes,''
JHEP {\bf 0404}, 064 (2004)
[arXiv:hep-th/0309043].

\bibitem{CMT2}
N.~R.~Constable, R.~C.~Myers and O.~Tafjord,
``Non-Abelian brane intersections,''
JHEP {\bf 0106}, 023 (2001)
[arXiv:hep-th/0102080].

\bibitem{CMT3}
N.~R.~Constable, R.~C.~Myers and O.~Tafjord,
``Fuzzy funnels: Non-abelian brane intersections,''
arXiv:hep-th/0105035.

\bibitem{Cook}
  P.~Cook, R.~de Mello Koch and J.~Murugan,
  ``Non-Abelian BIonic brane intersections,''
  Phys.\ Rev.\ D {\bf 68}, 126007 (2003)
  [arXiv:hep-th/0306250].

\bibitem{Berman}
D.~S.~Berman,
``Aspects of M-5 brane world volume dynamics,''
Phys.\ Lett.\ B {\bf 572} (2003) 101
[arXiv:hep-th/0307040].

\bibitem{Nambu}
  Y.~Nambu,
  ``Generalized Hamiltonian dynamics,''
  Phys.\ Rev.\ D {\bf 7}, 2405 (1973).
\bibitem{CZ}
  T.~L.~Curtright and C.~K.~Zachos,
  ``Branes, strings, and odd quantum Nambu brackets,''
  arXiv:hep-th/0312048.

\bibitem{Nogradi}
D.~Nogradi,
``M2-branes stretching between M5-branes,''
JHEP {\bf 0601} (2006) 010
[arXiv:hep-th/0511091].

\bibitem{KT}
  I.~R.~Klebanov and A.~A.~Tseytlin,
  ``Entropy of Near-Extremal Black p-branes,''
  Nucl.\ Phys.\ B {\bf 475}, 164 (1996)
  [arXiv:hep-th/9604089].

\bibitem{Kleb}
  I.~R.~Klebanov,
  ``World-Volume Approach To Absorption By Non-Dilatonic Branes,''
  Nucl.\ Phys.\ B {\bf 496}, 231 (1997)
  [arXiv:hep-th/9702076].

\bibitem{AGMOO}
  O.~Aharony, S.~S.~Gubser, J.~M.~Maldacena, H.~Ooguri and Y.~Oz,
  ``Large N field theories, string theory and gravity,''
  Phys.\ Rept.\  {\bf 323} (2000) 183
  [arXiv:hep-th/9905111].


\bibitem{baggerandlambert}
  J.~Bagger and N.~Lambert,
  ``Modeling multiple M2's,''
  arXiv:hep-th/0611108.

\bibitem{Acharya:1998en}
  B.~S.~Acharya, J.~M.~Figueroa-O'Farrill and B.~Spence,
  ``Branes at angles and calibrated geometry,''
  JHEP {\bf 9804} (1998) 012
  [arXiv:hep-th/9803260].

\bibitem{Hofman:2001zt}
C.~M.~Hofman and W.~K.~Ma,
``Deformations of closed strings and topological open membranes,''
JHEP {\bf 0106} (2001) 033
[arXiv:hep-th/0102201].

\bibitem{AlekseevRS}
  A.~Y.~Alekseev, A.~Recknagel and V.~Schomerus,
  ``Non-commutative world-volume geometries: Branes on SU(2) and fuzzy
  spheres,''
  JHEP {\bf 9909} (1999) 023
  [arXiv:hep-th/9908040].
  
\bibitem{CLT}
J.~Castelino, S.~M.~Lee and W.~I.~Taylor,
``Longitudinal 5-branes as 4-spheres in matrix theory,''
Nucl.\ Phys.\ B {\bf 526} (1998) 334
[arXiv:hep-th/9712105].

\bibitem{Ram1}
Z.~Guralnik and S.~Ramgoolam,
``On the polarization of unstable D0-branes into non-commutative odd
spheres,''
JHEP {\bf 0102}, 032 (2001)
[arXiv:hep-th/0101001].

\bibitem{Ram2}
S.~Ramgoolam,
``On spherical harmonics for fuzzy spheres in diverse dimensions,''
Nucl.\ Phys.\ B {\bf 610}, 461 (2001)
[arXiv:hep-th/0105006].

\bibitem{Ram3}
S.~Ramgoolam,
``Higher dimensional geometries related to fuzzy odd-dimensional spheres,''
JHEP {\bf 0210}, 064 (2002)
[arXiv:hep-th/0207111].


\bibitem{Papageorgakis:2006ed}
C.~Papageorgakis and S.~Ramgoolam,
``On time-dependent collapsing branes and fuzzy odd-dimensional spheres,''
arXiv:hep-th/0603239.


\bibitem{tfold1}
  C.~M.~Hull,
  ``Global aspects of T-duality, gauged sigma models and T-folds,''
  arXiv:hep-th/0604178.

  A.~Dabholkar and C.~Hull,
  ``Generalised T-duality and non-geometric backgrounds,''
  JHEP {\bf 0605} (2006) 009
  [arXiv:hep-th/0512005].

  J.~Shelton, W.~Taylor and B.~Wecht,
  ``Nongeometric flux compactifications,''
  JHEP {\bf 0510} (2005) 085
  [arXiv:hep-th/0508133].

  J.~Shelton, W.~Taylor and B.~Wecht,
  ``Generalized flux vacua,''
  arXiv:hep-th/0607015.

  C.~M.~Hull and R.~A.~Reid-Edwards,
  ``Flux compactifications of string theory on twisted tori,''
  arXiv:hep-th/0503114.

  K.~Becker, M.~Becker, C.~Vafa and J.~Walcher,
  ``Moduli stabilization in non-geometric backgrounds,''
  arXiv:hep-th/0611001.
  
\bibitem{Hull:2006qs}
  C.~M.~Hull,
  ``Global aspects of T-duality, gauged sigma models and T-folds,''
  arXiv:hep-th/0604178.

\bibitem{Cremmer:1997ct}
  E.~Cremmer, B.~Julia, H.~Lu and C.~N.~Pope,
  ``Dualisation of dualities. I,''
  Nucl.\ Phys.\ B {\bf 523}, 73 (1998)
  [arXiv:hep-th/9710119].
  
  
\bibitem{doublo}
  T.~Kugo and B.~Zwiebach,
  ``Target space duality as a symmetry of string field theory,''
  Prog.\ Theor.\ Phys.\  {\bf 87}, 801 (1992)
  [arXiv:hep-th/9201040].

  A.~A.~Tseytlin,
  ``Duality Symmetric Formulation Of String World Sheet Dynamics,''
  Phys.\ Lett.\ B {\bf 242}, 163 (1990).

  A.~A.~Tseytlin,
  ``Duality Symmetric Closed String Theory And Interacting Chiral Scalars,''
  Nucl.\ Phys.\ B {\bf 350}, 395 (1991).

  J.~Maharana and J.~H.~Schwarz,
  ``Noncompact symmetries in string theory,''
  Nucl.\ Phys.\ B {\bf 390}, 3 (1993)
  [arXiv:hep-th/9207016].

  C.~M.~Hull,
  ``Covariant Quantiztaion of Chiral Bosons and Anomaly Cancellation,''
  Phys.\ Lett.\ B {\bf 206}, 234 (1988).

  C.~M.~Hull,
  ``Chiral Conformal Field Theory And Asymmetric String Compactification,''
  Phys.\ Lett.\ B {\bf 212}, 437 (1988).

\bibitem{emily}
  E.~Hackett-Jones and G.~Moutsopoulos,
  ``Quantum mechanics of the doubled torus,''
  JHEP {\bf 0610}, 062 (2006)
  [arXiv:hep-th/0605114].
  
  
\bibitem{walcher}
  S.~Hellerman and J.~Walcher,
  ``Worldsheet CFTs for flat monodrofolds,''
  arXiv:hep-th/0604191.
  
  
\bibitem{Hitchin:2004ut}
  N.~Hitchin,
  ``Generalized Calabi-Yau manifolds,''
  Quart.\ J.\ Math.\ Oxford Ser.\  {\bf 54}, 281 (2003)
  [arXiv:math.dg/0209099].
  
\bibitem{Buscher:1987sk}
  T.~H.~Buscher,
  ``A Symmetry of the String Background Field Equations,''
  Phys.\ Lett.\ B {\bf 194}, 59 (1987).




  
\bibitem{ABMNV}
  L.~Alvarez-Gaume, J.~B.~Bost, G.~W.~Moore, P.~C.~Nelson and C.~Vafa,
  ``Bosonization on higher genus Riemann surfaces,''
  Commun.\ Math.\ Phys.\  {\bf 112} (1987) 503.
  
  
\bibitem{HNS}
M.~Henningson, B.~E.~W.~Nilsson and P.~Salomonson,
  ``Holomorphic factorization of correlation functions in  (4k+2)-dimensional
  (2k)-form gauge theory,''
  JHEP {\bf 9909} (1999) 008
  [arXiv:hep-th/9908107].

\bibitem{DVV}
  R.~Dijkgraaf, E.~P.~Verlinde and M.~Vonk,
  ``On the partition sum of the NS five-brane,''
  arXiv:hep-th/0205281.
  
\bibitem{moore}
  D.~Belov and G.~W.~Moore,
  ``Holographic action for the self-dual field,''
  arXiv:hep-th/0605038.


\bibitem{polchinski}
J.~Polchinski,
  ``Evaluation Of The One Loop String Path Integral,''
  Commun.\ Math.\ Phys.\  {\bf 104} (1986) 37.
  
\bibitem{ginsparg}
  P.~H.~Ginsparg,
  ``Applied conformal field theory,''
  arXiv:hep-th/9108028.




\bibitem{gerbe}
  N.~J.~Hitchin, ``Lectures on special Lagrangian submanifolds,'' arXiv:math.dg/9907034.


\bibitem{BGT}
  E.~A.~Bergshoeff, G.~W.~Gibbons and P.~K.~Townsend,
  ``Open M5-branes,''
  arXiv:hep-th/0607193.



\bibitem{Madore}
  J.~Madore,
  ``The fuzzy sphere,''
  Class.\ Quant.\ Grav.\  {\bf 9} (1992) 69.

\bibitem{dWHN}
  B.~de Wit, J.~Hoppe and H.~Nicolai,
  ``On the quantum mechanics of supermembranes,''
  Nucl.\ Phys.\ B {\bf 305}, 545 (1988).

\bibitem{Maldacena:1998bw}
  J.~M.~Maldacena and A.~Strominger,
  ``AdS(3) black holes and a stringy exclusion principle,''
  JHEP {\bf 9812}, 005 (1998)
  [arXiv:hep-th/9804085].

\bibitem{Jevicki:1998rr}
  A.~Jevicki and S.~Ramgoolam,
  ``Non-commutative gravity from the AdS/CFT correspondence,''
  JHEP {\bf 9904}, 032 (1999)
  [arXiv:hep-th/9902059].

\bibitem{Ho}
  P.~M.~Ho,
  ``Fuzzy sphere from matrix model,''
  JHEP {\bf 0012}, 015 (2000)
  [arXiv:hep-th/0010165].

\bibitem{SJ1}
M.~M.~Sheikh-Jabbari,
``Tiny graviton matrix theory: DLCQ of IIB plane-wave string theory, a
conjecture,''
JHEP {\bf 0409}, 017 (2004)
[arXiv:hep-th/0406214].

\bibitem{SJ2}
M.~M.~Sheikh-Jabbari and M.~Torabian,
``Classification of all 1/2 BPS solutions of the tiny graviton matrix theory,''
arXiv:hep-th/0501001.

\bibitem{Bha2}
  R.~Bhattacharyya,
  ``A short note on multi-bion solutions,''
  arXiv:hep-th/0505103.

\bibitem{Bhattacharyya:2005cd}
  R.~Bhattacharyya and R.~de Mello Koch,
  ``Fluctuating fuzzy funnels,''
  JHEP {\bf 0510} (2005) 036
  [arXiv:hep-th/0508131].

\bibitem{Bha3}
  R.~Bhattacharyya and J.~Douari,
  ``Brane intersections in the presence of a worldvolume electric field,''
  JHEP {\bf 0512} (2005) 012
  [arXiv:hep-th/0509023].
  
\bibitem{Thomas:2006ac}
  S.~Thomas and J.~Ward,
  ``Electrified fuzzy spheres and funnels in curved backgrounds,''
  arXiv:hep-th/0602071.

\bibitem{MPRS}
  S.~McNamara, C.~Papageorgakis, S.~Ramgoolam and B.~Spence,
  ``Finite N effects on the collapse of fuzzy spheres,''
  JHEP {\bf 0605} (2006) 060
  [arXiv:hep-th/0512145].

\bibitem{SJN}
 M.~M.~Sheikh-Jabbari,
 ``Inherent Holography In Fuzzy Spaces And An N-Tropic Approach To The
 Cosmological Constant Problem,''
 arXiv:hep-th/0605110.

\bibitem{Connes}
  A.~Connes,
  ``Noncommutative geometry,'' Academic Press (1994) 661p

\bibitem{GKP}
  H.~Grosse, C.~Klimcik and P.~Presnajder,
  ``Finite quantum field theory in noncommutative geometry,''
  Commun.\ Math.\ Phys.\  {\bf 180} (1996) 429
  [arXiv:hep-th/9602115].

\end{thebibliography}
